\documentclass[11pt,a4paper]{article}
\pdfoutput=1

\usepackage{lmodern}
\usepackage{verbatim,setspace}
\usepackage[usenames, dvipsnames, svgnames]{xcolor}
\usepackage[rawfloats=true]{floatrow}

\usepackage{dsfont}

\def\cN{{\mathcal{N}}}

\renewcommand\Im{\hbox{{\rm Im}}\,}
\renewcommand\Re{\hbox{{\rm Re}}\,}

\newcommand{\CR}{\nonumber \\*}

\def\cV{{\mathcal V}}
\def\cK{{\mathcal K}}
\newcommand{\ee}{\mathrm{e}}
\def\del{\partial}
\newcommand{\rr}{\mathbb{R}}
\newcommand{\cc}{\mathbb{C}}
\newcommand{\pp}{\mathbb{P}}

\newcommand{\Iprod}[2]{\langle {#1}, {#2} \rangle}
\newcommand{\im}{\mathrm{i}}

\newcommand{\se}{{\scalebox{0.8}{$\scriptscriptstyle SE$}}}
\newcommand{\Kah}{K\"ahler }
\newcommand{\KE}{K\"ahler--Einstein }
\newcommand{\nv}{n_{{\scriptscriptstyle V}}}

\newcommand{\Um}{\mathrm{U}}
\newcommand{\Ums}{{\scriptscriptstyle \Um}}
\newcommand{\tgd}{{\partial_{\scriptscriptstyle B}}}
\newcommand{\tgdb}{{{\bar{\partial}_{\scriptscriptstyle B}}}}
\newcommand{\stgd}{{{\scalebox{0.75}{${{\partial_{\scriptscriptstyle B}}}$}}}}
\newcommand{\stgdb}{{{\scalebox{0.75}{${{\bar{\partial}_{\scriptscriptstyle B}}}$}}}}

\newcommand{\?}{\;\!}
\newcommand{\jac}{{\mathsf J}}
\newcommand{\jst}{{\mathrm j_{\scriptscriptstyle *}}}

\def\La{\Lambda}
\def\Si{\Sigma}
\newcommand{\KahO}{\mathrm{K}_\Omega} 
\newcommand{\KahJ}{K_{\mkern-1.5mu J}} 

\newcommand{\HE}[1]{{\textcolor{blue}{#1}}}
\newcommand{\HI}[1]{{\textcolor{red}{#1}}}

\makeatletter
\newcommand{\thickhline}{%
    \noalign {\ifnum 0=`}\fi \hrule height 1.3pt
    \futurelet \reserved@a \@xhline
}

\usepackage[no-natbib-sort]{jheppub-mod}
\preprint{IPhT-T17/178}
\title{Gauged supergravities from M-theory reductions}
\author[a,b]{Stefanos Katmadas}  \author[c]{and Alessandro Tomasiello}
\affiliation[a]{
Institut de Physique Th\'eorique, Universit\'e Paris Saclay, CEA, CNRS, \\ 
\hspace{.14cm} 91191 Gif sur Yvette, France}
\affiliation[b]{Instituut voor Theoretische Fysica, Katholieke Universiteit Leuven\\
Celestijnenlaan 200D, B-3001 Leuven, Belgium}
\affiliation[c]{Dipartimento di Fisica, Universit\'a di Milano-Bicocca, I-20126 Milano, Italy \\ INFN, sezione di Milano-Bicocca, I-20126 Milano, Italy}

 \emailAdd{stefanos.katmadas [at] kuleuven.be}
 \emailAdd{alessandro.tomasiello [at] unimib.it}

\abstract{In supergravity compactifications, there is in general no clear prescription on how to select a finite-dimensional family of metrics on the internal space, and a family of forms on which to expand the various potentials, such that the lower-dimensional effective theory is supersymmetric. 
We propose a finite-dimensional family of deformations for regular Sasaki--Einstein seven-manifolds $M_7$, relevant for M-theory compactifications down to four dimensions. It consists of integrable Cauchy--Riemann structures, corresponding to complex deformations of the Calabi--Yau cone $M_8$ over $M_7$. 
The non-harmonic forms we propose are the ones contained in one of the Kohn--Rossi cohomology groups, which is finite-dimensional and naturally controls the deformations of Cauchy--Riemann structures. 
The same family of deformations can be also described in terms of twisted cohomology of the base $M_6$, or in terms of Milnor cycles arising in deformations of $M_8$. 
Using existing results on SU(3) structure compactifications, we briefly discuss the reduction of M-theory on our class of deformed Sasaki--Einstein manifolds to four-dimensional gauged supergravity.}

\begin{document}

 \maketitle

\section{Introduction}

Compactifications of superstring/M-theory to lower dimensions are often treated in terms of a reduction to lower-dimensional effective theories.
Focusing on the low-energy regime, where the massless modes are described in terms of supergravity theories in ten and eleven dimensions, there has been a long-standing effort to understand the possible reductions to lower-dimensional supergravities, which lend themselves to a simpler treatment.

The clearest example of this approach is given by Calabi--Yau compactifications. 
The fields of the higher-dimensional supergravity are expanded on the harmonic forms present on the Calabi--Yau manifold, leading to the various fields contained in the multiplets of the lower-dimensional supergravity. Here, the number of such harmonic forms fixes the number of multiplets in the reduced theory. The common origin of all such compactifications in the assumption of vanishing internal fluxes implies strong restrictions on the lower dimension theory: all multiplets are uncharged under the gauge fields and the supergravity potential vanishes.

When fluxes are added, it is not straightforward to give a principle that determines the type of internal manifold. The existence of a supersymmetric vacuum leads to conditions on the internal manifold; e.g. for Type II compactifications to four dimensions, the internal manifold should admit an SU(3) structure \cite{gurrieri-louis-micu-waldram} (or its $T \oplus T^*$ counterpart \cite{grana-louis-waldram}), together with a system of differential conditions that generalize the special holonomy condition. To obtain a lower-dimensional supergravity, however, one wants a family of internal metrics, which may or may not contain a metric leading to a supersymmetric vacuum. This is sometimes called a ``nonlinear'' reduction, as opposed to a ``linearised'' one, which only looks at infinitesimal fluctuations around a given solution. It is natural to require again the presence of a G-structure, but this by itself is a very weak constraint, which leads to an infinite-dimensional family of metrics. It is not a priori clear how to select a finite-dimensional subfamily. 
Related to this, the space of forms to be used in the reduction is now no longer restricted to the space of harmonic forms, but has to be enlarged to include non-closed forms. The appropriate space of forms has to obey some stringent constraints (which were spelled out in \cite{kashanipoor-minasian} for type II theories), but in general there is no clear strategy on how to solve those.

In spite of all these difficulties, there are some examples where the approach based on SU(3) structures does work, and one indeed obtains reductions consistent with supersymmetry and the known structure of gauged supergravity. One idea is to take as expansion forms the forms defining the SU(3) structure themselves; it works well when they satisfy some simple differential conditions relating them to each other. For type II this means \cite{kashanipoor} taking a real two-form $J$ and a complex three-form $\Omega$, related to each other in a so-called nearly K\"ahler structure (namely, a manifold whose cone has $G_2$ holonomy). For M-theory \cite{gauntlett-kim-varela-waldram}, the SU(3) structure requires also a one-form $\eta$, and the differential conditions define a Sasaki--Einstein manifold (a manifold whose cone is a Calabi--Yau manifold). The family of metrics is very simple in this case, corresponding to a change of the overall volume.

Any strategy to obtain compactifications where finer data on the internal manifold are probed needs to include a prescription for the appropriate additional forms on which to expand the SU(3) structure forms. 
One possibility is to take the internal space to be a coset $G/H$; the relevant forms are then identified with the set of left-invariant forms. This was done in \cite{cassani-kashanipoor} for type II, and in 
\cite{cassani-koerber-varela} for M-theory, see also \cite{Bena:2010pr, Cassani:2010na} for a five-dimensional example.
Unfortunately, it is difficult to infer from such examples an intrinsic characterization of a more general class of manifolds sharing the same properties.
Another point of view is to remain agnostic on the characterisation of the internal manifold, and only assume that it admits a set of two- and three-forms with convenient properties; this essentially boils down to the requirement that they close under the exterior differential and Hodge duality. The reduction can then proceed in a very similar way as for the Calabi--Yau case, where the parameters describing the non-closure of the forms are viewed as charges, or gauging parameters, for the multiplets in the lower-dimensional theory. This approach was taken for example in \cite{gurrieri-louis-micu-waldram, DAuria:2004kwe, house-palti} for IIA compactifications and in \cite{Micu:2006ey,cassani-koerber-varela} for M-theory.

In this paper, we propose another class of solutions to this supersymmetric family problem. We will put forward a concrete proposal for a class of seven-dimensional manifolds with an SU(3) structure and a natural finite-dimensional set of forms defined on them. They are deformations of regular Sasaki--Einstein manifolds, i.e.~those that can be described as a U(1) fibration over a \KE base $M_6$. In other words, the Sasaki--Einstein metric is a point in our family of metrics. 

Our class can be described in several equivalent ways. The strong presence of algebraic-geometric techniques makes it in a sense an AdS analogue of a Calabi--Yau reduction. One way to describe our supersymmetric family is as ``links'', $M_7$, around complex deformations of a non-compact eight-dimensional Calabi--Yau cone $M_8$, constructed as a complete intersection; namely, $M_7$ is obtained by intersecting the deformed $M_8$ with a large sphere. The Sasaki--Einstein metric on $M_7$ is recovered when the complex deformation is turned off and $M_8$ is conical. The complex deformations of $M_8$ are naturally described in terms of its middle-dimensional cohomology, which can be computed in a simple algebraic fashion from the properties of the singularity. In particular, the overall number of such deformations is equal to the Milnor number, $\mu$, of the singularity. The induced deformations on the link $M_7$ also admit an intrinsically seven-dimensional description, in terms of so-called Cauchy--Riemann (CR) structures. 

The set of forms we propose can also be described in several ways. 
One is as a finite-dimensional Kohn--Rossi (KR) cohomology group of the CR structure on $M_7$. 
(Other KR cohomologies, of infinite dimension, were used in earlier work on KK reduction on Sasaki--Einstein manifolds \cite{eager-schmude-tachikawa,eager-schmude}.) Another is as a sum of twisted cohomologies on $M_6$. Finally, our forms are related to de Rham and relative middle-dimensional cohomologies on $M_8$. All these descriptions are useful in different ways.

While we make no claim of achieving full mathematical rigor, using these complementary points of view we provide strong evidence that the forms we propose satisfy the M-theory analogues of the conditions in \cite{kashanipoor-minasian}. That means they can be used in compactifications to a four-dimensional gauged $\cN=2$ supergravity, thus generalizing the constructions reviewed above. In particular we provide an extension of the Sasaki--Einstein compactifications where several charged hypermultiplets are present. 

On the other hand, one issue we do not address in this paper is the consistency of our compactification. This is the property that every solution of the four-dimensional theory can be lifted to a solution of the higher-dimensional one. It means that the equations of motion of any modes that have not been kept go to zero on the ``supersymmetric family'' one is considering. Consistent compactifications were once rare, but are now a lot more common; 
the truncations in both \cite{gauntlett-kim-varela-waldram} and \cite{cassani-koerber-varela} are consistent. 
This issue is not so pressing for compactifications that have Minkowski vacua, which are usually not consistent but which can be physically justified by arguing that the modes which have been kept in the compactification are much lighter than those which have not.
For compactifications with AdS vacua, however, the spectrum usually has no ``separation of scales'', and the usefulness of a non-consistent compactification is debatable. 

In our case, the modes that are kept can be viewed as deformations of the six-dimensional base of the internal manifold, making our compactification analogous to that on a Calabi--Yau, which in the Minkowski case does not lead to a consistent reduction. We argue that, even if strict consistency is not achieved in our reductions, one may use this structure to organise the eigenmodes of the internal Laplacian in terms of the Laplacian associated to the CR structure, with the modes considered in this paper belonging to the lowest eigenvalue.

It would of course be very interesting to clarify this point. As a limited piece of evidence that our reduction does capture some of the eleven-dimensional physics in a useful way, we notice that in our compactification all complex deformations of the K\"ahler--Einstein base $M_6$ are automatically moduli of the solution; this can be confirmed using recent mathematical results \cite{szekelyhidi}. In any case, we hope that the kind of techniques we are introducing in this paper will be useful for other reductions as well.

This paper is organized as follows. Section \ref{sec:CS-man} serves as an extended introduction, as it contains a general discussion of the issues arising in nonlinear reduction of higher-dimensional theories and introduces the particular class of SU(3) structures on regular Sasaki--Einstein manifolds we aim to realize in this paper. 
In particular, in section \ref{sec:3form-defs} we give an overview of our approach to the three-form deformations of regular Sasaki--Einstein manifolds, presented in sections \ref{sec:cmplx-defs}, \ref{sec:CR-cone} and \ref{sec:CR-struct} through three different and complementary routes. Section \ref{sec:cmplx-defs} focuses on the definition of a Sasaki--Einstein manifold as a U(1) bundle over a \KE space and deals with the possible charged $(2,1)$ forms on a class of such manifolds. In view of the definition of a Sasaki--Einstein manifold as a link around an isolated singularity of a Calabi--Yau cone, section \ref{sec:CR-cone} provides a concise discussion on the deformations of isolated singularities and the (co)homology of the resulting geometries. In section \ref{sec:CR-struct} we consider the implications of these structures on the deformations induced on the link around the singularity and provide their description in terms of CR structure deformations. The results are shown to agree with the ones derived in section \ref{sec:cmplx-defs} at the Sasaki--Einstein point. In section \ref{sec:reduction} we consider the reduction of M-theory on the manifolds described in the previous sections, 
and briefly discuss the resulting four-dimensional $\cN=2$ gauged supergravity. 
Finally, Appendix \ref{app:ex} discusses some examples of Sasaki--Einstein manifolds arising from complete intersections, while Appendix \ref{app:GH} applies some of the concepts introduced in section \ref{sec:CR-cone} to the Gibbons--Hawking metrics, that provide a useful nontrivial example in four dimensions.

\section{Deformations of Sasaki--Einstein manifolds}
\label{sec:CS-man}

We are interested in reductions of M-theory to $\cN\!=\!2$ four-dimensional gauged supergravity theories with AdS$_4$ vacua. In particular we will focus on the case where the internal manifold corresponding to that vacuum is a seven-dimensional Sasaki--Einstein space, while the various fields in the four-dimensional supergravity theory
correspond to deformations away from the Sasaki--Einstein point. Moreover we will assume the Sasaki--Einstein to be regular, meaning that it is a circle fibration over a  six-dimensional manifold, which is then required to be K\"ahler--Einstein.

In general one may consider various deformations away from the Sasaki--Einstein point. However, it is not easy to identify a class of deformations that leads to an ${\cal N}=2$ supersymmetric theory in four dimensions. We will first review in section \ref{sub:gen} some of the general obstacles one finds, and then describe in later sections our strategy to overcome them, for a restricted family of SU(3) structures defined in section \ref{sub:su3-fam}. In section \ref{sec:Kah-defs} we discuss deformations of the two-form, while in section \ref{sec:3form-defs} we present a short summary of the treatment of three-form deformations in later sections.

\subsection{General issues with nonlinear reductions} 
\label{sub:gen}

We want to perform a so-called ``nonlinear reduction''. Namely, we want to evaluate the action of eleven-dimensional supergravity on a certain slice of the space of all fields, and in particular of all metrics on $M_7$, such that the resulting evaluated action is an ${\cal N}=2$  supergravity in four dimensions. The task is non-trivial because of this last requirement. 

A well-understood case of nonlinear reduction consists in considering special holonomy metrics on $M_7$. An example is SU(3) holonomy, which leads to spaces of the form CY$_6\times S^1$. It is natural (as in the literature on supersymmetry-preserving solutions) to try more generally to use SU(3) structures on $M_7$. These are described by any choice of tensors whose common stabilizer in SO(7) is SU(3); for example  by a real one-form $\eta$, a real two-form $J$, and a complex three-form $\Omega$, with some algebraic conditions --- namely, that $\Omega$ is decomposable (i.e.~it can be expressed locally as a wedge of three one-forms); and that 
\begin{equation}\label{eq:alg-SU3-1}
J\wedge \Omega=0\,, \qquad \frac1{3!}\?J\wedge J\wedge J = \frac1{(2\?\im)^3}\? \Omega \wedge \bar \Omega \,. 
\end{equation}
Given a set of such forms one can then define a metric $g$ on $M_7$, with the property that
\begin{equation}\label{eq:alg-SU3-2}
\eta\cdot J=\eta\cdot \Omega=0 \,. 
\end{equation}

There are however infinitely many SU(3) structures on a given $M_7$, and one wants to pick a  finite-dimensional family ${\cal F}$.\footnote{Formally it is not even necessary to do this; one can simply rewrite the higher-dimensional action as a four-dimensional action with infinitely many fields, as done for example in \cite{grana-louis-waldram} for type II theories. However, one sometimes ends up with puzzling features such as a K\"ahler potential that also depends on the coordinates of the internal manifold.} As we will now see, this choice of slice is severely restricted by supersymmetry, since on top of the choice of the family ${\cal F}$, we also need to pick a set of forms along which we can expand the fluxes and gauge potentials. Now, the action contains the exterior differential $d$ and the Hodge $*$, so that one needs to impose that the set of forms be closed under the action of $d$ and $*$. The simplest possibility consists of picking harmonic forms; this is particularly appropriate for reductions on special holonomy manifolds. As explained in the introduction, however, for various physical applications we need to keep forms that are not harmonic. This would suggest, for example, to keep eigenspaces of the Laplacian with higher eigenvalue.
 
The constraints imposed by supersymmetry arise by the fact that the choice of the family ${\cal F}$ of SU(3) structures enters in the kinetic terms of the scalars, while the choice of forms enters in the kinetic terms of the vectors. The requirement of supersymmetry on the lower-dimensional theory relates these two. 
The clearest way of achieving four-dimensional supersymmetry is to take the forms defining the SU(3) structure to be linear combinations of the forms defining the expansion for the fluxes and potentials. For example, denoting the relevant forms as $\alpha_\Lambda$ and $\beta^\Lambda$, corresponding to electric and magnetic fluxes respectively, we can mimic the Calabi--Yau case and consider the following expansion for the three-form
\begin{equation}\label{eq:Oab}
	\Omega = X^\Lambda \alpha_\Lambda - F_\Lambda \beta^\Lambda\, 
\end{equation}
for some constant coefficients $X^\Lambda$ and $F_\Lambda$, half of which will become the ``moduli'' (namely coordinates for the family ${\cal F}$). Let us stress once again that the forms $\alpha_\Lambda$ and $\beta^\Lambda$ are in general not harmonic. 

An expansion as in (\ref{eq:Oab}) is a nontrivial requirement. The forms $\alpha_\Lambda$ and $\beta^\Lambda$ span a vector space $V_{\cal F}$, which depends on the point on the family ${\cal F}$. It is not unusual to demand that a variation of an object belongs to a vector space; but here we are demanding that the full finite object belongs to a vector space, even though the vector space depends on the point. Imagine following a path in ${\cal F}$ and computing the finite variation $\Delta\Omega$ from many small variations $\delta \Omega$, each of which belonging to $V_{\cal F}$.\footnote{There is an additional subtlety in making such statement: varying (\ref{eq:Oab}) to get $\delta \Omega$ one sees that the variations $\delta \alpha_\Sigma$ and $\delta \beta^\Sigma$ also appear. As emphasized in \cite{kashanipoor-minasian}, these terms have to be taken care of somehow, in order for the supersymmetric reduction to work.} If these vector spaces were really completely unrelated to each other, each small variation $\delta\Omega$ would take us in a direction completely independent from the previous one, and the finite variation $\Delta \Omega$ could not possibly belong to a finite-dimensional vector space. The only option seems to be that the vector spaces $V_{\cal F}$ are really all related to each other in some fashion.

Several such examples are provided by compactifications on coset manifolds, in which case the vector space of forms $V_{\cal F}$ is identified as the space of invariant forms.
A larger class is provided by Calabi--Yau compactifications, where $V_{\cal F}$ is the space of harmonic three-forms and the deformations $\delta \Omega$ are all $(3,0)$ and $(2,1)$ forms inside this space \cite{candelas-delaossa-moduli}. In other words, in this case (\ref{eq:Oab}) is consistent because Dolbeault cohomology is contained inside de Rham cohomology. The formal description of the total space resulting from varying $V_{\cal F}$ in this case was given in \cite{DUBROVIN1992627}, which identified the flat structure on the total space as a manifold with a Frobenius structure.

In this paper, we will use a  $V_{\cal F}$ that does not consist of harmonic forms, but that still has a geometrical meaning. This is perhaps most clear in terms of an auxiliary eight-dimensional manifold $M_8$, of which $M_7$ is the boundary; a natural space of forms on $M_8$ with a Frobenius structure is known to arise in the class of manifolds we discuss. In addition, there are other perspectives that involve only the geometry of $M_7$, or the geometry of a lower-dimensional manifold, $M_6$, such that $M_7$ is a fibration over it, which will be introduced in due course.


\subsection{A family of SU(3) structures} 
\label{sub:su3-fam}

As anticipated in the introduction, the family of SU(3) structures we will consider contains the Sasaki--Einstein structure as a particular case. Recall that a Sasaki--Einstein structure is given by a one-form $\eta$, a two-form $J_\se$ and a three-form $\Omega_\se$, satisfying \eqref{eq:alg-SU3-1} and 
\begin{align}\label{eq:SE-point}
d\eta = &\, 2 \, J_\se \,, \CR
d \Omega_\se= &\, 4\,\im\?\eta\wedge \Omega_\se \,.
\end{align}
This is a particular example of an SU(3) structure in seven dimensions; in general, an SU(3) structure is again given by a triplet $(\eta,\,J,\,\Omega)$, whose derivatives are parametrised by several tensors, known as torsion classes \cite{Dall'Agata:2003ir, Behrndt:2005im}.

More specifically, we will restrict our attention to \emph{regular}  Sasaki--Einstein manifolds: this implies that the orbits of the vector $\xi$ dual to $\eta$ should be closed, and they should define a U(1) fibration over a manifold $M_6$:
\begin{equation}\label{eq:7D-man}
S^1\hookrightarrow M_7 \to M_6 \,.
\end{equation}
The metric will then be
\begin{equation}
ds^2(M_7) = \eta^2 + ds^2(M_6)\, ,
\end{equation}
and the vector dual to $\eta$ is an isometry acting on the fibre of (\ref{eq:7D-man}). 
It also follows that the base $M_6$ is a K\"ahler--Einstein manifold, with K\"ahler form $J_\se$ proportional to the Ricci form, $\rho$, with a canonical proportionality constant:
\begin{equation}\label{eq:KE-Ric}
\rho = 6\? J_\se \,.
\end{equation}
While the list of compact K\"ahler--Einstein manifolds is relatively short in dimension 4, in dimension 6 there are many examples, some of which will appear in our discussion in the main text (e.g. section \ref{sub:hypersurf}) and some of which we review in appendix \ref{app:ex}. In fact this subject is undergoing rapid development, due to the recent proof of the K-stability conjecture \cite{chen-donaldson-sun}, that relates the existence of K\"ahler--Einstein metrics to an algebraic-geometrical condition --- which makes this class a bit like the Calabi--Yau case.

Having described a Sasaki--Einstein structure, let us now describe the more general family of SU(3) structures relevant for this paper, of which the Sasaki--Einstein will be a particular point. We will take most of the torsion classes to vanish: in other words, even though $d \eta$, $dJ$ and $d \Omega$ will be more general than in (\ref{eq:SE-point}), they will still be severely restricted. In particular, we assume that the base remains symplectic, meaning that the symplectic form $J$ is always closed, ultimately leading to ungauged vector multiplets in the four-dimensional supergravity theory obtained by compactification of M-theory on $M_7$. Similarly, we only allow for a small subset of the torsion classes parametrising the derivative of the complex structure. 

More concretely, we focus on SU(3) structures satisfying\footnote{In terms of the general SU(3) structures in seven dimensions and using the notation of \cite{Behrndt:2005im}, our assumption that $dJ=0$ sets $W_3=W_4=T_2=V_2=0$. The quadratic constraints can be solved by $W_1=W_2=0$, while different possible choices impose restrictions on the K\"ahler moduli. The assumption on $d\eta$ reparametrises two more classes through $R\,J + T_1=\rho$, while we further assume that $W_0=W_5=V_1=0$.}
\begin{align}\label{eq:su3-struct}
d\eta = &\, \frac{1}{3}\, \rho \,, \cr
dJ = &\, 0 \,, \\
d \Omega= &\, \eta\wedge \left( \mathrm{i}\,E\,\Omega + S \right)\,. \nonumber
\end{align}
Here, $E$ is an SU(3)-singlet, while $S$ is a complex $(2, 1)$-form in the $\bf{6}$ of SU(3). The two-form $\rho$ stands for the Ricci form on the K\"ahler--Einstein base $M_6$, which is fixed to this value when deforming away from the Sasaki--Einstein metric. This is a particular assumption on the SU(3) structure, since $d\eta$ is parametrised by several non-constant torsion classes in general.

These simplifications make it possible to show that the set of forms required for the reduction of both the K\"ahler and three-form sectors do indeed exist for a class of manifolds. It turns out that the \Kah sector is simpler and can be described in terms of harmonic forms on the K\"ahler--Einstein base $M_6$ at the Sasaki--Einstein point of the family, as discussed in some detail in the following subsection \ref{sec:Kah-defs}. A similar discussion at a general point is given in section \ref{sec:Kah-deformations}, after introducing the relevant mathematical background. On the other hand, the discussion of the deformations for the three-form is significantly more complicated and spans sections \ref{sec:cmplx-defs}, \ref{sec:CR-cone} and \ref{sec:CR-struct}. For the convenience of the reader, section \ref{sec:3form-defs} provides an overview of the main results shown in these sections.

\subsection{K\"ahler deformations}
\label{sec:Kah-defs}

We start with the K\"ahler deformations, whose description turns out to be only a slight deviation from the well established Calabi--Yau case. In view of \eqref{eq:su3-struct}, it is clear that any deformation of the K\"ahler form $J$, is uncharged with respect to the U(1) bundle over the six-dimensional K\"ahler--Einstein base, $M_6$. It follows that it is sufficient to consider the K\"ahler deformations of a manifold with constant Ricci form $\rho$, so that the U(1) fibration described by $\eta$ remains unchanged. 

Consider a basis $\{\Gamma_a\}$, of $H_2(M_6,\mathbb{Z})$, in terms of which we expand the Ricci form as 
\begin{equation}\label{eq:magn-gaug}
m^a = \frac{1}{3}\int_{\Gamma_a} \rho \,,
\end{equation}
where the somewhat unconventional normalisation factor of is added for later convenience. 
The components parametrised by the constants $m^a$ will be taken as fixed throughout this paper and will turn out to correspond to a gauging in the lower-dimensional supergravity obtained after reduction.
Similarly, a set of coordinates $t^a$ on the space of K\"ahler classes can be introduced as
\begin{equation}
t^a = \int_{\Gamma_a} J  \,,
\end{equation}
for $J$ an arbitrary representative of the K\"ahler class $[J]$. 

By the Calabi conjecture, later proven by Yau, given a complex structure on $M_6$ and a K\"ahler class specified by the $t^a$, one can find a metric with associated K\"ahler form $J(t)$ within this K\"ahler class, such that it leads to a fixed Ricci form $\rho$. Note that the statement of this result does not put any restriction on the Ricci form. This is exactly the same as in the familiar setting of a vanishing Ricci form for Calabi--Yau manifolds, where this result has been used extensively in the physics literature. In this paper, we make use of the Calabi conjecture in the more general case of an arbitrary but fixed Ricci form, parametrised by the constants $m^a$. As it turns out, the steps required are very similar to the Calabi--Yau case, so we consider these in some detail.

In order to study the moduli space defined by the $t^a$, we consider a basis $\{[\omega_a]\}$ of integral cohomology $H^2(M_6,\mathbb{Z})$, dual to the basis $\{ \Gamma_a \}$ introduced above, satisfying
\begin{equation}
\int_{\Gamma_a}\omega_b = \delta^a_b \,.
\end{equation}
Assuming a complex structure on $M_6$, the $\{[\omega_a]\}$ can be taken to be $(1,1)$ forms and provide a basis for the K\"ahler class and the Ricci class as
\begin{equation}\label{eq:J-rho}
J = t^a \omega_a\,, \qquad \rho = 3\,m^a \omega_a\,.
\end{equation}
By Yau's theorem, these data uniquely determine a metric via
\begin{equation}\label{eq:metr-J}
\im\? g_{m \bar{n}} = J_{m\bar{n}} \,,
\end{equation}
which in turn specifies the harmonic representatives $\omega_a(t)$ in each cohomology class.
This puts constraints on the moduli dependence of the base of (1,1)-forms, as one can verify by considering a variation of the metric \eqref{eq:metr-J} with respect to the $t^a$. Using \eqref{eq:J-rho}, this reads
\begin{equation}\label{metricvariation}
\frac{\partial J_{m \bar{n}}}{\partial t^a} = \omega_{a\?m\bar{n}} +  t^b \frac{\partial}{\partial t^a}\omega_{b\?m\bar{n}}  \,.
\end{equation}
By the above construction, the moduli dependence in the $\omega_a(t)$ only modifies each form by possible exact pieces, required by Yau's theorem, so that the basis $\{[\omega_a]\}$ is constant and $\partial_a\omega_{b}$ is an exact form. Therefore, \eqref{metricvariation} describes the relation between the form $\partial_a J_{m\bar{n}}$ and its harmonic representative, $\omega_{a\,m\bar{n}}$, identifying $t^b \partial_a \omega_{b\?m\bar{n}}$ as the relevant exact form connecting the two. 

In order to restrict this exact contribution, we turn to a direct generalisation of the standard computation leading to the Lichnerowicz equation for Calabi--Yau manifolds. Since we are to keep the Ricci form fixed, one may set the variation of the Ricci tensor to zero to obtain the constraint 
\begin{equation} \label{eq:Lichnerowicz}
 \delta R_{M N} = 0 \quad\Rightarrow\quad 
 \nabla^2 \delta g_{M N} - 2\, R_M{}^P{}_N{}^Q\,\delta g_{P Q} + 2\, R_{(M}{}^P\delta g_{N) P}=0\,,
\end{equation}
where we used the standard coordinate condition $\nabla^M \delta g_{M N}=0$ and we discarded a term proportional to $\nabla_{M}\nabla_N \mathrm{tr}(\delta g)$. Note that the last term of \eqref{eq:Lichnerowicz} vanishes for Calabi--Yau manifolds, while the first two terms are the Lichnerowicz equation in that case. One can now rewrite this equation in (anti-)holomorphic indices and specialise to the case of K\"ahler variations, $\delta g_{m\bar{n}}$, to find 
\begin{equation} \label{eq:Weitzenbock}
 \nabla^2 \delta g_{m\bar{n}} + R_{m\bar{n}}{}^{p\bar{q}}\,\delta g_{p\bar{q}} 
 - R_{m}{}^p\delta g_{p \bar{n}}-  R_{\bar{n}}{}^{\bar{p}}\delta g_{m \bar{p}}=0\,.
\end{equation}
The constraint \eqref{eq:Weitzenbock} can be recognised as the standard Weitzenb\"ock identity, which relates the scalar Laplacian to the Laplacian acting on a $(1,1)$ form. It then follows that \eqref{eq:Weitzenbock} simply imposes that $\delta g_{m\bar{n}}$ is a harmonic $(1,1)$ form.

Returning to \eqref{metricvariation}, we note that it relates two harmonic forms, namely $\partial_a J_{m\bar{n}} = \im\?\partial_a g_{m\bar{n}}$ and $\omega_a$, by an exact piece. However, the standard Hodge decomposition for compact \Kah manifolds implies that harmonic forms are unique, so we conclude
\begin{equation}\label{eq:consistent-Kah}
 t^b \frac{\partial}{\partial t^a}\omega_{b\?m\bar{n}} = 0\,.
\end{equation}
This condition is important for the reduction of the Ricci tensor over $M_6$, as has been stressed in the literature (see \cite{kashanipoor-minasian} for details).

In summary, we have shown that the \Kah moduli space for a compactification on a regular Sasaki--Einstein manifold, as in \eqref{eq:7D-man}, can be described by a standard expansion of the \Kah form over the harmonic $(1,1)$ forms on the base $M_6$, under the assumption of a fixed Ricci form $\rho$ on $M_6$ appearing in \eqref{eq:su3-struct}. Relying on the natural complex structure available for a Sasaki--Einstein manifold, the Calabi conjecture, proven by Yau's theorem, guarantees the existence of a unique metric in each \Kah class. In the following sections we will describe the deformations of the three-form $\Omega$ away from the Sasaki--Einstein point, so that a generalisation of the discussion above is required. We return to this point in section \ref{sec:Kah-deformations}.

\subsection{Overview of three-form deformations}
\label{sec:3form-defs}

We now turn to the deformations of the three-form $\Omega$ away from the Sasaki--Einstein point \eqref{eq:SE-point}, within the class given in \eqref{eq:su3-struct}. This task is more complicated than the description of \Kah deformations in the previous subsection and extends over the following sections, where three complementary approaches are discussed. Here, we give an overview of the main ideas, to be used as a road map for what follows.

Drawing inspiration from the previous subsection, it is natural to treat the deformations of a Sasaki--Einstein manifold using its description as a U(1) bundle over a \KE base,  $M_6$, as in \eqref{eq:7D-man}. However, unlike the the situation for the \Kah form above, the three-form $\Omega$ is not a well-defined form on $M_6$ but only on the total space, as signalled by its nontrivial Lie derivative along the circle, computed by \eqref{eq:SE-point} as
\begin{equation}
	\pounds_\xi \Omega_\se = 4\,\im\, \Omega_\se\,,
\end{equation}
where $\xi$ is the vector dual to the one-form $\eta$. If we define the coordinate $\psi$ such that $\xi=\del_\psi$, we may however write
\begin{equation}\label{eq:OmegaSE}
	\Omega_\se = \ee^{4 \? \im\? \psi} \Omega_0\ ,
\end{equation}
where $\Omega_0$ is not a three-form on $M_6$, but can be viewed as a section of the anticanonical bundle, $K^*$, over the base. This fact makes it clear that the standard deformation theory on the base is not sufficient, but also suggests that one may consider deformations charged under $K^*$, since the non-closure of $\Omega$ is due to a nontrivial charge with respect to that bundle.

Taking this point of view, we proceed in the next section to construct deformations of \eqref{eq:OmegaSE} by allowing a sum over appropriate charges, schematically
\begin{equation}\label{eq:Omega-sum}
 \Omega = \sum_k \ee^{\im\?k\? \psi} \omega_k\ .
\end{equation}
Here, the $\omega_k$ stand for appropriate sections of the $k$-th power\footnote{This can be relaxed in some cases, as will be made more precise in section \ref{sub:tw-coh}.} of the anticanonical bundle $K^*$ over $M_6$. The range of the sum is not infinite, as one might be afraid, as we show in section \ref{sub:hypersurf} for some simple concrete examples of hypersurfaces in $\cc\pp^4$. Using standard techniques in algebraic geometry, we define the appropriate twisted cohomology to which the forms $\omega_k$ in \eqref{eq:Omega-sum} belong. We find that the relevant cohomology groups are controlled by a finite set of monomials of the coordinates of $\cc\pp^4$, thus restricting the sum in \eqref{eq:Omega-sum} to a finite range. 

More precisely, we find that for a hypersurface in $\cc\pp^4$, specified by a homogeneous polynomial 
\begin{equation} \label{eq:p-def}
 f(z_i)=0\,, \qquad z_i \in \cc^5 \,, 
\end{equation}
the forms in \eqref{eq:Omega-sum} are in one to one correspondence with the monomials parametrising the \emph{Jacobi ring}, $\jac$, of the polynomial. The latter is defined as the quotient of all polynomials in the $z_i$, denoted $\cc[z_i]$, modulo the ideal generated by the derivatives of the polynomial, $\langle\?\del_i f\?\rangle$, as 
\begin{equation} \label{eq:J-def}
\jac \equiv \frac{\cc[z_i]}{\langle\?\del_i f\?\rangle}\ .
\end{equation}
Effectively, $\jac$ contains all polynomials that are not functionally dependent on the derivatives $\del_i f$ and is finite by definition for any polynomial $f(z_i)$. The occurrence of the Jacobi ring \eqref{eq:J-def} signals a connection between the deformations described in terms of twisted cohomology on the \KE base of a regular Sasaki--Einstein manifold and the cone over it. The latter is described as a complex isolated singularity and the deformations of such varieties are also known to be described by \eqref{eq:J-def}. The description of deformations away from the conical singularity is the subject of section \ref{sec:CR-cone}.

More concretely, the cone over a Sasaki--Einstein manifold $M_7$, denoted as $M_8$, is defined as
\begin{equation}\label{eq:CY-cone}
 ds^2{\scalebox{0.91}{$(M_8)$}} = dr^2 + r^2 ds^2{\scalebox{0.91}{$(M_7)$}} \,,
\end{equation}
and is a noncompact Calabi--Yau manifold by construction. The tip of the cone features an isolated singularity, except for the case where the cone is simply $\mathbb{C}^4$ and $M_7=S^7$. More general examples are provided by considering the homogeneous equation \eqref{eq:p-def} as a hypersurface in $\cc^5$ rather than in $\cc\pp^4$. This is equivalent to constructing the complex cone over $M_6$, which is indeed Calabi--Yau if $M_6$ is assumed to be K\"ahler--Einstein.

One may now define the deformation space of this non-compact Calabi--Yau hypersurface singularity, which turns out to be identified exactly as \eqref{eq:J-def}. The homogeneity of \eqref{eq:p-def} is crucial in this respect, as it allows to lift the the Reeb U(1) isometry of the Sasaki--Einstein manifold to an isometry of the Calabi--Yau hypersurface that acts nontrivially on the deformations. In fact, the deformations of the CY cone carry exactly the same grading of charges under this isometry as the deformations constructed in terms of twisted cohomology on the K\"ahler--Einstein base. 

The description in terms of the cone provides the advantage of a geometrical picture, due to a famous result of Milnor identifying the homology structure of the deformed cone with a bouquet of topological four-spheres. In simpler terms, this implies that upon deformation the cone develops $\mu\!=\!\dim \jac$ nontrivial four-cycles (see Figure \ref{fig:cone} on page~\pageref{fig:cone}); $\mu$ is known as the Milnor number. An example of this deformation is given by the Stenzel metrics in dimension $n$, also referred to as $n$-dimensional conifolds in the physics literature. These arise from deformations of the polynomial 
\begin{equation}
f_{\scriptscriptstyle S}(z_i)=\sum_{i=1}^{n+1}(z_i)^2\,, \quad\Rightarrow\quad \jac\equiv \frac{\cc[z_i]}{\langle\del_i f_{\scriptscriptstyle S} \rangle} = \{ 1 \}\,,
\end{equation}
so that the Jacobi ring is one-dimensional, describing the deformation induced by adding a constant to $f_{\scriptscriptstyle S}$. One therefore expects to find a deformed metric featuring a single nontrivial $n$-cycle, which indeed exists and is known explicitly \cite{candelas-delaossa-comments, Stenzel1993, Cvetic:2000db}. Similar metrics are expected to arise from the deformations of isolated singularities with $\mu>1$, but obtaining these explicitly is a rather difficult task.

The deformations of the cone can be used to define deformations away from the Sasaki--Einstein point independently of the twisted cohomology described above, by viewing $M_7$ as a hypersurface enclosing (or as the link over) the singularity. Since the deformed cone features nontrivial four-cycles, it admits a metric that is only asymptotically conical, so that the seven-dimensional metric on the hypersurface is also deformed. This new metric on the link differs from the Sasaki--Einstein metric by a change of complex structure, $\Omega_4$, of the embedding space, parametrised in terms of $2\?(\mu+1)$ complex parameters $(X^\La,F_\La)$ as
\begin{equation}\label{eq:OmegaM8}
\Omega_4 = X^\Lambda \alpha_\Lambda - F_\Lambda \beta^\Lambda \ .
\end{equation}
Here, the $\beta^\Lambda$ and $\alpha_\Lambda$ are two sets of four-forms belonging to ordinary and relative de Rham cohomology respectively, which are naturally dual to the compact and relative homology four-cycles arising on the deformed cone. Upon restriction to the link $M_7$, the parametrisation in \eqref{eq:OmegaM8} leads to a family of SU(3) structures in the class specified in \eqref{eq:su3-struct}, expanded on the basis forms $\alpha^\Lambda$ and $\beta_\Lambda$.

The restriction of the complex structure on $M_8$ to $M_7$ allows to view our family of SU(3) structures as a family of induced Cauchy--Riemann (CR) structures. CR structures are the odd-dimensional analogue of complex structures, in the same way as a Sasaki structure is analogous to a \Kah structure. One can define intrinsically seven-dimensional deformations based on CR-structures, which turn out to be equivalent to the deformations constructed through twisted cohomology for the case of regular Sasaki--Einstein manifolds considered in this paper. However, the approach based on \eqref{eq:OmegaM8} has the advantage of providing a definition of an induced CR structure for a general deformation of the cone, and it therefore allows to describe the family of SU(3) structures $V_{\cal F}$ around any such point. In fact, the deformations of the cone are known to admit a flat structure \cite{Saito-universal, DUBROVIN1992627}, confirming the relation between the vector spaces at different points of this family anticipated in section \ref{sub:gen}.

We close this section with some comments on the application of our reasoning to more general cases. In this paper, we restrict ourselves to the case of hypersurfaces, so that both the \KE base and the singular cone over the Sasaki--Einstein spaces we consider are given by a single equation in $\cc\pp^4$ and $\cc^5$ respectively. A wider class of examples is given by complete intersections of $s$ functions in $\cc\pp^{3+s}$ and $\cc^{4+s}$ respectively. The techniques used to argue towards the finiteness of twisted cohomology over \KE spaces in section \ref{sec:cmplx-defs} are readily available in the case of complete intersections, so that the argument can be extended in principle. In terms of the cone, an extension of deformation theory to complete intersection singularities is known, as discussed briefly in Appendix \ref{app:ex}.

Finally, we point out that the ideas developed in the following sections do not depend crucially on the dimensionality, so that it is straightforward to consider deformations of regular Sasaki--Einstein manifolds in five or more dimensions in exactly the same way.

\section{Three-form deformations from twisted cohomology}
\label{sec:cmplx-defs}

In this section we describe how to obtain deformations of the three-form $\Omega$ away from the Sasaki--Einstein point \eqref{eq:SE-point} in the more general class \eqref{eq:su3-struct} we have identified, using the picture of the seven-dimensional manifold as a circle fibration over a \KE base space. We do this by first describing an appropriately twisted Dolbeault cohomology on the \KE manifold in section \ref{sub:tw-coh}, which is then worked out in detail for some examples in section \ref{sub:hypersurf}, while a real version is briefly explored in section \ref{sub:derham}. Finally, in section \ref{sub:fam} we proceed to use this twisted cohomology to define a family of deformations for the three-form.

\subsection{Deformations from twisted cohomology} 
\label{sub:tw-coh}

In order to motivate the deformations of the three-form $\Omega_\se$, we first need some background on its properties. As explained in section \ref{sec:3form-defs}, the nontrivial Lie derivative of this form along the U(1) Reeb isometry implies that it cannot be viewed as an honest form on the base $M_6$. Rather, one may write it as in \eqref{eq:OmegaSE}
\begin{equation}\label{eq:OmegaSE-2}
	\Omega_\se = \ee^{4 \? \im\? \psi} \Omega_0\ ,
\end{equation}
where $\Omega_0$ is not a three-form but a section of the line bundle $K^*$, the anticanonical bundle. A $(3,0)$-form without zeros does not always exist, because the bundle $\Lambda^3 T^*_{(1,0)}\equiv \Lambda^3 \Omega$ is in general nontrivial; but $\Omega_0$ is viewed as a $(3,0)$-form valued in $K^*$, i.e.~a section of $\Lambda^3 \Omega \otimes K^*=K\otimes K^*=\cal O$, the trivial line bundle, which does have global sections. 
As a consequence, $\Omega_0$ is not closed under the ordinary Dolbeault $\bar\del$ but under a ``twisted'' $\bar \del + 4w$, where $w$ is a connection on $K$. All in all, we have 
\begin{equation}\label{eq:OmegaSE-d}
d \Omega_\se= 4\?\im\?(d \psi+ w)\wedge \Omega_\se=4\?i\? \eta \wedge \Omega_\se\,,  
\end{equation}
which is \eqref{eq:SE-point}.

From this point of view, it is clear how one should satisfy our more general Ansatz \eqref{eq:su3-struct}: one can just add 3-forms $\omega_k$ which are valued in different powers of the anticanonical bundle $K^*$, as in \eqref{eq:Omega-sum}. Actually, a slightly more general possibility sometimes exists. For most K\"ahler--Einstein manifolds, the anticanonical bundle $K^*$ does not admit a ``root'', in the sense that there is no positive line bundle ${\cal L}$ such that ${\cal L}^\jst=K^*$ for some integer $\jst$; however, in some cases that is possible, and $\jst$ is called the ``index'' of $M_6$. (A famous case is $\cc\pp^3$, for which $\jst=4$.) To take this possibility into account, we will consider powers of ${\cal L}$ rather than $K^*$, although one should bear in mind that often $\jst=1$ and ${\cal L}=K^*$. So a first rough idea is that we should write
\begin{equation}\label{eq:Omegasum}
	\Omega = \sum_k \ee^{i k \hat \psi} \omega_k \ ,\qquad	 \hat \psi = \frac{4}{\jst} \psi\ ,  
\end{equation}
where $\omega_k$ will be 3-forms valued in ${\cal L}^k$, which will also be sometimes called "twisted forms''. We will denote the differential on the formal sum $\oplus {\cal L}^k$ by $\bar\del_0$; it is the usual Dolbeault differential, plus a connection term which depends on the twisting of the form we are acting on. On $M_6$, the expression \eqref{eq:Omegasum} is a sum of sections of different bundles, which ordinarily we would not want to consider; but on $M_7$ it is a perfectly sensible three-form.

We should specify, however, the range of $k$ in the sum \eqref{eq:Omegasum}. A priori, it seems there are infinitely many possible values for $k$, and one might think there is no natural way of truncating the sum to finitely many forms. In fact, however, some of the Dolbeault cohomologies that describe these three-forms are non-zero only in a finite range of $k$'s.

In general, if $E$ is a holomorphic bundle with a connection $A$, the $(0,2)$ part of the curvature can be taken to vanish, and thus $\bar\del_E\equiv \bar\del+ A_{0,1}$ is a differential. In the case at hand, we identify $E={\cal L}^k$ for each $k$ and define $H^{p,q}(M_6,E)$ as the space of $(p,q)$-forms which are closed under $\bar\del_E$, modulo those that are exact. An alternative way of writing this space is $H^q(M_6,\Lambda^p\Omega\otimes {\cal L}^k)$, where recall that $\Omega\equiv T^*_{1,0}$ is the holomorphic cotangent bundle. Then our statement is that 
\begin{equation}\label{eq:tw21}
	H^{2,1}(M_6, {\cal L}^k)= H^1(M_6,\Lambda^2 \Omega \otimes {\cal L}^k)
\end{equation}
is non-zero only for a finite range of $k$. (Serre duality also gives us $H^{1,2}(M_6,{\cal L}^k)^*\cong H^{2,1}(M_6,{\cal L}^k)$.) In fact, we will also see that the total space $\oplus_k H^{2,1}(M_6, {\cal L}^k)$ has a natural algebraic interpretation; these statements will be given a topological interpretation in terms of the cone over $M_7$, in section \ref{sec:CR-cone}.

We will often also consider the closely related twisted Beltrami differentials, $\mu_k$, namely elements of the cohomology
\begin{equation}\label{eq:tw11}
\mu_k \in H^1(M_6, T\otimes {\cal L}^k)\ .
\end{equation}
An element of this space, acting on $\Omega_0 \in H^{3,0}(M_6,K^*)=H^0(M_6,\Lambda^3 \Omega\otimes K^*)= \cc$, produces an element 
\begin{equation}
\mu_k\cdot \Omega_0\in H^{2,1}(M_6,{\cal L}^k\otimes K^*)= H^{2,1}(M_6,{\cal L}^{k+\jst})\,.
\end{equation}
This is entirely analogous to the action of an ordinary Beltrami differential (an infinitesimal change in complex structure) on the three-form $\Omega$ of a Calabi--Yau manifold; the only new element is that both $\mu$ and $\mu\cdot \Omega_0$ carry a twist, in other words they take values in a line bundle.

\subsection{An example: hypersurfaces in \texorpdfstring{$\cc\pp^4$}{CP4}} 
\label{sub:hypersurf}

We will illustrate all this in a couple of simple examples, which should hopefully also give an idea of how the general story works. We will namely consider homogeneous degree $d$ Fermat hypersurfaces in $\cc\pp^4$: 
\begin{equation}\label{eq:fermat}
	\{f(z_i)=\sum_{k=1}^5 z_k^d=0 \}  \subset P\equiv \cc\pp^4\ .
\end{equation}
These are known \cite{nadel} to be positive curvature K\"ahler--Einstein for $d=2,3,4$; they are in a sense the natural counterpart in our setting of the case $d=5$, which is the famous quintic Calabi--Yau. The case $d=2$, the quadric, can be viewed as the quotient $\frac {{\rm SO}(5)}{{\rm SO}(3)\times {\rm SO}(2)}$; the corresponding Sasaki--Einstein manifold is the coset $M_7=\frac {{\rm SO}(5)}{{\rm SO}(3)}$ and is known as the Stiefel manifold $V_{5,2}$. This manifold has also appeared in the physics literature, as the reduction of M-theory on $V_{5,2}$ was considered in \cite{cassani-koerber-varela}, using its coset structure, while the cone over it and its deformation belong to the class of Stenzel spaces  \cite{Stenzel1993,Cvetic:2000db}. 

For these examples, one can identify the line bundle ${\cal L}$ as the restriction to $M_6$ of ${\cal O}(1)$ on $\cc\pp^4$, so that the anticanonical is $K^*={\cal O}(5-d)|_{M_6}$ and the index is $\jst=5-d$.  We can thus identify \eqref{eq:tw11} as $H^1(M_6,T(k))$, where $T(k)\equiv T\otimes {\cal O}(k)$. We will compute this cohomology class by using the definition of the normal bundle $N$ as a quotient of the tangent bundle of $\cc\pp^4$ by the one of $M_6$. This is commonly expressed as an exact sequence: 
\begin{equation}\label{eq:N}
	0 \to TM_6 \buildrel i \over \to TP \buildrel \phi \over\to N \to 0 \ .
\end{equation}
The adjunction formula also tells us $N={\cal O}(d)|_{M_6}$. If we take the tensor product of (\ref{eq:N}) by ${\cal O}(k)$ and consider the associated long exact sequence,
\begin{align}\label{eq:long-N-0}
0 \to\,&\, H^0 (M_6, T(k)) \to H^0 (M_6, TP(k)|_{M_6}) \buildrel \phi \over\to H^0 (M_6, {\cal O}(d+k)) \buildrel \pi\over \to
\CR
\buildrel \pi\over \to \,&\, H^1 (M_6, T(k)) \to H^1 (M_6, TP(k)|_{M_6}) \buildrel \phi \over\to \dots \ ,
\end{align}
the desired cohomology (\ref{eq:tw11}) appears, along with other cohomologies that we have to evaluate. In particular, we aim to show that $H^1 (M_6, TP(k)|_{M_6})=0$, so that \eqref{eq:long-N-0} truncates.

To do this we can use the ``resolution'' 
\begin{equation}\label{eq:res}
	0\to {\cal O}(k-d) \to {\cal O}(k) \to {\cal O}_{M_6}(k) \equiv {\cal O}(k)|_{M_6} \to 0 \,,
\end{equation}
and the Euler exact sequence
\begin{equation}\label{eq:euler}
	0 \to\, {\cal O} \to\, 5\? {\cal O}(1) \to\, TP\to 0 \ .
\end{equation}
The long exact sequence associated to (\ref{eq:res}) shows that the holomorphic functions of degree $k$ on $M_6$ are those of degree $k$ on $\cc\pp^4$, modded out by those that can be factorized as $f\? p_{d-k}$ and thus vanish on $M_6$ ($p_{d-k}$ being a polynomial of degree $d-k$, and $f(z)$ from (\ref{eq:fermat}) being the polynomial that defines $M_6$). In particular there are
\begin{equation}\label{eq:Ok}
	{\rm dim}H^0(M_6, {\cal O}_{M_6}(k) )=\binom{k+4}{4} - \binom{k-d+4}{4}
\end{equation}
such functions. Since $H^1(\cc\pp^4,{\cal O}(k))=0$, the same long exact sequence also shows $H^1(M_6, {\cal O}_{M_6}(k))=0$. This implies that the long exact sequence associated to (\ref{eq:euler}) truncates:
\begin{equation}\label{eq:long-euler}
	0 \to H^0(M_6,{\cal O}_{M_6}(k)) \to\, 5\? H^0 ( M_6, {\cal O}_{M_6}(k+1))\,\buildrel e\over\to\, H^0(M_6,TP(k)|_{M_6})\to 0\ .
\end{equation}
The map $e$ in (\ref{eq:long-euler}) takes five degree $k+1$ polynomials $a_i$ on $M_6$ to a section of the restricted tangent bundle $TP(k)|_{M_6}$; intuitively this simply means that such a section $\alpha$ can be written as
\begin{equation}\label{eq:aidi}
	\alpha=\sum_{i=1}^5 a_i \del_i\ .
\end{equation}
The exactness of (\ref{eq:long-euler}) also tells us that if the $a_i= z_i a$, for $a$ a degree $k$ polynomial, then the corresponding section of $TP(k)|_{M_6}$ should vanish: this is indeed the case, since $z_i \del_i=0$ on $P=\cc\pp^4$. In addition, \eqref{eq:long-euler} goes on to show that $H^1(M_6,TP(k)|_{M_6})=0$, which indeed truncates the long exact sequence \eqref{eq:long-N-0} to:
\begin{align} \label{eq:long-N}
0 \to\,&\, H^0 (M_6, T(k)) \to H^0 (M_6, TP(k)|_{M_6}) \buildrel \phi \over\to H^0 (M_6, {\cal O}(d+k)) \buildrel \pi\over \to
\CR
\buildrel \pi\over \to \,&\, H^1 (M_6, T(k)) \to 0 \ .
\end{align}
It follows that the space we want to understand, $H^1 (M_6, T(k))$, has thus been expressed as a cokernel of the map denoted $\phi$ in \eqref{eq:long-N}. Recalling that a section, $\alpha$, of $TP(k)|_{M_6}$ can be written as (\ref{eq:aidi}), the map $\phi$ can be written as
\begin{equation}\label{eq:phi}
	\phi:\,\,\, \sum_{i=1}^5 a_i \del_i\ \mapsto\ \sum_{i=1}^5 a_i \del_i f\ ,
\end{equation}
where once again $f$ from (\ref{eq:fermat}) is the polynomial defining $M_6$. Note that this is formally the same map as in (\ref{eq:N}), only restricted to elements of $H^0$. Thus our problem has been reduced to an algebraic one: computing the cokernel of (\ref{eq:phi}). 

Let us study this for $d=3$, namely for the case where $M_6$ is a cubic. From (\ref{eq:Ok}) we see that both the source and target vector spaces of $\phi$ are zero unless $k\ge -3$. For $k=-3$ and $-2$ actually $H^0 (M_6, TP(k)|_{M_6})$ still vanishes (there are no nonzero $a_i$), while the dimension of $H^0 (M_6, {\cal O}(d+k))$ is 1 and 5 respectively. These are simply degree 0 and 1 polynomials; see Table \ref{tab:jacobi}, left column. The first interesting case is $k=-1$, where source and target have dimensions respectively 5 and 15. $\phi$ takes a choice of 5 constants $a_i$ to $\sum_{i=1}^5 a_i \del_i f= 3\sum_{i=1}^5 a_i z_i^2$. The map has no kernel, so the cokernel has dimension 10. We can also think of it this way: it is the space of quadratic polynomials which cannot be written as $\sum_{i=1}^5 a_i z_i^2$ for any choice of $a_i$. So we can take it to be generated by monomials $z_i z_j$, $i\neq j$; there are 10 such monomials.

The next case is $k=0$; here the dimensions of the source and target are 24 and 34, and again there is no kernel, so the cokernel has dimension 10. It is now the space of degree 4 polynomials that cannot be written as $\sum_{i=1}^5 a_i z_i^2$, with the $a_i$ some linear polynomials. This time we can take it to be generated by $z_i z_j z_k$, $i\neq j\neq k \neq i$. Going to higher $k$, the source and target vector spaces for $\phi$ have higher and higher dimensions; it is easier to describe the cokernel directly as a space of monomials. For $k>2$, there is no such monomial; for $k=2$ there is only one, $z_1 z_2 z_3 z_4 z_5$.\footnote{This single monomial is the analogue of the monomial called $Q$ in \cite{candelas-yukawa}, where it played an important role in the computation of the Yukawa couplings for a Calabi--Yau model.} We summarized the results in Table \ref{tab:jacobi} on the left. On the right we have also shown the results for the quartic, which proceeds in exactly the same way, but involves a more extended set of monomials. Notice that in both cases there is a duality that takes a monomial $m$ to $Q_d/m$, where $Q_d=(z_1 z_2 z_3 z_4 z_5)^{d-2}$. 

\begin{table}[ht]
	\centering
\begin{tabular}{|ccc|}
\hline
$k$ & dimension & generators\\
\hline
2 & 1 & $Q_3 \equiv z_1 z_2 z_3 z_4 z_5$\\
1 & 5 & $Q_3/z_i$\\
0 & 10 & $Q_3/z_i z_j$, $i\neq j$\\
-1 & 10 & $z_i z_j$, $i\neq j$\\ 
-2 & 5 & $z_i$\\
-3 & 1 & 1 \\ \hline
\end{tabular}
\begin{tabular}{|ccc|}
\hline
$k$ & dimension & generators\\
\hline
6 & 1 & $Q_4 \equiv z_1^2 z_2^2 z_3^2 z_4^2 z_5^2$\\
5 & 5 & $Q_4/z_i$\\
4 & 15 & $Q_4/(\text{all quadr.})$\\
3 & 30 & $Q_4/(\text{all cubic except}\ z_i^3)$\\
2 & 45 & $Q_4/(\text{all quartic except}\ z_i^3 z_j)$ \\
1 & 51 & all quintic except $z_i^3 \cdot$ quadr.\\
0 & 45 & all quartic except $z_i^3 z_j$\\
-1 & 30 & all cubic except $z_i^3$\\ 
-2 & 15 & all quadr. monomials\\
-3 & 5 & $z_i$ \\ 
-4 & 1 & 1\\\hline
\end{tabular}
\caption{Dimension and generators of $H^1(M_6,T(k))$ for all the $k$ for which it is non-zero. Left, $d=3$; right, $d=4$.}
\label{tab:jacobi}
\end{table}

All in all, we see that  the twisted Beltrami differentials are in one-to-one correspondence with the elements of the generators of the \emph{Jacobi ring}, 
\begin{equation}\label{eq:J}
	\jac\equiv \frac{\cc[z_i]}{\langle\del_i f \rangle}\ .
\end{equation}
The numerator denotes the ring of all polynomials in the $z_i$; the denominator is the ideal generated by the derivatives of $f$, namely the set of all polynomials that can be written as $p_i\del_i f$ for some polynomials $p_i$. Its total dimension 
\begin{equation}\label{eq:mu-def}
	\mu \equiv \dim(\jac)\ 
\end{equation}
is called the \emph{Milnor number}.\footnote{Unfortunately it is traditional to call this number by the letter $\mu$, which is also a traditional name for a Beltrami differential.}
In the hypersurface case of the present subsection,  
\begin{equation}\label{eq:J-dim}
	\mu=(d-1)^5
\end{equation}
which can indeed be seen to be the sum of all the numbers of each of the two columns in Table \ref{tab:jacobi}.  
In section \ref{sec:singularities} we will also see an alternative (and quicker) way of obtaining the numbers in that table. 

\begin{figure}[ht]
	\centering
		\includegraphics[width=14cm]{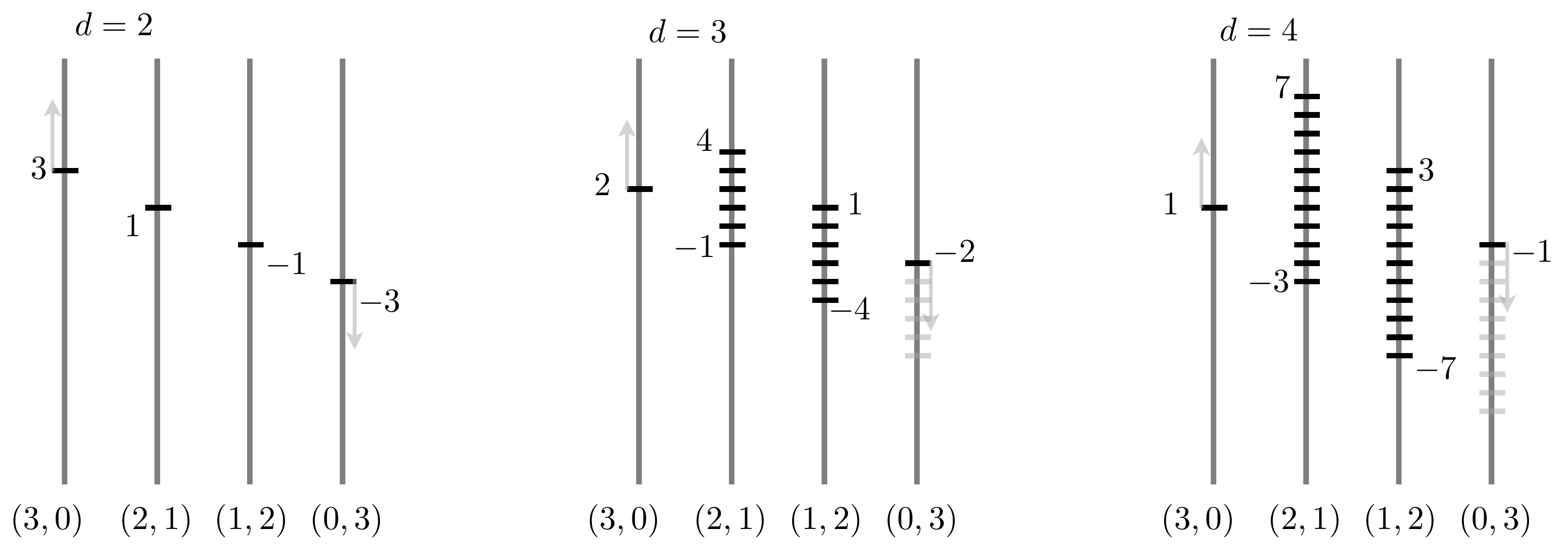}
	\caption{\small The range of twisted cohomologies for the Fermat quadric, cubic, quartic ($d=2,3,4$ respectively). In degrees $(2,1)$ and $(1,2)$, a tick denotes a degree for which twisted cohomology is non-vanishing; the actual dimensions are equal to the ones for $H^1(M_6,T(k))$ in Table \ref{tab:jacobi}. In degree $(3,0)$, cohomology is non-zero for $k\ge$ the black tick; in $(0,3)$, for $k\le$ the black tick.}
	\label{fig:fermat}
\end{figure}

Using the result of the above computation of (\ref{eq:tw11}), one can now obtain \eqref{eq:tw21} by acting on $\Omega_0$ with the twisted Beltrami differentials. For example, for $d=3$, the index is $\jst=2$, so $\Omega_0 \in H^{3,0}(M_6, {\cal L}^2)$. Acting with the $\mu$'s in Table \ref{tab:jacobi}, we find that $H^{2,1}(M_6,{\cal L}^k)\neq 0$ for $-1\le k\le 4$. For $d=4$, a similar computation shows that  $H^{2,1}(M_6,{\cal L}^k)\neq 0$ for $-3\le k\le 7$. We summarize the situation in figure \ref{fig:fermat}.

In figure \ref{fig:fermat} we have also shown the situation for the quadric, the case $d=2$. In that case, we see that there is only one element of the Jacobi ring, which is the constant 1. Further inspection reveals that this single element has charge $-2$: in other words, the only twisted Beltrami is the single generator of $H^1(M_6,T(-2))$. Acting on $\Omega_0$, which has charge 3, this gives one element $\omega\in H^{2,1}(M_6,{\cal L})$. The real and imaginary parts of $\Omega_0$ and $\omega$ can be identified, up to a linear redefinition, with the basis of three-forms called $\alpha_A$ and $\beta^A$, for $A=0,1$ in \cite{cassani-koerber-varela}, obtained using the coset structure of the manifold mentioned below \eqref{eq:fermat}. The twisted $(2,1)$-form also appears in \cite[App.~D]{eager-schmude}. We will come back to this case later. 

There are other instructive cross-checks. The $k=0$ case, $H^1(M_6,T)$, represents untwisted Beltrami differential. We have just seen that there are none in the $d=2$ case, while for $d=3$ and 4 we see from Table \ref{tab:jacobi} that there are respectively 10 and 45. These numbers can also be obtained by recalling that complex structure deformations can be obtained by deforming the polynomial $p$ away from the Fermat point. There are $\binom{d+4}{4}$ such deformations. However, some of these can be undone by linear redefinitions of the coordinates $z_i$ of $\cc\pp^4$, which make up the group $GL(5)$, of dimension 25. Thus we arrive at $\binom{d+4}{4}-25$. For $d=3$ and 4 this is indeed equal to 10 and 45 respectively. For $d=2$ it is negative, signalling that there are indeed no complex deformations. 

Another important case is when the $(2,1)$ forms obtained from the Beltrami are untwisted. For $d=2$, this cannot happen, so ${\rm dim} H^{2,1}(M_6)=0$. For $d=3$, since $\Omega_0$ has charge 2, we can obtain untwisted $(2,1)$-forms by acting on $\Omega_0$ with a $\mu$ of charge $k=-2$; from Table \ref{tab:jacobi} we see that there are 5 such differentials, so ${\rm dim} H^{2,1}(M_6)=5$. Finally, for $d=4$ we can act on $\Omega_0$ (of charge 1) with 30 $\mu$'s of charge $-1$, so ${\rm dim} H^{2,1}(M_6)=30$. These numbers match with \cite[p.215]{EMS-AGV}, and with a simple index calculation (taking into account that ${\rm dim} H^{1,1}=1$ in all these cases). 

Notice that we have considered only $d<5$ because in this paper we are interested in the case of K\"ahler--Einstein manifolds with positive curvature; however, the computations in this section also apply to the case $d=5$, and in particular, recalling (\ref{eq:J-dim}), lead to a $4^5$-dimensional twisted cohomology. This includes the familiar $101$-dimensional untwisted $(2,1)$ cohomology, which in this case is also identified with the space of complex deformations. It might be interesting to consider reductions in which this cohomology needs to be considered. Line-bundle-valued cohomology was already used (for line bundles whose $c_1=0$) in heterotic reductions on Calabi--Yau's in \cite{blesneag-buchbinder-candelas-lukas}. 

Let us also have a look at the explicit expressions for the twisted forms and Beltrami differentials described above. We obtain these by explicitly computing the map called $\pi$ in (\ref{eq:long-N}), following the definitions leading from the short exact sequence (\ref{eq:N}) to the associated long exact sequence (\ref{eq:long-N}). Consider a section $s$ in $H^0(M_6,{\cal O}(d+k))$. Since $\phi$ in (\ref{eq:N}) is surjective,\footnote{Note that this is not the case with the related map $\phi$ in \eqref{eq:long-N}, which is not surjective.} we can define its preimage, $s'$, a section of the bundle $TP(k)|_{M_6}$. Consider now the derivative $\bar \del s'$ and compute
\begin{equation}
 \phi \bar \del s'=\bar\del \phi s'=\bar \del s=0 \,,
\end{equation}
where we used the fact that $\bar\del s=0$. (For notational simplicity, we denote by $\bar\del$ all the Dolbeault differentials in the various bundles; they should all be understood as being appropriately twisted.) Since (\ref{eq:N}) is exact, the kernel of $\phi$ should equal the image of $i$. Thus there should exist a section $s''$ of the bundle $TM_6$ such that 
\begin{equation}
 \bar\del s'=i s'' \,.
\end{equation}
This will now satisfy $0=\bar \del i s''=i \bar\del s''$, and since $i$ has zero kernel it follows $\bar \del s''=0$; in other words, $s''$ will be in $H^1(M_6,TM_6\otimes {\cal O}(k))= H^1(M_6, T(k))$, as desired. This is the general idea: let us now find out what these sections are explicitly. We can take
\begin{equation}\label{eq:def-vecs}
s'=\frac{s}{|\del f|^2}\sum_{i=1}^5\overline{\del_i f}\?\del_i\,,
\quad\text{where}\quad
|\del f|^2 \equiv \sum_{i=1}^5 \overline{\del_i f} \?\del_i f \ . 
\end{equation}
Indeed we can check from (\ref{eq:phi}) that $\phi s'= s$. Now we can compute (using repeated indices convention)
\begin{equation}\label{eq:s''}
	\mu=\pi s= \bar\del \left(\frac{s}{|\del f|^2}\overline{\del_i f}\?\del_i\right)= 
	s\? P_{ik}\? \overline{\del_j \del_k f}\? d\bar z^j \?\del_i \ ,
\end{equation}
where
\begin{equation}
	 P_{ik}\equiv \frac1{|\del f|^2} \left(\delta_{ik}-  \frac1{|\del f|^2}\? \overline{\del_i f}\?\del_k f\right)
\end{equation}
is a projector on $TM_6$. (Indeed it satisfies $P_{ik}\del_i f=0$.) We have omitted the map $i$, because it just instructs us to consider a vector in $TP$ which has no normal components (as $s''$ is) as a vector in $TM_6$.

Thus (\ref{eq:s''}) is the explicit expression of the twisted Beltrami associated to $s$ under the map $\pi$ in (\ref{eq:long-N}). Notice that it is proportional to $s$, which can be thought of as a polynomial in the Jacobi ring (\ref{eq:J}). The expression (\ref{eq:s''}) was also found for Calabi--Yau's in \cite{candelas-yukawa} using the differential geometry of the manifold $P$.

So far we have not made any statements about twisted $(3,0)$ cohomology. As it turns out, that is not finite-dimensional. For the examples in figure \ref{fig:fermat}, the ticks in $(3,0)$ cohomology merely represent the \emph{lowest} allowed absolute value of the charge, not the only one. In other words, $H^{3,0}(M_6,{\cal L}^k)\neq 0$ for all $k\ge 5-d$. Dually, all $H^{0,3}(M_6,{\cal L}^k)\neq 0$ for all $k\le d-5$. (For $d=2$, the dimensions can be found in \cite[App.~D]{eager-schmude}.)

\medskip

To summarize, in this section we have seen that for our hypersurfaces the space of twisted Beltrami differentials (\ref{eq:tw11}) and the space of twisted $(2,1)$-forms (\ref{eq:tw21}) are finite-dimensional, and in one-to-one correspondence with the Jacobi ring (\ref{eq:J}).


\subsection{Analogue of de Rham cohomology} 
\label{sub:derham}

Ordinary Dolbeault cohomology on a compact K\"ahler space is related to de Rham cohomology by the Hodge decomposition. As we remarked in section \ref{sub:gen}, this plays a crucial role in Calabi--Yau compactifications. We will now see that some analogue of this is also available for twisted cohomology. 

For ordinary Dolbeault cohomology, one chooses harmonic representatives, namely forms which are not only annihilated by $\bar\del$ but also by $\bar\del^\dagger$. One then observes that the Dolbeault Laplacian $\Delta_{\bar\del}=\bar\del \bar\del^\dagger + \bar\del^\dagger \bar\del$ is proportional to the ordinary Laplacian $\Delta= d d^\dagger + d^\dagger d$. Hence the harmonic representatives are in fact also annihilated by the Laplacian, and by a standard argument on a compact manifold they are also annihilated by $d$ and $d^\dagger$. In particular, they are in de Rham cohomology.

For twisted Dolbeault cohomology, the argument is a little different, but similar in spirit. We will use the following result\footnote{We could also invoke more directly \cite[Thm.~2.3]{akahori-garfield}.} for the Laplacian acting on primitive $p$-forms on a Sasaki--Einstein space \cite{schmude}
\begin{equation}\label{eq:Lapl-SE}
\Delta = 2\? \Delta_{\bar\del} - \pounds_{\xi}^2 -2\?\im\?(3-p)\?\pounds_{\xi}\,,
\end{equation}
where $\Delta_{\bar\del}= \bar\del \bar\del^\dagger + \bar\del^\dagger \bar\del$ is the Laplacian constructed out of the twisted Dolbeault differential $\bar\del$. We again select representatives which are ``harmonic'', namely belonging to the space 
\begin{equation}
 H^{2,1}_k \equiv \{\? \omega_k \in H^{2,1}(M_6, {\cal L}^k)\,\, |\,\, \bar\del^\dagger \omega_k=0\?\}\,.
\end{equation}
As pointed out for example in \cite[App.~C]{eager-schmude}, for $k\neq 0$ (which is the case of interest) we can assume such representatives to be primitive, namely annihilated by contraction with $J$. Moreover, as noted in section \ref{sub:tw-coh}, to such forms $\omega_k$ we can associate a well-defined form $\hat \omega_k \equiv \ee^{\im\?k \hat \psi} \omega_k$ on $M_7$. This has the feature that
\begin{equation}
	\pounds_{\xi} \hat\omega_k = \im\? k\? \hat \omega_k\,.
\end{equation}
and that $\iota_\xi \hat \omega_k = 0 $. Using these properties, \eqref{eq:Lapl-SE} reduces to 
\begin{equation}\label{eq:Dok}
	\Delta \omega_k = k^2 \hat\omega_k\,.
\end{equation}

Consider now the operator
\begin{equation}\label{eq:dk}
	d_k \equiv d - \im\?k \eta\? \wedge \,, 
\end{equation}
which obeys
\begin{equation}
	\{ d_k,d_k^\dagger\}= \Delta + \im\?k\? \pounds_\xi -\im\?k *\! \pounds_\xi \! * +k^2\,.
\end{equation}
On $\omega_k\in H^{2,1}_k$, recalling its primitivity, we have $*\omega_k = \im\? \omega_k $; so 
\begin{equation}
	\{d_k,d_k^\dagger\}\? \hat\omega_k = (\Delta -k^2)\? \hat\omega_k\,.
\end{equation}
 Comparing with (\ref{eq:Dok}) we see that 
\begin{equation}
	d_k \hat\omega_k = d_k^\dagger \hat\omega_k = 0 \,.
\end{equation}
In particular, $\hat\omega_k$ is annihilated by the operator $d_k$, which generalizes the usual de Rham differential. Notice that it is not in general a differential, given that $d_k^2 = -\im\?k\? d\eta\wedge$; but it is on forms annihilated by $d\eta\wedge$, which is the space of primitive forms when $d \eta = J$.


\subsection{Family of three-form deformations} 
\label{sub:fam}

In subsection \ref{sub:hypersurf} we have performed a detailed computation of the twisted $(2,1)$ cohomology defined in (\ref{eq:tw21}) for Fermat hypersurfaces; we have found that the result is in one-to-one correspondence with the Jacobi ring (\ref{eq:J}). In fact this is a general conclusion for hypersurfaces. It is also not hard to imagine how to generalize both our computation and our conclusion for complete intersections, i.e.~manifolds $M_6$ defined by $s$ equations in $\cc\pp^{s+3}$. In fact, the incarnation of the same structures on the complex cone over $M_6$, which is by definition a Calabi--Yau manifold, are known to have an extension to the case of complete intersections, as explained briefly in section \ref{sec:CR-cone} and in Appendix \ref{app:ex}.

Let us now try to make (\ref{eq:Omegasum}) more precise. It is instructive to first look back at deformations of ordinary complex structures, which are associated to untwisted Beltrami differentials. (In our hypersurface examples of section \ref{sub:hypersurf}, those are present for $d=3$ and 4.) On the base $M_6$, $\Omega_0$ in \eqref{eq:OmegaSE-2} satisfies $d \Omega_0= 4w\wedge \Omega_0$: this equation is in fact equivalent to integrability of the complex structure $I_0$ associated to $\Omega_0$.\footnote{The connection $4w$ is usually called $W_5$ in the literature about SU(3) structures in six dimensions.} Acting with an untwisted Beltrami $\mu$ at the infinitesimal level produces a $(2,1)$ deformation $\mu\cdot \Omega_0$; at the finite level, one can integrate this as $\Omega=\ee^{\mu \cdot} \Omega_0$. A check that this is the correct finite expression is that $\Omega$  satisfies
\begin{equation}\label{eq:dO}
	d \Omega = 4\? w\wedge \Omega
\end{equation}
if and only if $\mu$ satisfies the Kodaira--Spencer equation. 

For twisted Beltrami differentials, a similar story applies, with some changes. We have several different possible charges; as we did earlier, we denote these charges by an index, so that for example $\mu_k \in H^1(M_6, T\otimes {\cal L}^k)$. We can formally collect the twisted Beltrami with all the different allowed charges in a single object, $\mu$, defined as 
\begin{equation}\label{eq:mu-tot}
\mu= \sum_k \ee^{\im k \hat \psi} \mu_k \,,
\end{equation}
where we recall that we introduced $\hat\psi= \frac{4}{\jst} \psi$ in \eqref{eq:Omegasum} to take into account the possibility that the canonical $K$ might have a root. When viewed as a form on the base (by fixing $\psi$ to 0, for example), \eqref{eq:mu-tot} is a formal sum of sections of different bundles on $M_6$. We can think of the full expression \eqref{eq:mu-tot} as tensor on $M_7$, analogous to a Beltrami differential on a complex manifold.

We can use the $\mu$ in \eqref{eq:mu-tot} to deform the three form at the Sasaki--Einstein point, $\Omega_\se$ in \eqref{eq:OmegaSE-d}, as:
\begin{equation}\label{eq:mu.O}
\mu \cdot \Omega_\se= 
\mu \cdot \left( \ee^{\im\? \jst\hat \psi}\Omega_0 \right) = 
\sum_k \ee^{\im\?(k + \jst)\hat \psi}\mu_k \cdot \Omega_0 \ .
\end{equation}
Again this can be seen as a form on $M_7$, or on $M_6$ as a twisted $(2,1)$-form with mixed charges; in other words, a section of
\begin{equation}\label{eq:Htw}
	H^{2,1}_{\rm tw} \equiv \oplus_k H^{2,1}(M_6,{\cal L}^k)\ ,
\end{equation}
where each value of the charge $k$ corresponds to a set of forms $\omega_k=\ee^{\im\?(k + \jst)\hat \psi}\mu_k \cdot \Omega_0$ as in \eqref{eq:Omegasum}, satisfying
\begin{equation}
d \omega_k = 4\?\left(\frac{k}{\jst} +1 \right)\? (d\psi+w)\wedge \omega_k \,.
\end{equation}
In the hypersurface examples of last subsection, the allowed values for $k$ are depicted in figure \ref{fig:fermat} (the ticks in the second vertical line) and the form in \eqref{eq:mu.O} can be thought of as having components along this entire range.

At the finite level, (\ref{eq:mu.O}) should get integrated to
\begin{equation}\label{eq:Omu}
	\Omega= \ee^{\mu\cdot }\Omega_\se\ ,
\end{equation}
again in analogy with the untwisted case.
The Kodaira--Spencer equation is now equivalent not to (\ref{eq:dO}) but to {\it the last equation} in (\ref{eq:su3-struct}).
Again, the crucial conceptual difference with the untwisted case is that now acting with $\mu$ on $\Omega_\se$ results in an object like (\ref{eq:Omegasum}), which on $M_6$ would be a sum of sections of different bundles, while on $M_7$ is a perfectly well-defined three-form.\footnote{Alternatively to (\ref{eq:Omu}), it would also be possible to parametrize deformations by ${\rm {\rm Re} }\Omega$, and use the techniques in \cite{hitchin-gcy} to obtain ${\rm Im} \Omega$ from it. The main features of the discussion below would not change.}

We have already commented on the $(3,0)$ and $(2,1)$ parts of (\ref{eq:Omu}). The $(1,2)$-part $\frac12 (\mu\cdot)^2 \Omega_0$ would have components even outside the finite cohomology range; those components exist, but are necessarily exact. For example, in figure \ref{fig:fermat} for $d=3$, if we consider an element $\mu_2\in H^2(M_6,T\otimes {\cal L}^2)$, then $\mu_2\cdot \Omega_0$ is in $H^{2,1}(M_6,{\cal L}^4)\neq 0$, while $(\mu_2\cdot)^2 \Omega_0 \in H^{1,2}(M_6,{\cal L}^6)$. This cohomology vanishes because 6 is outside the range $[-4,1]$; so $(\mu_2\cdot)^2 \Omega_0$ is exact. 

Finally, the $(0,3)$ part of (\ref{eq:Omu}) is $\frac16 (\mu\cdot)^3 \Omega_0$. This has a component proportional to $\bar\Omega_0$, which is in $H^{0,3}(M_6,{\cal L}^{\jst})$. However, it also has components in lower degrees, not all of which vanish. For example, for $d=3$ in figure \ref{fig:fermat}, $ (\mu\cdot)^3 \Omega_0$ has components in degrees $\in [-7,-2]$; these degrees are denoted in the figure by gray ticks.

These $(0,3)$ components are a priori not exact. While the presence of this ``tail'' is a departure from the more familiar Calabi--Yau setting, it has in fact little effect. Recall that all these $(0,3)$ components are in fact proportional to each other. In the hypersurface case, we can see this explicitly from (\ref{eq:s''}): $s$, a polynomial in the Jacobi ring (\ref{eq:J}), appears there multiplicatively.
Multiplying $\Omega$ by a suitable function $f$, one can then make sure that the $(3,0)$ and $(0,3)$ parts of $f \Omega$ are actually proportional, up to $\bar\del_0$-exact terms. We can moreover choose $f$ such that $\bar\del_0 f=0$. (Recall that $\bar\del_0$ is the twisted Dolbeault differential on $\oplus_k {\cal L}^k$.)

Let us show this for our example of hypersurfaces. The $(0,3)$-part of $\Omega$, $(\mu\cdot)^3 \Omega_0$, can be written as $s^3 \mu_{-d}^3\cdot\Omega_0$, where $\mu_{-d}$ is the single generator of $H^1(M_6,T\otimes{\cal L}^{-d})$; $\mu_{-d}^3\cdot \Omega_0$ is the lowest gray tick in figure \ref{fig:fermat}, and the various terms in $s^3$, where $s$ is the polynomial mentioned above, generate all the forms of higher degree. The form $Q \mu_{-d}^3\cdot \Omega_0$, where $Q$ is the single generator of the Jacobi ring of highest degree (as in Table \ref{tab:jacobi}), is equal to $\bar \Omega_0$ up to exact terms. We now need to find an $f$ such that 
\begin{equation}\label{eq:f-resc-Om}
f\? s^3 \?\mu_{-d}^3\!\cdot\! \Omega_0=f\? Q\? \mu_{-d}^3 \!\cdot\! \Omega_0\,,
\end{equation}
up to exact terms. For any polynomial $q$ of degree higher than $Q$, the form $q\? \mu_{-d}^3\!\cdot\!\Omega_0$ is exact. So a possible $f$ is given by $\frac{q}{s^3-Q}$. 

In the following two sections, we will consider the deformations from the point of view of the cone, naturally leading to the so-called Kohn--Rossi cohomology $H^{p,q}_{\stgdb}$ on $M_7$, that generalizes the twisted cohomology $H^{p,q}_{\rm tw}\equiv\oplus_k H^{p,q}(M_6,{\cal L}^k)$ we considered so far. This new cohomology is ``adapted'' to the deformed $\Omega$, in the sense that $\Omega$ will belong to $H^{3,0}_{\stgdb}$; the variation $\delta \Omega$ will then be in $H^{2,1}_{\stgdb}$ and the function $f$ in \eqref{eq:f-resc-Om} will be an element of $H^{0,0}_{\stgdb}$, fixed by appropriate requirements on the three-form. The corresponding theory of deformations of CR structure is described in terms of these objects.


\section{Isolated singularities and their deformations}
\label{sec:CR-cone}

In this section, we provide some mathematical background on cones described as singular hypersurfaces in $\cc^5$, focusing on the algebraic structure of their deformations and their (co)homology. In this paper, we are interested in hypersurfaces that are in addition Calabi--Yau,\footnote{This excludes toric CY cones, which cannot be described in this way in general, but allows for a straightforward extension to complete intersection CY's.} whose base is by definition Sasaki--Einstein, but most of the considerations in this section are independent of this requirement. 
Section \ref{sec:singularities} discusses the algebraic description of the deformations away from a conical singularity, providing several examples, including the cones over the compact \KE hypersurfaces of section \ref{sub:hypersurf}. We then proceed to discuss a similar description of the (co)homology of the deformed manifolds in section \ref{sec:monodromy}. Finally, section \ref{sub:fixed} discusses the definition of a fixed, real, basis of (co)homology, near the boundary of the deformed manifold, which will be useful in the next section.

\subsection{Isolated singularities}
\label{sec:singularities}

In this section we collect useful facts about isolated singularities, focusing on the Milnor fibration and the monodromy operator. The noncompact Calabi--Yau manifolds we consider in the rest of this paper are a subset of this class, since isolated singularities need not be Ricci flat. For simplicity, we refer only to hypersurface singularities, described by a single complex equation in $\mathbb{C}^{n+1}$, commenting on the generalization to complete intersections in Appendix \ref{app:ex}. In this subsection we will keep the dimension $n$ general, while from the next one we will focus on the case of interest to us, $n=4$.

Consider a function $f(z)$, where $z\in \mathbb{C}^{n+1}$, such that it has an isolated singularity at the origin, i.e.~$df\bigr|_{z=0}=0$ and we assume that $f(0)=0$. In order to study the resulting hypersurface, we consider a deformation away from the singular value for $f$ and define the manifolds
\begin{equation}\label{eq:disk-def}
 M_\lambda = \{\, z \in \mathbb{C}^{n+1}\,\, |\,\,\, f(z)= \lambda \, \}\,;
\end{equation}
$M_0$ is then the $n$-dimensional manifold containing the singularity. We also introduce the link, by intersecting $M_\lambda$ with the $2\?n+1$-dimensional sphere, as
\begin{equation}\label{eq:M7}
	M_{2n-1}= M_\lambda \cap \{\, \sum_i |z_i|^2 =1\, \}\,.
\end{equation}
We will also sometimes work with $\bar M_\lambda=M_\lambda\cap \{ \sum_i |z_i|^2 \le 1\, \}$, the part of $M_\lambda$ which is ``inside'' $M_{2n-1}$, such that $\del \bar M_\lambda = M_{2n-1}$. 

\begin{figure}[ht]
	\centering
		\includegraphics[scale=.35]{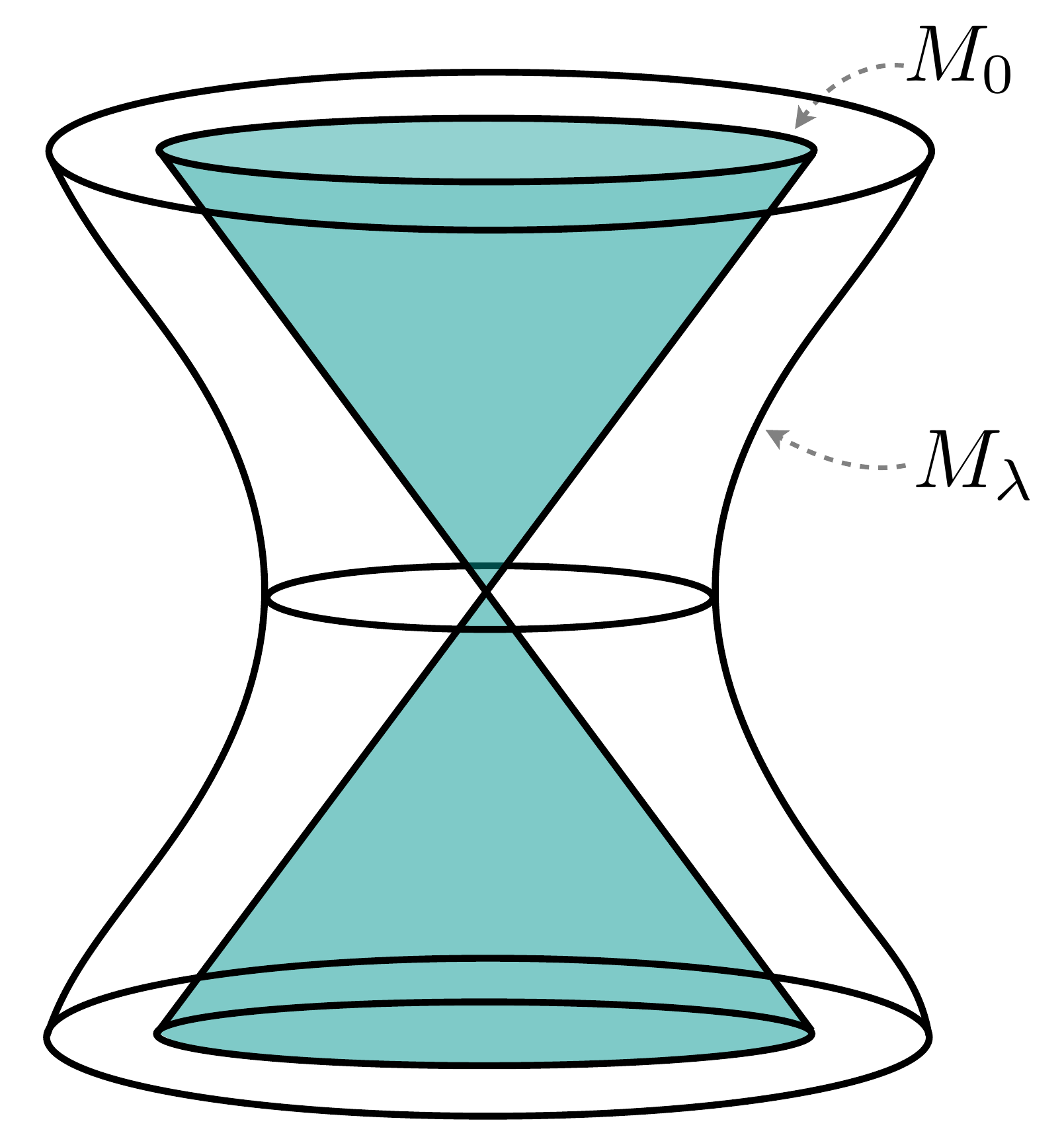}
	\caption{\small The singular manifold described by $f(z)=z_1^2+z_2^2$ is the cone $M_0$. When the small real deformation $\lambda$ in \eqref{eq:disk-def} is turned on, the resulting smooth manifold $M_\lambda$ features a finite one-cycle with size controlled by the deformation parameter $\lambda$.}
	\label{fig:MilFib}
\end{figure}

All this is illustrated for a simple example in Figure \ref{fig:MilFib}; in this case a $\lambda \neq 0$ makes a single finite one-cycle emerge. By a famous result of Milnor, for general hypersurface (later extended to complete intersection) singularities, $M_\lambda$ has the homology type of a bouquet of $n$-dimensional spheres: $H_n(M_\lambda)$ is its only nontrivial (co-)homology group. The number of spheres and the dimension of the middle cohomology group are identified with the Milnor number of the singularity $\mu$, which we defined already in (\ref{eq:mu-def}) as the dimension of the Jacobi ring (\ref{eq:J}). (Note that this quotient can in principle be infinite-dimensional, but in this paper we will deal exclusively with polynomials, for which \eqref{eq:mu-def} is always finite.)

\begin{figure}[t]
	\centering
		\includegraphics[width=14cm]{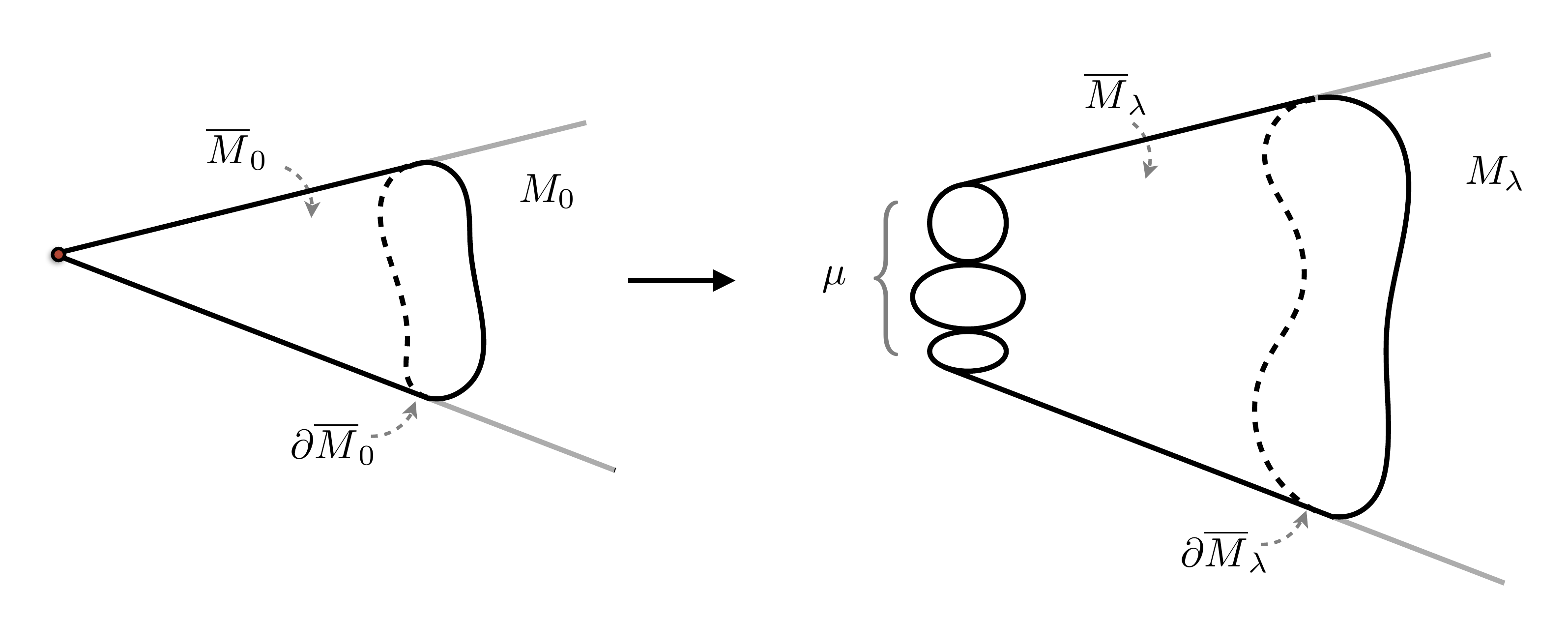}
	\caption{\small An illustration of the deformation of a conical singularity to a bouquet of $\mu$ spheres, when passing from the homogeneous function $f(z)$ to the unfolding $F(z,t)$ in \eqref{eq:unfolding}. The point-like singularity is replaced by a deformed space that features nontrivial cycles, which vanish when the deformation parameters $t^a$ are switched off.}
	\label{fig:cone}
\end{figure}

An important example class, which includes the explicit CY examples considered later, are the Brieskorn--Pham manifolds, defined by a sum of monomials, as
\begin{equation} \label{eq:Brie-def}
f_{\scriptscriptstyle\sf BP} = \sum_{i=1}^{n+1} z_i^{d_i}\,, 
\end{equation}
where the $d_i\ge2$ are integers and the Milnor number turns out to be 
\begin{equation} \label{eq:Brie-mil}
\mu_{\scriptscriptstyle\sf BP} = \prod_{i=1}^{n+1} (d_i-1)\,. 
\end{equation}
A well known example in this class are the $A_\mu$ singularities, for which 
\begin{equation} \label{eq:Ak-def}
f_{\scriptscriptstyle\sf A_\mu} = z_1^{\mu+1}+\sum_{i=2}^{n+1} z_i^{2}\,.
\end{equation}
In four dimensions, i.e.~for $n=2$, these manifolds admit the well known Gibbons--Hawking metrics \cite{Gibbons:1979zt}. In this case, the deformation to a non-singular manifold is easy to obtain by decomposing the $\mu$-th order singularity in \eqref{eq:Ak-def} to a metric which explicitly features $\mu+1$ non-singular centres and $\mu$ spheres defined between them, as reviewed in more detail in Appendix \ref{app:GH}. This situation is illustrated in Figure \ref{fig:cone}.

In higher dimensions, the explicit metrics are not easy to obtain. Nevertheless, one may describe the deformed manifolds algebraically, using the $\mu$-dimensional space of functions defined by \eqref{eq:J} to deform the function $f(z)$, in such a way that the resulting manifold is non-singular. Denoting the basis elements of the Jacobi ring $\jac$ in \eqref{eq:mu-def} by $\jac_\alpha$, where $\alpha=1,\ldots,\mu$ and introducing corresponding complex parameters $t^\alpha$, one can define this generic deformation as
\begin{equation}\label{eq:unfolding}
 M_{t^\alpha}= \{\, z \in \cc^{n+1} \,\,| \,\,\, F(z,t) \equiv f(z) +  t^\alpha \? \jac_\alpha = 0 \, \} \,,
\end{equation}
which is usually called an unfolding. For example, for the case of Brieskorn--Pham manifolds \eqref{eq:Brie-def}, a basis for the Jacobi ring (\ref{eq:J}) is given by the monomials
\begin{equation}\label{eq:Jac-ring}
 \jac^{\scriptscriptstyle\sf BP} = 
 \{\, 1\,, z_i\,, z_{i}\,z_{j}\Bigr|_{i\neq j} \,, \dots \,, \prod_{i=1}^{n+1} z_i^{d_i-2}\, \}\,.
\end{equation}
For $d_i=d=3,4$ and $n=4$, these are identical to the ones shown in Table \ref{tab:jacobi}. 
The unfolding \eqref{eq:unfolding} corresponds to a generic deformation of the singularity that preserves the topological properties, in particular the middle (co)homology.

In general, the deformation \eqref{eq:unfolding} is not equivalent to the complex deformations of the singular manifold, whose number is given by the so-called Tjurina number
\begin{equation}\label{eq:tau-def}
 \tau \equiv \dim\frac{\cc[z_i]}{\langle\? f, \partial_i f \?\rangle}\,. 
\end{equation}
This is clearly similar to the expression for the Milnor number in \eqref{eq:mu-def}, and in fact one finds that $\mu\ge\tau$. The inequality is saturated for the so-called quasihomogeneous singularities, which are invariant under rescaling the coordinates as
\begin{equation}\label{eq:Cs-action}
 z_i \rightarrow \zeta^{d/d_i} z_i \quad \Rightarrow \quad f(z) \rightarrow \zeta^d f(z) \,,
\end{equation}
for $\zeta\in\mathbb{C}$ and some integer $d$ and $d_i$. For example, the Brieskorn-Pham class in \eqref{eq:Brie-def} satisfies this requirement. It follows that for quasihomogeneous singularities, one may identify the space of complex deformations \eqref{eq:tau-def} with the general deformations in \eqref{eq:J} and use the unfolding in \eqref{eq:unfolding} to describe them. In this paper, we will only consider quasihomogeneous singularities, as our examples in section \ref{sec:cmplx-defs} belong to the Brieskorn-Pham class. 

The $\mathbb{C}^*$ action defined by \eqref{eq:Cs-action} leaves $M_0=\{f=0\}$ invariant. At the same time, it maps the deformed manifolds $M_\lambda=\{f= \lambda\}$ into one another. This leads to the concept of monodromy.\footnote{The monodromy is fundamental in the study of general singularities, but it is particularly simple for quasihomogeneous singularities.}
Consider for example the path in the moduli space of the deformations $M_\lambda$ defined by
\begin{equation}\label{eq:disk-def-2}
  \lambda = \delta\, \ee^{\im\?\theta}\, ,\qquad 0\le\theta\le2\pi\,.
\end{equation}
Although $M_\delta=M_{\delta \ee^{2 \pi \im}}$, the $\mu$ cycles in the middle homology will be mixed nontrivially. This operation defines a $\mu\times\mu$ matrix, $\mathfrak{m}$, acting on $H_n(M_{\delta})$, known as the monodromy operator.

By a fundamental theorem of singularity theory, the eigenvalues of the monodromy are of unit absolute value, usually parametrised as $\exp(2\pi\im\?\nu_\alpha)$ for a set of real $\nu_\alpha$ with $\alpha=1,\ldots,\mu$, called the spectrum of the singularity. The diagonalisation is justified in this paper as $\mathfrak{m}$ admits a semisimple representation for singularities arising from quasi-homogeneous polynomials $f(z)$.
One can compute both the eigenvalues $\nu_\alpha$ and the corresponding multiplicities $b_\alpha$ for a hypersurface in the class \eqref{eq:Brie-def} using the so-called spectral polynomial \cite[Prop. 7.27]{dimca1987topics}
\begin{equation}\label{eq:spec-pol}
 S(T) \equiv \frac{1}{T^2}\prod_{i=1}^{n+1}\frac{T^{1/d_i} -T}{1-T^{1/d_i}} 
 = \sum_{[\alpha]}\, b_{[\alpha]}\, T^{[\nu_\alpha]}\,,
\end{equation}
where in the second equality we expand $S(T)$ in monomials and the sum over $[\alpha]$ is over the distinct values of the $\nu_\alpha$.
Whenever explicit values for the quantities $\nu_\alpha$ and $b_\alpha$ are given below, they are understood as arising from \eqref{eq:spec-pol}, unless stated otherwise.

We now give some details on the example hypersurfaces considered in Section \ref{sub:hypersurf} from the point of view of the cone. These examples are contained in the Brieskorn-Pham class \eqref{eq:Brie-def}, upon setting $n=4$ and all exponents equal, i.e.~$d_i=d$, with $d=2,3,4$. The resulting noncompact manifolds are identified with the Calabi--Yau manifolds constructed as complex cones over the hypersurfaces \eqref{eq:fermat}. The Milnor number is given by \eqref{eq:Brie-mil} as $\mu=(d-1)^5$, which indeed matches with the total number of twisted Beltrami differentials in \eqref{eq:J-dim}. The spectral polynomials for these three singularities read
\begin{align}\label{eq:ex-def}
 d=2:& \, \quad S(T)= T^{1/2} \cr 
 d=3:& \, \quad S(T)= T^{-1/3} + 5 + 10\, T^{1/3} + 10\, T^{2/3}+ 5\,T + T^{4/3} \\
 d=4:& \, \quad S(T)= T^{-3/4} + 5\,T^{-1/2} + 15\,T^{-1/4} + 30 + 45\,T^{1/4} + 51\,T^{1/2} \cr
     & \hspace{1.6cm} + 45\,T^{3/4} + 30\,T + 15\,T^{5/4} + 5\,T^{3/2} + T^{7/4}\,. \nonumber
\end{align}
Notice that the coefficients of in each monomial, corresponding to the multiplicities of each eigenvalue, reproduce precisely the dimensions of the cohomologies in Table \ref{tab:jacobi}. The ranges for the eigenvalues $\nu_\alpha$ also reproduce the ones presented in figure \ref{fig:fermat}, up to a normalization factor of $d$. The hypersurface for $d=2$ with the single deformation in \eqref{eq:ex-def} turned on is described by the well known 8-dimensional Ricci-flat Stenzel metric \cite{Stenzel1993,Cvetic:2000db}. The corresponding metrics for the deformed hypersurfaces for $d=3,\,4$ are not known explicitly, but it was shown in \cite[Rem.~5.3--5.4]{Conlon-Hein-1} that the subset of the deformations in \eqref{eq:Jac-ring} that preserve the order of the defining polynomial also admit Ricci-flat metrics.

As we mentioned in section \ref{sub:hypersurf}, it is also interesting (although not immediately relevant for the present paper) to consider the case $d=5$, for which the spectral polynomial reads:
\begin{align} \label{eq:quintic-forms}
 S(T)=&\, T^{-1} + 5\,T^{-4/5} + 15\,T^{-3/5} + 35\,T^{-2/5} + 65\,T^{-1/5} + 101 
 \cr
     & + 135\,T^{1/5} + 155\,T^{2/5} + 155\,T^{3/5} + 135\,T^{4/5} + 101\,T
 \cr
     &  + 65\,T^{6/5} + 35\,T^{7/5} + 15\,T^{8/5} + 5\,T^{9/5} + T^{2}\,, 
\end{align}
One indeed sees here the appearance of the 101 elements of untwisted cohomology in the vanishing eigenvalue/exponent sector.

\subsection{Cohomology and monodromy}
\label{sec:monodromy}

We now turn to a discussion of the cohomology of $M_\lambda$, and the way monodromy acts on the corresponding representatives.
To this end, we first define the vector field corresponding to the $\mathbb{C}^*$ action in \eqref{eq:Cs-action} on the hypersurface $f(z)=0$ as 
\begin{equation}\label{eq:boost-cone}
 X_0 = \sum_{i=1}^{5}\frac1{d_i}\?z_i\? \frac{\partial}{\partial z_i} \?.
\end{equation}
This action makes $M_0$ a complex cone, since the absolute value $|\zeta|$ of the dilatation parameter in \eqref{eq:Cs-action} corresponds to dilatations along the cone, while the phase $\arg{\zeta}$ corresponds to the so-called Reeb Killing vector $\xi_0 = {\rm Im} X_0$.

Let us now assume that the canonical bundle of $M_0$ is trivial, which is the case of interest for us. There exists then a globally defined holomorphic 4-form $\Omega_4^0$. (The subscript ${}_4$ is to remind us that it is a four-form on $M_0$, related to but not to be confused with the three-form $\Omega$ of section \ref{sec:cmplx-defs}; the superscript ${}^0$ marks the fact that it is the one appropriate for the conical case.) Like any such form on a complex manifold, integrability of the complex structure is equivalent to (\ref{eq:dO}). When $H^{0,1}=0$, the one-form $w$ can actually be taken to vanish, so that $d \Omega_4^0=0$.

For the case at hand, $\Omega_4^0$ can be constructed from the top holomorphic form on $\mathbb{C}^{5}$ as the form satisfying
\begin{equation}\label{eq:top-hol-res}
 dz^1\wedge\dots \wedge dz^5 = df\wedge\Omega_4^0\,.
\end{equation}
On a patch where $\partial_{1} f\neq 0$, for example, this can be represented as
\begin{equation}\label{eq:top-hol-res2}
 \Omega^0_4 = \frac{dz^2\wedge\dots \wedge dz^5}{\partial_1 f}\,,
\end{equation}
while similar representations can be written on other patches, with holomorphic transition functions between them. These representations fit into a globally defined holomorphic $4$-form without zeros or poles away from the tip of the cone. Note that $\Omega^0_4$ defined as above has a well-defined weight $[\Omega^0_4]=\sum_i \tfrac{d}{d_i}-d$ under \eqref{eq:Cs-action}. 

The above construction is known as the \emph{residue map} ${\rm Res}$, which takes a 5-form $\sigma$ on $\cc^5$ to a 4-form $ {\rm Res\?}\sigma$ on $M_0$, defined by: 
\begin{equation}
\int_C \sigma = \int_c {\rm Res\?}\sigma\,, 
\end{equation}
where $c$ is a 4-cycle in $M_0$ and $C$ is a 5-cycle obtained as a tube over $c$ (namely by considering the $S^1$ fibre of the normal bundle to $M_0$ over each point of $c$). Returning to $\Omega^0_4$ in (\ref{eq:top-hol-res2}), this form is obtained as 
\begin{equation}\label{eq:Om-res}
{\rm Res\?}\frac1f dz^1\wedge\ldots \wedge dz^5= {\rm Res\?} \frac{df}f \Omega^0_4 = \Omega^0_4\,.
\end{equation}
Intuitively, this is because one integrates over the ``coordinate'' $f$ around the $f=0$ locus. 

For a noncompact manifold, the holomorphic top form $\Omega^0_4$ is actually not in cohomology, since the volume form $\Omega^0_4\wedge\bar\Omega^0_4$ is exact. There is however a description in terms of residues for the cohomology as well. 
One can obtain cohomology representatives $\Omega_\alpha$ for the middle cohomology as
\begin{equation}\label{eq:Oa}
\Omega^0_\alpha = {\rm Res\?} \frac{\jac_\alpha}{f^2} \?dz^1\wedge\ldots \wedge dz^5\,.
\end{equation}
Since these forms are defined in terms of a basis $\{\jac_\alpha\}$ for the Jacobi ring, there are $\mu$ of them; thus they are in one to one correspondence with the elements of the set of deformations and the 4-cycles in $M_\lambda$. They also have well-defined weights  under \eqref{eq:Cs-action}, since 
\begin{equation}\label{eq:weights-1}
[\Omega^0_\alpha]=\sum_i \frac{d}{d_i} -2\, d + [\jac_\alpha]\,.  
\end{equation}
Note the similarity with the twisted Beltrami differentials \eqref{eq:s''}, for the particular examples in section \ref{sub:hypersurf}, which are also proportional to the elements of the Jacobi ring.

The weights in \eqref{eq:weights-1} are yet another manifestation of the monodromy eigenvalues $\nu_\alpha$ for a quasihomogeneous polynomial: 
\begin{equation}
[\Omega^0_\alpha] =\nu_\alpha  \,,
\end{equation}
as one can check using \eqref{eq:spec-pol}. They represent the action of the Reeb vector $\xi_0={\rm Im} X_0$ on the basis $\{\Omega^0_\alpha\}$ of the middle cohomology $H^4(M_0)$:
\begin{equation}\label{eq:Lie-defs}
\pounds_{\xi_0} \Omega^0_\alpha = \im\?\lambda_\alpha\?\Omega^0_\alpha\,.
\end{equation}
The constants $\lambda_\alpha$ in \eqref{eq:Lie-defs} are determined by the $\nu_\alpha$ once a convenient normalisation convention for the Killing vector $X_0$ is chosen. In this paper, we fix this convention to match the normalisation in \eqref{eq:SE-point} for the three-form on the Sasaki--Einstein base of the cone, implying that the holomorphic 4-form $\Omega^0_4$ has the canonical charge 4 under the action of $\xi_0$: 
\begin{equation}\label{eq:Lie-omega}
 \pounds_{\xi_0} \Omega^0_4 = 4\?\im\?\Omega^0_4\,.
\end{equation}
For an $n$-dimensional hypersurface as in \eqref{eq:Brie-def}, this choice results in the eigenvalues $\lambda_\alpha$ given by
\begin{equation}\label{eq:lambda-kappa}
 \lambda_\alpha = \frac{4}{\sum_i\tfrac{1}{d_i} - 1}\, \nu_\alpha \,.
\end{equation}
In what follows, we prefer to use the eigenvalues $\lambda_\alpha$ to characterise the deformations and the cohomology, with the understanding that they are given in terms of the more fundamental eigenvalues of the monodromy, through \eqref{eq:lambda-kappa}.

We have seen that the eigenvalues of the monodromy are related in a simple fashion to the eigenvalues of the Reeb action on the cohomology of $M_0$. It is perhaps not too surprising, then, that the latter can be extended to an action on the unfolding \eqref{eq:unfolding}. We can deform the vector field \eqref{eq:boost-cone} to one that leaves $M_{t^\alpha}$ invariant:\footnote{Note that the second term in \eqref{eq:boost-cone-def} is directly analogous to the vector fields \eqref{eq:def-vecs}, which determine the deformations of the \KE base.}
\begin{equation}\label{eq:boost-cone-def}
 X = X_0 + \frac{t_*^\alpha\? \jac_\alpha}{\sum_j\del_j F(z,t)^2}\sum_{i=1}^{5}\del_i F(z,t)\?\frac{\partial}{\partial z_i}\,,
\end{equation}
where the parameters $t_*^\alpha$ are determined by the linear equation
\begin{equation}\label{eq:t-star}
 X_0(t^\alpha\? \jac_\alpha) + t_*^\alpha\? \jac_\alpha = d\, t^\alpha\? \jac_\alpha \?.
\end{equation}
Note that $X_0(t^\alpha \jac_\alpha)$ is by definition an element of the Jacobi ring, if $f(z)$ is polynomial, so that \eqref{eq:t-star} can be solved as a linear system. The imaginary part of \eqref{eq:boost-cone-def} again identifies a U(1) vector field on the deformed hypersurface, for any value of the $t^\alpha$. 

The existence of such a vector field is intimately tied with the fact that away from special loci in the moduli space of $\{t^\alpha\}$, where some cycles of $M_{t^\alpha}$ collapse, the manifolds $M_{t^\alpha}$ are diffeomorphic to each other \cite[10.3.1]{Arnold-Gusein-Varchenko}. Since the identification of the action of the Reeb vector with the eigenvalues of the monodromy was done for a small $\delta$ in \eqref{eq:disk-def-2}, it then follows that such an identification can be extended for an appropriate vector for a more general deformation. Similarly, the cohomology bundle $H^p(M_\lambda)$ is naturally defined on the punctured disc where $\lambda$ takes values \cite[10.2]{hertling}, so that one again finds a $\mu$-dimensional cohomology bundle by extension to $M_{t^\alpha}$. A formal basis of forms $\Omega_\alpha$ on each $M_{t^\alpha}$ can then be constructed as in \eqref{eq:Oa}, upon replacing $f(z)$ by the unfolding $F(z,t)$. The action of $\xi$, the imaginary part of \eqref{eq:boost-cone-def}, can then be arranged to satisfy
\begin{equation}\label{eq:Lie-defs-fin}
\pounds_{\xi} \Omega_\alpha = \im\?\lambda_\alpha\?\Omega_\alpha\,,
\end{equation}
with the same eigenvalues $\lambda_\alpha$.

The statements of the previous paragraph form the basis for the construction of so-called Frobenius manifolds associated to quasi-homogeneous hypersurface singularities \cite[11.1]{hertling}. In particular, this relies on the extension of the forms in \eqref{eq:Om-res}--\eqref{eq:Oa} to the deformed manifold \eqref{eq:unfolding}, in terms of the so called primitive form \cite{Saito-universal, Saito-residue}. While this overarching structure will not be directly relevant for the considerations of this paper, we point out that this property identifies the cohomology spaces at the various points in the deformation family parametrised by the $t^\alpha$ and guarantees that the space of deformations is controlled by a single holomorphic function. This construction is analogous to what is known for the moduli spaces of Calabi--Yau compactifications, and in fact recently this approach has been applied to the computation of the \Kah potential on the complex structure deformations of the quintic (the subsector of vanishing charge in \eqref{eq:quintic-forms}), in  \cite{Aleshkin:2017fuz}.

The fact that one can choose de Rham representatives $\Omega_\alpha$ which are $(4,0)$ is unlike what happens in compact Calabi--Yau's, where the cohomology representatives have various $(p,q)$ degrees, and in fact $H^{k,0}$ is one-dimensional for $k=0,d$ and zero otherwise. But for a non-compact manifold the situation is a little different. Indeed for smooth Stein manifolds (complex submanifolds of $\cc^N$) it is known that Dolbeault cohomology vanishes for all degrees except $(p,0)$:
\begin{equation}\label{eq:cartanB}
	H^{p,q}=0 \, ,\qquad q\neq 0\,.
\end{equation}
(See for example \cite[Th.~2.4.6]{forstneric}; this is sometimes called ``Cartan's Theorem B''). De Rham cohomology is still related to Dolbeault, but by taking the $\del$ cohomology on the complex $H^{\bullet,0}$. Namely, 
\begin{equation}\label{eq:holdR}
	H^p_h\equiv H^{p,0}/\del H^{p-1,0}\,,
\end{equation}
which is often called \emph{holomorphic de Rham} group, is isomorphic to the usual de Rham cohomology. So in fact in the smooth case forms will have $(k,0)$ representatives. For example, the complex deformations of $\Omega_4$ will generate $(3,1)$-forms. But these forms will be equivalent in cohomology to a linear combination of the $(4,0)$ representatives $\Omega_\alpha$.\footnote{In particular, around a smooth point all infinitesimal complex deformations are in fact trivial, since they are parametrized by $H^{3,1}$, which vanishes. However, finite complex deformations can still be nontrivial, in general. We thank J.~Stoppa for an illuminating conversation on this point.} On the special loci of the moduli space where the singularity has not been completely smoothed out, the situation is a little more complicated, as we will explain in section \ref{sub:KR}.

\subsection{Fixed basis for homology and cohomology} 
\label{sub:fixed}

\begin{figure}[ht]
	\centering
	\includegraphics[scale=.15]{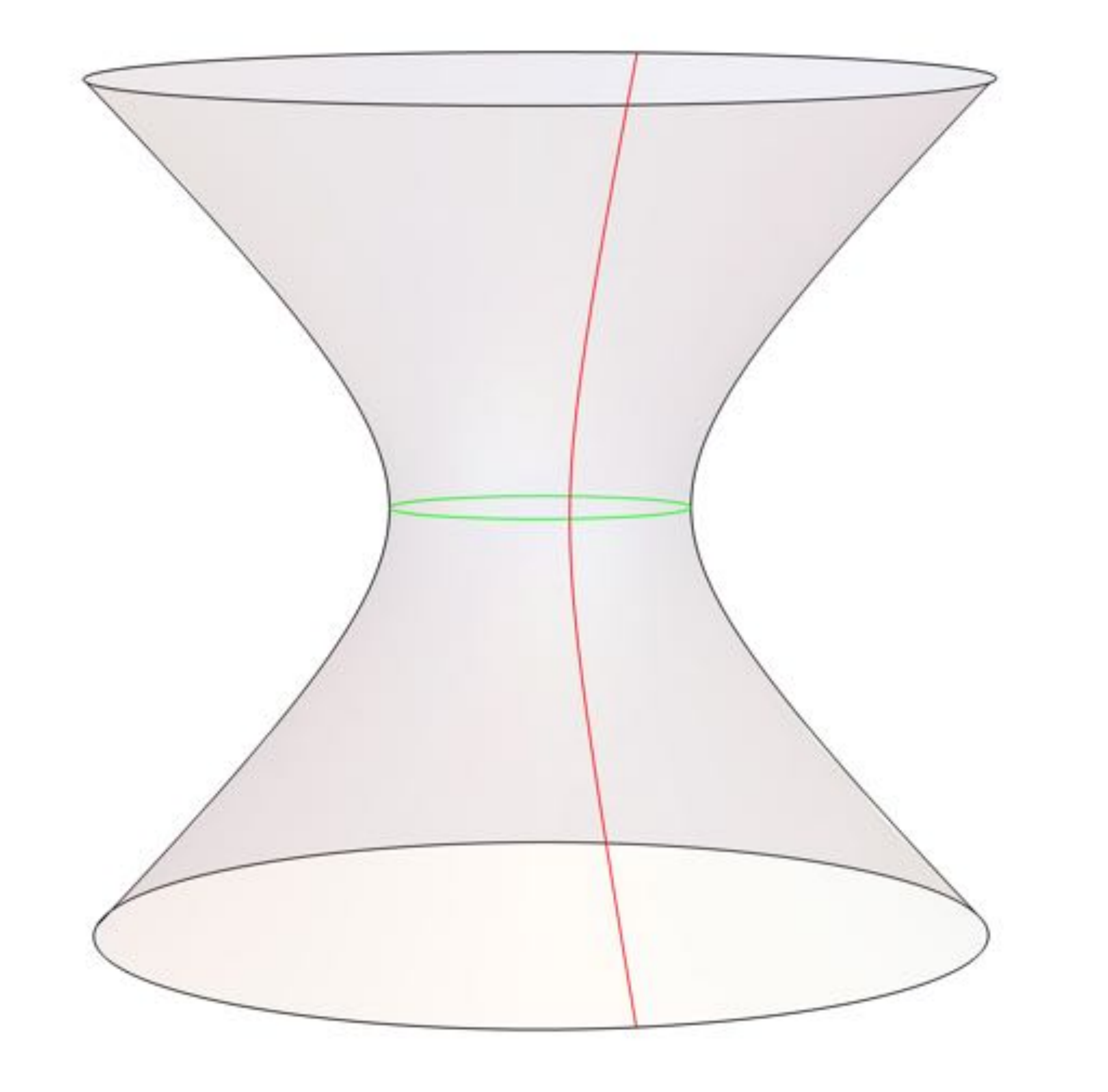}
	\includegraphics[scale=.15]{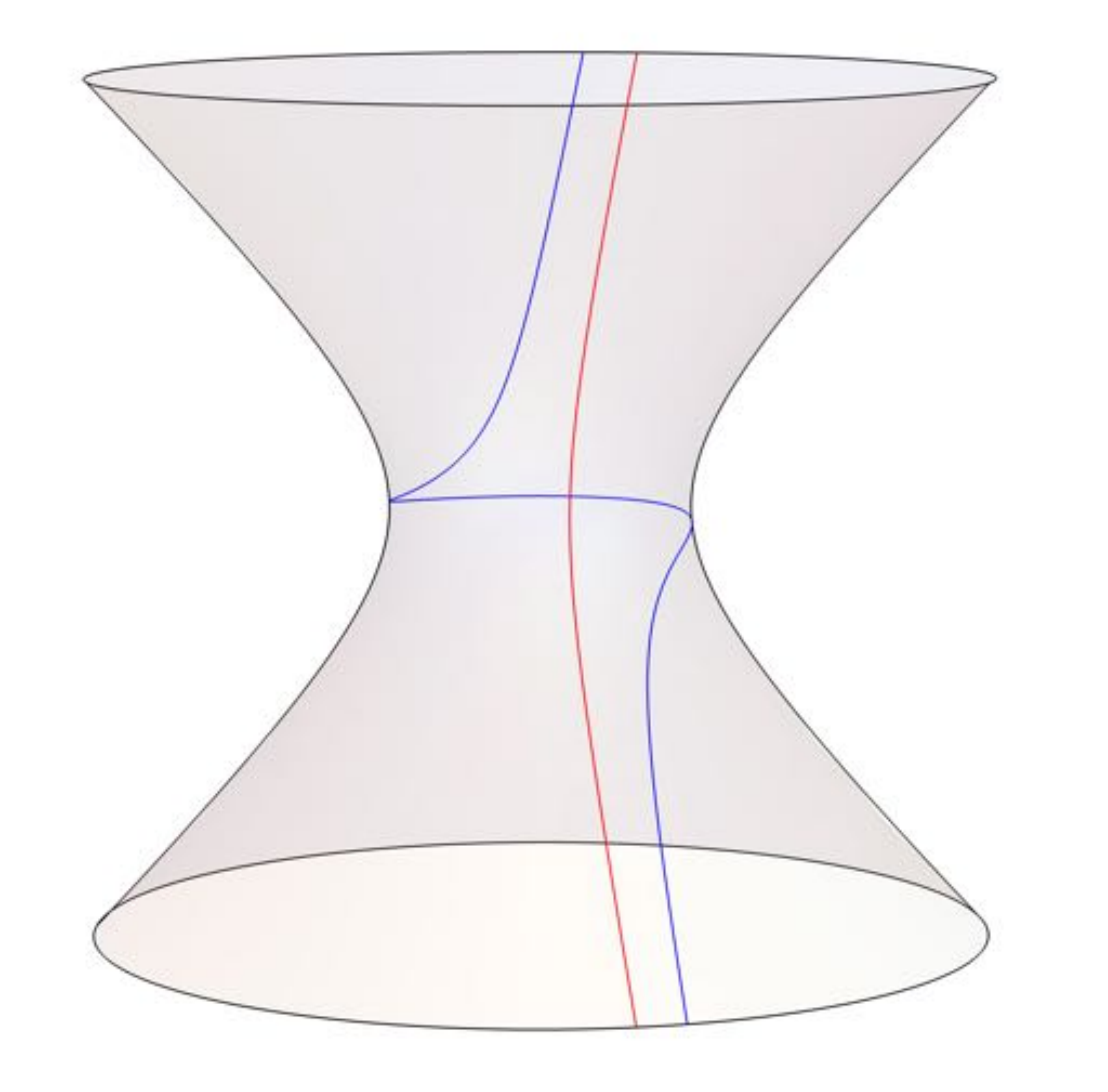}
	\caption{\small Left: a basis of one compact and one noncompact cycle for the singularity in Figure \ref{fig:MilFib}. Right: the noncompact cycle before and after a monodromy; the difference of the two noncompact cycles is homologous to the compact cycle.}
	\label{fig:monodromy}
\end{figure}

For our purposes, we also need a fixed basis of forms, adapted to the topological cycles of the manifold, similar to the basis one uses for reductions on compact Calabi--Yau's. Let us start by looking at the cycles. Denote by $A_\alpha$ the $\mu$ topological compact vanishing cycles described by the Milnor spheres, shown in figure \ref{fig:cone}; these are a basis for the homology group $H_n(M_{\lambda})$. One can in fact also introduce dual non-compact \emph{covanishing} cycles $B^\alpha$, defined so that 
\begin{equation}\label{eq:ABd}
\Iprod{A_\alpha}{B^\alpha} = \delta_\alpha^\beta\ .
\end{equation}
The cycles $B^\alpha$ can be seen as a basis for the relative homology group of $\bar M_{\lambda}$ (defined just below (\ref{eq:M7})):
\begin{equation}\label{eq:rel-hom}
H_n(\bar M,\partial \bar M)= \frac{\{\,C_n| \del C_n \subset \del \bar M \,\}}{\{\, \del C_{n+1}\, \}} \supset H_n(\bar M) \ ,
\end{equation}
where we omit the subscript $\lambda$ in this section for simplicity.
Since $H_n(\bar M) \subset H_n(\bar M,\partial \bar M)$, the $A_\alpha$ themselves also belong to $H_n(\bar M,\partial \bar M)$, so that the two homology bases are related: 
\begin{equation}\label{eq:ACB}
A_\alpha = C_{\alpha \beta} B^\beta\,,
\end{equation}
for some matrix $C_{\alpha \beta}$ with entries in $\mathbb{Z}$. The latter is then identified with the intersection matrix of the $A_\alpha$, as we have
\begin{equation}\label{eq:Cab}
	\Iprod{A_\alpha}{A_\beta} = C_{\alpha \beta}\ .
\end{equation}
A related concept is the \emph{variation operator} ${\rm Var}$, which describes the action of the monodromy operator $\mathfrak{m}$ on the covanishing cycles. The monodromy acts as the identity far from the singularity; so $\mathfrak{m} B^\alpha$ differs from $B^\alpha$ only by a compact cycle, as illustrated in Figure \ref{fig:monodromy} for the cone singularity in Figure \ref{fig:MilFib}. In other words, 
\begin{equation}\label{eq:var}
	{\rm Var} B^\alpha \equiv \mathfrak{m}B^\alpha - B^\alpha = v^{\alpha \beta} A_\beta
\end{equation}
for an integer-valued matrix $v^{\alpha \beta}$, which happens to be triangular and invertible over the integers. Thus ${\rm Var}$ gives another way of relating the two sets of cycles, which is in fact related to (\ref{eq:Cab}):
\begin{equation}\label{eq:vvC}
	\left(v^{-1}+(v^{-1})^t\right)_{\alpha \beta} = C_{\alpha \beta}\ .
\end{equation}
For more details see \cite[Ch. 1-2]{Arnold-Gusein-Varchenko}, \cite[Ch. 5]{Ebeling-book}. In appendix \ref{app:GH} we work out these concepts explicitly for Gibbons--Hawking spaces ((\ref{eq:Ak-def}) for $n=2$).

We now introduce $2\mu$ real $n$-forms, $ \alpha_\alpha$, $ \beta^\alpha$, dual to the two sets of cycles, in the sense that
\begin{equation}
	\int_{A_\alpha} \alpha_\beta=\int_{B_\beta} \beta^\alpha=\delta^\alpha_\beta\,, \qquad\quad \int_{A_\alpha} \beta^\beta=\int_{B_\alpha} \alpha^\beta=0\,.
\end{equation}
Both sets of forms belong to the usual de Rham cohomology $H^n(\bar M)$. We can also define the more restrictive relative cohomology
\begin{equation}
	H^n(\bar M, \del \bar M)\equiv \frac{\{\, \omega_n | d \omega_n =0 , \,\, \omega_k{}_{|\del \bar M} = d \lambda_{k-1} \,\} }{\{\, d \omega_{k-1}\,\}} \subset H^n(\bar M)\,,
\end{equation}
in other words, it consists of the forms that are closed, and whose restriction to $\del\bar M$ is exact as a form on that manifold. While the $\alpha_\alpha$ also belong to $H^n(\bar M, \del \bar M)$, the $\beta^\alpha$ do not;\footnote{We could also take both the $\alpha_\alpha$ and the $\beta^\alpha$ to belong in $L^2$ cohomology, for which also quite a bit is known (see for example \cite{albin}), but the characterization we have just described will be more useful for us.} this is dual to the fact we saw below (\ref{eq:rel-hom}), that the the $B^\alpha$ only belong to relative homology, while the $A_\alpha$ also belong to ordinary homology. 
Similarly, the variation operator can be shown to provide an isomorphism between the two cohomologies.

We will also take the $\alpha_\alpha$ and $\beta^\alpha$ to be \emph{harmonic}, in the sense of being closed and co-closed. For compact manifolds, one can use the Hodge decomposition, which implies that every de Rham cohomology class has a harmonic representative. The analogue statement for noncompact manifolds is a bit more subtle: as summarized for example in \cite{cappell-deturck-gluck-miller}, the space of all forms, $\Lambda^p$, and the space of harmonic forms ${\cal H}$ decompose as\footnote{In \cite{cappell-deturck-gluck-miller}, the word ``harmonic'' is reserved for zero-modes of the Laplacian, while forms that are harmonic in our sense are simply called ``closed and co-closed'', and their space is denoted ${\rm CcC}^p$.}
\begin{equation}\label{eq:hodge-dec}
\Lambda^p={\rm cE}^p_{\rm N}\oplus {\cal H}^p\oplus {\rm E}^p_{\rm D} \,,\qquad {\cal H}^p= {\cal H}^p_{\rm N}\oplus {\rm EcC}^p = {\cal H}^p_{\rm D}\oplus {\rm CcE}^p \,. 
\end{equation}
Here ${\rm cE}^p$, ${\rm E}^p$, ${\rm EcC}^p$, ${\rm CcE}^p$ denote respectively the spaces of co-exact, exact, exact and co-closed, closed and co-exact $p$-forms. The summands are all orthogonal to each other. The subscripts ${}_{\rm N}$ and ${}_{\rm D}$ stand for ``Neumann'' and ``Dirichlet'', in the following sense:  introducing a local coordinate $\rho$ near the link, so that $\rho=\rho_0$ describes the link itself, we decompose locally the metric as 
\begin{equation}\label{eq:metr-near-link}
 ds^2 = d\rho^2 + ds^2_{M_7} \,,
\end{equation}
where $ds^2_{M_7}$ is allowed to depend on $\rho$ in principle and reduces by definition to the metric on the link for $\rho=\rho_0$. Neumann forms are purely tangential,
\begin{equation}\label{eq:KR-restr}
 \imath_{\del_\rho}\chi = 0\,, 
\end{equation}
while Dirichlet forms $\omega_{\rm D}$ are such that $\imath_{\del_\rho} \omega_{\rm D} \neq 0$. Now, de Rham and relative cohomology can be represented as Neumann and Dirichlet harmonic forms: 
\begin{equation}
H^p(\bar M)\cong {\cal H}^p_{\rm N} \, ,\qquad H^p(\bar M, \del\bar M) \cong {\cal H}^p_{\rm D}\ .
\end{equation}
Using this, we can take 
\begin{equation}\label{eq:aDbN}
	\alpha_\alpha \in {\cal H}^p_{\rm D} \, ,\qquad \beta^\alpha \in {\cal H}^p_{\rm N}\, .
\end{equation}
From (\ref{eq:aDbN}) we see that $* \alpha_\alpha \in {\cal H}^p_{\rm N}$ and $* \beta^\alpha \in {\cal H}^p_{\rm D}$, since the Hodge star $*$ exchanges the Dirichlet and Neumann properties. So in particular the $*$ closes on that basis. 

One would now like to expand the four-form $\Omega_4$, that defines the volume form $\Omega_4\wedge \bar{\Omega}_4$ of the general deformed $\bar M_\lambda$, on the basis we have just described, just as one does for the compact case. One subtlety, however, arises from the behaviour for $r\to \infty$. Our forms were so far defined on the $\bar M$ with boundary $\del \bar M$, but there is a similar formulation where one considers $L^2$ cohomology on the whole of $M$ instead of relative cohomology on $\bar M$. The analogues of the $\alpha_\alpha$ and $\beta^\alpha$ in this formulation are both decaying at infinity. But $\Omega_4$ does not behave like that, and in fact it is asymptotic to $\Omega_4^0$ at $r\to \infty$. This suggests that we enlarge our basis to include an additional pair of real forms $\alpha_0$ and $\beta^0$, which we define as the real and imaginary part of the holomorphic volume form $\Omega_4^0$ for the conical Calabi--Yau $M_0$:
\begin{equation}\label{eq:ab0}
	\Omega_4^0 = \alpha_0 + i \beta^0 \ .
\end{equation}
We can also introduce an index $\Lambda$ that includes both 0 and $\alpha$, so that 
\begin{equation}
\alpha_\Lambda=(\?\alpha_0,\? \alpha_\alpha\?)\,, \qquad\quad \beta^\Lambda = (\?\beta^0,\? \beta^\alpha\?) \,,
\end{equation}
a notation that will be used throughout the remainder of the paper.

Let us now consider $\Delta \Omega_4 \equiv \Omega_4 - \Omega_4^0$. Since this form is closed, according to \eqref{eq:hodge-dec} it can be written as a form in ${\cal H}_{\rm N}$ plus an exact form. Let us redefine $\Omega_4^0$ so as to include this exact form. Now we have that $\Delta \Omega_4 \in {\cal H}^4$. We could now use either of the two decompositions for ${\cal H}$ in (\ref{eq:hodge-dec}). Since $\Omega_4$ is self-dual, however, we expect it to have components both along ${\cal H}_{\rm N}$ and along ${\cal H}_{\rm D}$. In other words, we would like to expand $\Delta\Omega_4$ on a basis for ${\cal H}_{\rm N}\oplus {\cal H}_{\rm D}$, where now the two summands are non-orthogonal to each other and in fact are maximally so \cite{Shonkwiler}. The remainder can be taken to lie in the space that is orthogonal to both ${\cal H}_{\rm N}$ and ${\cal H}_{\rm D}$; since ${\rm EcC}={\cal H}_{\rm N}^\perp$ and ${\rm CcE}={\cal H}_{\rm D}^\perp$, this space is ${\rm EcC}\cap {\rm CcE} = {\rm EcE}$, the space of forms that are exact and co-exact. To sum up, we can refine (\ref{eq:hodge-dec}) as 
\begin{equation}\label{eq:H-dec-M8}
{\cal H}= {\cal H}_{\rm N}\oplus {\cal H}_{\rm D}\oplus {\rm EcE}\, .
\end{equation}
A basis for forms in ${\rm EcE}$ can in fact be taken to be forms that are $\del$-exact. So we have that $\Delta \Omega_4$ is a sum of the $\alpha_\alpha$, the $\beta^\alpha$, and a $\del$-exact term, and then we can write 
\begin{equation}\label{eq:Omega4XaFb}
\Omega_4 = X^\Lambda \alpha_\Lambda - F_\Lambda \beta^\Lambda + \del(\ldots)\ .
\end{equation}
We can notice already at this stage the similarity with the analogous expansion of the three-form on compact Calabi--Yau manifolds. 

We should also stress, however, that (\ref{eq:Omega4XaFb}) is different from the usual expression in the compact case in a crucial respect: as we saw in (\ref{eq:cartanB}) and in the discussion below, at a generic point of the moduli space, where $M_8$ is smooth, all de Rham cohomology can be represented by $(4,0)$ forms. As also noticed there, this might seem to be in apparent tension with the fact that infinitesimal complex deformations are generated by Beltrami differentials, which live in $H^1(M_8,T)$ and produce $(3,1)$ forms upon acting on $\Omega$. There is no contradiction: (\ref{eq:Omega4XaFb}) is a statement about the finite deformation, which is perhaps less commonly done in mathematics but will be useful for us. On the other hand, infinitesimal complex deformations of an affine manifold (i.e.~around a smooth point in moduli space) are indeed always trivial, in the sense that they can always be seen as induced by a diffeomorphism. This is not true at points in the moduli space where the singularity is only partially resolved, at which points part of the de Rham cohomology migrates to $(3,1)$ degree, as we will see more quantitatively in (\ref{eq:31hol}).

After this caveat, we can also simply reabsorb the $\del(\ldots)$ terms in (\ref{eq:Omega4XaFb}) again in the definition of $\alpha_0$ and $\beta^0$. If we do this, we can complete the action of the Hodge star $*$ as follows. We know already that the $* \alpha_\alpha$ is a linear combination of the $\beta^\alpha$, and vice versa. Since $\Omega$ is self-dual, we can see also that $* \alpha_0$ and $* \beta_0$ should be a linear sum of all the forms $\alpha_\Lambda$, $\beta^\Lambda$. Thus, if we assemble all the forms as
\begin{equation}\label{eq:SigmaDef1}
\Sigma_8 = \begin{pmatrix} \beta^\La \\ \alpha_\La   \end{pmatrix} \,,
\end{equation} 
the nontrivial intersection matrix \eqref{eq:Cab}, or equivalently the variation matrix, is then encoded in the ``twisted selfduality'' property
\begin{equation}\label{eq:Hodge-4}
 *\!\?\Sigma_8 = \mathcal{M}\,\Sigma_8 \,,
\end{equation}
where $\mathcal{M}$ is a constant symmetric matrix. 

We have not assumed a particular point in deformation space, parametrised by the $t^\alpha$, in writing the conditions \eqref{eq:SigmaDef1}--\eqref{eq:Hodge-4}, so that one can make such a choice for any $t^\alpha$ for which no cycles degenerate. It then follows that the matrix $\mathcal{M}$ in \eqref{eq:Hodge-4} can be taken to depend on the deformation parameters and thus corresponds to a period matrix, familiar from compact Calabi--Yau manifolds.


\section{Deformation space for Sasaki--Einstein manifolds}
\label{sec:CR-struct}

In this section, we describe the space of deformations away from the Sasaki--Einstein metric for a manifold constructed as a link around a Calabi--Yau conical singularity. We will take here the odd-dimensional point of view, using the language of Cauchy--Riemann structures, at the same time connecting with results and ideas from sections \ref{sec:cmplx-defs} and \ref{sec:CR-cone}, where the deformations were viewed in terms of the base of and the cone over the Sasaki--Einstein manifold.

We consider a seven-dimensional regular Sasaki--Einstein manifold $M_7$, i.e.~a circle fibration over a six-dimensional \KE base. The latter is assumed to be a hypersurface in most considerations, as in the two previous sections, for simplicity. On $M_7$ there exists an SU(3) structure, described by a \Kah form $J_\se$, a complex three-form $\Omega_\se$,  and a one-form $\eta$, dual to the Killing vector along the circle, satisfying \eqref{eq:SE-point}--\eqref{eq:KE-Ric}, with $\rho$ standing for the Ricci tensor on the \KE base. We wish to describe more general SU(3) structures on such manifolds of the restricted type given in \eqref{eq:su3-struct}, which reduce back to the SU(3) structure at the Sasaki--Einstein metric, for particular choices of $J$ and $\Omega$. Based on the discussion in sections \ref{sec:cmplx-defs} and \ref{sec:CR-cone}, we have established the existence of a finite set of forms that can be used to describe deformations for the complex structure, while in section \ref{sec:Kah-defs} a set of two-forms relevant for the \Kah deformations around the Sasaki--Einstein metric was introduced.

Our discussion of three-form deformations will hinge on the notion of Cauchy--Riemann (CR) structures and of  Kohn--Rossi (KR) cohomology, which we will introduce in sections \ref{sub:CR} and \ref{sub:KR} respectively.
In section \ref{sec:CR-deformations} we discuss deformations of CR structure, which are governed by the Kohn--Rossi cohomology. We then go on in section \ref{sec:Kah-deformations} to introduce the description of \Kah deformations using a general CR structure, extending the discussion in section \ref{sec:Kah-defs}. In each case, we discuss the special geometry in the space of deformations, which is crucial both for performing the reduction and for matching to an $\cN=2$ supergravity in four dimensions, following the standard treatment of \cite{grana-louis-waldram,kashanipoor-minasian,hitchin-67} which is itself analogous to the corresponding construction for Calabi--Yau compactifications \cite{candelas-delaossa-moduli}.

\subsection{Cauchy--Riemann manifolds}
\label{sub:CR}

In this subsection, we will describe the geometry of the link $M_7$ in a more intrinsic way, in terms of the so called Cauchy--Riemann structure, which arises generically on real hypersurfaces in a complex manifold. We refer to \cite{dragomir-tomassini} for a comprehensive introduction to the subject.

A \emph{Cauchy--Riemann (CR) structure} on an odd-dimensional manifold $M_{2\?n+1}$ is defined as a $n+1$-dimensional subbundle $T$ of the complexified tangent bundle $T\subset TM_{2\?n+1}\otimes \cc$, such that $T\cap \bar T= \{ L_\xi \}$, where $L_\xi$ is a one-dimensional real subbundle of $T M_{2\?n+1}$. In other words, the complexified tangent bundle of $M_{2\?n+1}$ can be decomposed as 
\begin{equation}\label{eq:CR-str}
T M_{2\?n+1} \otimes \cc = T \oplus \bar T_0 = T_0 \oplus \bar T_0 \oplus L_\xi\,,
\end{equation} 
where $T_0$ is the complement of $T$ with respect to $L_\xi$. This structure is the odd-dimensional analogue of an almost complex structure on an even-dimensional manifold. Here, we specify $n=3$, since we are mainly interested in seven-dimensional manifolds, noting that all statements hold for any $n>1$.

On an $M_7$ which is the boundary of an almost complex manifold $M_8$, such as for example the link over a quasihomogeneous isolated singularity, one can naturally define a CR structure $T$. Consider a local radial coordinate $\rho$ on a patch of $M_8$ such that $M_7$ is the real hypersurface $\{\rho=\rho_0\}$, as discussed around \eqref{eq:metr-near-link}. The normal space of $M_7$ inside $M_8$ is along $d\rho$. Let us fix a Hermitian metric $g_8$ on $M_8$. We can now define $T$ as the subbundle of holomorphic forms, $\omega\in (TM_8)_{1,0}$, which are orthogonal to $d\rho$ under $g_8$, i.e.~those satisfying $\omega \cdot d\rho=0$. In this situation one can also define the Reeb vector as $\xi \equiv I_8d\rho$, where $I_8$ is the almost complex structure on $M_8$. $L_\xi$ in (\ref{eq:CR-str}) is then generated by this $\xi$. If the Reeb vector has closed orbits, we can view $M_7$ as a U(1)-fibration, as for example for regular Sasaki--Einstein manifolds arising when $M_8$ is a Calabi--Yau cone and $\rho$ is the corresponding radial coordinate.

A CR structure also induces a decomposition on the complexified cotangent bundle and on the exterior algebra, $\Lambda_{\mathbb{C}}^*$. Denoting by $\eta$ the \emph{contact form} dual to the vector $\xi$, such that $\imath_\xi \eta=1$, this decomposition reads
\begin{equation}\label{eq:CR-Om-decomp}
  \Lambda_{\mathbb{C}}^* = \bigoplus_{p,q} \Lambda^{p,q}\,.
\end{equation}
Here, $\Lambda^{p,q}$ stands for the bundle of forms with $p$ legs along $T$ and $q$ legs along $\bar{T}_0$. This asymmetry in the two types of indices is made to accommodate the legs along $\eta$, which are then conventionally counted as holomorphic, and is convenient when relating forms on $M_7$ to forms on $M_8$, see \eqref{eq:KR-from-cone} below.

The restriction of the Hermitian form $J_8$ on $M_8$ to $M_7$ is a two-form $J$ which satisfies
\begin{equation}\label{eq:CR-Kah}
d J=0\, \qquad \imath_\xi J= 0 \,,
\end{equation}
and is $(1,1)$ in the sense of (\ref{eq:CR-Om-decomp}). Similarly, the almost complex structure $I_8$ induces a corresponding tensor $I$ of type $(1,1)$ (one index up and one down) such that $I^2 = -1 + \xi \otimes \eta$, so that it provides a CR structure by definition. Note that the derivative of the contact form, $d\eta$, known as the Levi form, also has the properties \eqref{eq:CR-Kah}; it induces a metric on $M_6$ which is widely used in the CR literature. In general $d\eta$ may be degenerate, but when $M_7$ admits a Sasaki--Einstein structure, it is by definition given in terms of the canonical Einstein metric by \eqref{eq:SE-point} and is therefore of maximal rank. CR manifolds for which the Levi metric $d\eta$ is non-degenerate are known as strongly pseudoconvex.

A CR structure is said to be \emph{integrable} if the holomorphic subbundle $T$ is closed under the Lie bracket, $[T_0,T_0]\subset T_0$. This is the analogue of integrability for a complex structure. In fact, a CR structure induced on a boundary $M_7$ of a complex manifold, as in our case, is integrable. Conversely, an integrable CR manifold is the boundary of a complex manifold if $d \eta$ is non-degenerate \cite{plateau}. 

There is also an analogue of a top-holomorphic form $\Omega$; it is a form that can be written locally as
\begin{equation}
	\eta \wedge \Omega = \eta \wedge w_1 \wedge w_2 \wedge w_3 \ ,
\end{equation}
where $w_i$ are $(1,0)$-forms in the sense of (\ref{eq:CR-Om-decomp}). Just like for complex manifolds, integrability of a CR structure can be reformulated as the equation 
\begin{equation}\label{eq:CR-int}
	d (\eta\wedge\Omega) = w\wedge \eta\wedge\Omega
\end{equation}
for some 1-form $w$. In fact, one may obtain such a form by restriction of the holomorphic $(4,0)$-form $\Omega_4$ on the ambient complex manifold. From the earlier definition of the $\xi= I_8 d\rho$ we see that
\begin{equation}\label{eq:Ofour-dimensionalre}
	\Omega_4 = (d\rho + i \eta) \wedge \Omega\,
\end{equation}
on the points of $M_8$ at its boundary. In particular we can write
\begin{equation}\label{eq:CR-form}
  	\eta\wedge\Omega = \Omega_4|_{M_7} \ . 
\end{equation}
Since $d \Omega_4=0$ on the complex manifold $M_8$, $\eta\wedge\Omega$ is closed, i.e.~$w=0$ in \eqref{eq:CR-int}; in particular this implies integrability of the CR structure, as mentioned above.

\subsection{Kohn--Rossi cohomology}
\label{sub:KR}

On a complex manifold, besides the de Rham differential $d$, one can also define a Dolbeault differential $\del$, such that $d=\del +\bar \del$, and $\del^2=0$. Similarly, on an integrable CR manifold one can use the decomposition \eqref{eq:CR-Om-decomp} to define a differential operator $\tgd$ on the space of $(p,q)$-forms, such that 
\begin{equation}\label{eq:ext_d_dec}
  d = \tgd + \tgdb + \eta \wedge\!\pounds_\xi \,,  
\end{equation}
and $\tgd^2 = \tgdb^2 = 0$. Each of $\tgd$ and $\tgdb$ increases the (anti-)holomorphic degree of a form by one. One can then define the \emph{Kohn--Rossi (KR) cohomology} groups $H^{p,q}_\stgdb$ as the cohomology of the complex
\begin{equation}\label{eq:KR-def}
  \dots \xrightarrow{\tgdb} \Lambda^{p,q-1} \xrightarrow{\tgdb}
 \Lambda^{p,q} \xrightarrow{\tgdb} \Lambda^{p,q+1} \xrightarrow{\tgdb} \dots\ ,
\end{equation}
in exactly the same way as the anti-holomorphic differential $\bar \partial$ on a complex manifold gives rise to Dolbeault cohomology. 

The cohomology groups $H^{p,q}_\stgdb$ are known to be finite-dimensional for values of $(p,q)$ that do not lie at the edge of the resulting Hodge diamond \cite{kohn-rossi}, which reads
\begin{equation}
  \label{eq:hodge}
  \arraycolsep=1.4pt\def\arraystretch{1.1}
\begin{array}{c}
\HE{H^{4,3}_\stgdb} \hspace{1cm} \phantom{\HE{H^{4,3}_\stgdb}} \\ 
\HE{H^{4,2}_\stgdb} \hspace{1cm} \HE{H^{3,3}_\stgdb} \hspace{1cm} \phantom{\HE{H^{2,3}_\stgdb}}  \\ 
\HE{H^{4,1}_\stgdb} \hspace{1cm} \HI{H^{3,2}_\stgdb} \hspace{1cm} \HE{H^{2,3}_\stgdb} \hspace{1cm} \phantom{\HE{H^{1,4}_\stgdb}} \\
\HE{H^{4,0}_\stgdb} \hspace{1cm} \HI{H^{3,1}_\stgdb} \hspace{1cm} \HI{H^{2,2}_\stgdb}  \hspace{1cm} \HE{H^{1,3}_\stgdb}  \hspace{1cm} \phantom{\HE{H^{0,4}_\stgdb}} \\
\HE{H^{3,0}_\stgdb} \hspace{1.cm} \HI{H^{2,1}_\stgdb} \hspace{1cm} \HI{H^{1,2}_\stgdb}  \hspace{1.cm} \HE{H^{0,3}_\stgdb} \\
\HE{H^{2,0}_\stgdb} \hspace{1cm} \HI{H^{1,1}_\stgdb}  \hspace{1cm} \HE{H^{0,2}_\stgdb} \\
\HE{H^{1,0}_\stgdb} \hspace{1cm}  \HE{H^{0,1}_\stgdb} \\
\HE{H^{0,0}_\stgdb} \\
\end{array}
\end{equation}
The groups along the edge (in blue) are in general infinite-dimensional. This is because, unlike compact complex manifolds, there exist non-constant holomorphic functions, i.e.~functions satisfying 
\begin{equation}\label{eq:hol-CR-fun}
\tgdb f= 0\,,
\end{equation} 
on a compact CR manifold. The solutions to \eqref{eq:hol-CR-fun} are elements of $H^{0,0}_\stgdb$ and are descended from holomorphic functions on the ambient noncompact manifold. On the other hand, one can consider the further cohomology induced by $\tgd$ on the KR cohomology groups, and define
\begin{equation}\label{eq:hdr}
	H^p_\stgd \equiv H^{p,0}/\tgd H^{p-1,0}\,,
\end{equation}
in analogy with the holomorphic de Rham groups (\ref{eq:holdR}) of $M_8$.

In fact, KR cohomology can also be defined via $M_8$ (for example see \cite[Sec.~2]{plateau}). Namely, one can consider the quotient 
\begin{equation}\label{eq:KR-from-cone}
 B^{p,q}\equiv \Lambda_8^{p,q}/C^{p,q} \,,
\end{equation}
of $(p,q)$-forms $\Lambda_8^{p,q}$ on $M_8$ by the space of forms $C^{p,q}$ that can be written as $\bar\del \rho \wedge (\ldots)$. Since $\bar\del C^{p,q}\subset C^{p,q+1}$, one finds that $\bar\del$ gives a well-defined differential on $B^{p,q}$. Its cohomology is another possible definition of the KR cohomology $H^{p,q}_\stgdb$, upon restriction to $M_7$. This higher-dimensional origin of the Kohn--Rossi differential complex clarifies the asymmetry of the diamond in \eqref{eq:hodge}.

With some more work, it is possible to relate $H^{p,q}_\stgdb$ and $H^p_\stgd$ more directly to Dolbeault cohomology of $M_8$. More precisely, it can be shown (see e.g.~\cite[Sec.~3]{plateau} and \cite{luk-yau}) 
\begin{equation}\label{eq:KRdol}
	H^{p,q}_\stgdb = H^{p,q}(M_8-Z) \, ,\qquad H^p_\stgd = H^p_h(M_8-Z)\,,
\end{equation}
where $Z$ is the singular locus of $M_8$. Using an exact sequence, $H^{p,q}(M_8-Z)$ can also be related to a certain Dolbeault cohomology on $M_8$ ``with support'' on $Z$, which can be evaluated using algebraic methods. For the conical hypersurface case, this allows to compute 
\begin{equation}
	{\rm dim}H^{3,1}_\stgdb={\rm dim}H^{2,2}_\stgdb = {\rm dim}H^{2,1}_\stgdb = {\rm dim}H^{1,2}_\stgdb = \mu\,.
\end{equation}
(For $p+q \neq 3,4$, $q=1,2$, ${\rm dim}H^{p,q}_\stgdb=0$.) We see the Milnor number appearing again. More generally, away from the conical point, one can show that (\cite{naruki}; see also \cite[Eq.~(N.1)]{tanaka})
\begin{equation}\label{eq:31hol}
	{\rm dim} H^{3,1}_\stgdb+ {\rm dim} H^4_\stgd  = \mu\,, 
\end{equation} 
holds.\footnote{Note that \eqref{eq:31hol} is true in the case of interest in this paper, where $M_8$ arises as a deformation of an isolated singularity, for which the (co)homology groups are nontrivial only in the middle dimension. The formula of \cite{naruki} contains one more term, ${\rm dim}H^3_\stgd$, in the general case.}
For a generic deformation, one finds a \emph{smooth} $M_8$, so that the arguments we just described also lead to \cite[Thm.~C]{plateau} that $H^{3,1}_\stgdb=0$ and \eqref{eq:31hol} then implies that $H^4_\stgd =\mu$.  This situation can be thought of as the counterpart of the result (\ref{eq:cartanB}) for noncompact complex manifolds, and of the statement that the de Rham representatives sit in holomorphic de Rham cohomology. At the other extreme, if $M_7$ is the Sasaki--Einstein link over a conical $M_8$, then all $H^p_\stgd=0$ (by \cite[Thms.~A,B]{luk-yau} and \cite[Rem.~2.5]{huang-luk-yau}), so that \eqref{eq:31hol} implies ${\rm dim}H^{3,1}_\stgdb= \mu$. In between, there are situations where the singularity is partially resolved; both $H^{3,1}_\stgdb$ and $H^4_\stgd$ will have dimensions between 0 and $\mu$.
 
When $M_7$ is a regular Sasaki--Einstein, there is also a third point of view for Kohn--Rossi cohomology. In this case, the decomposition \eqref{eq:ext_d_dec} can be interpreted as a twist of the holomorphic differential $\partial$ on the base by the U(1) action on the total space, to obtain the twisted differential $\tgd$. It then follows that for a Sasaki--Einstein manifold, the Kohn--Rossi cohomology $H^{p,q}_\stgdb$ is simply the twisted cohomology of section \ref{sec:cmplx-defs}. As we saw in this section, $H^{p,q}_\stgdb$ is defined for any boundary of a complex space, and in particular also for the deformations of the CY manifolds we are considering. In this sense, KR cohomology generalizes twisted cohomology, since it is adapted to the CR structure obtained for any deformation, as promised at the end of section \ref{sub:fam}. This brings us to the issue of deformations of CR structure, to which we now turn our attention.

\subsection{CR deformations of Sasaki--Einstein manifolds}
\label{sec:CR-deformations}

\subsubsection{CR deformations}
When Kohn--Rossi cohomology is nontrivial, a CR structure admits deformations based on it, in exactly the same way as deformations of a complex structure are described by Dolbeault cohomology. This construction was pioneered by Kuranishi \cite{Kuranishi}, in a program aiming to characterise isolated complex singularities in terms of intrinsic properties of their links. The result is a theory of deformations for CR manifolds with a non-degenerate Levi metric (defined under (\ref{eq:CR-Kah})), a class that includes by definition those manifolds that admit a Sasaki--Einstein metric, see \cite{Akahori, Akahori-Miyajima,akahori-garfield,Akahori-Kahler, 2001math4056A} for a partial list of references. For this class, all CR deformations on the link descend from the deformations of the singularity and this connection is generic for singularities of complex dimension three and higher, corresponding to deformations of Sasaki--Einstein manifolds in five or more dimensions \cite{2001math4056A} (note that all the KR cohomology groups are a priori infinite-dimensional for three-dimensional CR manifolds). We again restrict to the seven-dimensional case for clarity.

Deformations of CR structure are described by four-forms with one leg along $\eta$, more precisely in terms of the form bundle
\begin{equation}\label{eq:Fpq-bund}
 F^{p+1,q} = \{\, \varphi \,\, :\,\, \varphi=\eta\wedge\Lambda^{p,q}\,, \,\,\, d\eta\wedge\varphi = 0 \,\}\,,
\end{equation}
where the second property ensures primitivity with respect to the Levi form. The KR cohomology can be realised on these spaces, i.e.~$H^{p,q}_\stgdb\subseteq F^{p,q}$, \cite{2001math4056A,Akahori-Kahler,Akahori-Poly}. The space of deformations is related to $F^{3,1}$, as
\begin{equation}\label{eq:Z1-def}
 Z^1 = \{\, \varphi \in F^{3,1} \,\,\, :\,\,\, \tgd\varphi=\tgdb\varphi=0 \, \}\,,
\end{equation}
where $Z^1\subseteq H^{3,1}_\stgdb$ by definition.  
Whenever \eqref{eq:Z1-def} is a basis for KR cohomology, so that $Z^1\cong H^{3,1}_\stgdb$ and in addition the group $H^{0,1}_\stgdb=0$, the deformations are unobstructed and one may vary the CR structure $\Omega$ as
\begin{equation}
\delta \left( \eta\wedge\Omega \right) = \varphi  \,,
\end{equation}
with $\varphi \in Z^{1}$, where the one-form $\eta$ is modified at most by exact pieces. Note that these conditions are entirely analogous to those arising for complex structure deformations. We refer to the works cited above for more details on the description of the deformations, which applies to more general cases than the ones of interest in this paper.

In the case of a Sasaki--Einstein manifold, (\ref{eq:Z1-def}) is made of forms of the type $\hat \omega_k = \ee^{ik\hat\psi}\omega_k$, with $\omega_k\in H^{2,1}_k$ as defined in section \ref{sub:derham}. Around a generic point, on the other hand, recall from the discussion around (\ref{eq:31hol}) that $H^{2,1}_\stgdb$ vanishes, and thus the deformations will be trivial, in the sense that they simply mix a set of holomorphic forms into eachother; this parallels the discussion for $M_8$ after (\ref{eq:Omega4XaFb}). However, one can still define (2,1)-forms by expanding around a generic point and the procedure above can be applied, even though these forms can be represented by forms in holomorphic de Rham.

\subsubsection{Links over deformed singularities}
We now turn to the case of links over deformations of quasihomogeneous isolated singularities, which are of interest in this paper. Since these are defined as boundaries of complex manifolds as in \eqref{eq:metr-near-link}, they carry the natural CR structure discussed in section \ref{sub:CR}. The restriction map $M_8 \to M_7$ is consistent with the action of the Reeb vector in \eqref{eq:boost-cone}--\eqref{eq:boost-cone-def}, so that a form with a well-defined charge restricts to a form with the same property. So, for example, in the conical case we can use \eqref{eq:CR-form} to define a three-form $\Omega^0$ associated to a CR structure, by restriction of $\Omega^0_4$. The relations \eqref{eq:Lie-omega} and \eqref{eq:CR-form} then lead to
\begin{equation}\label{eq:Lie-omega-se}
 \pounds_{\xi_0} \Omega^0 = 4\,\im\,\Omega^0 \quad \Rightarrow \quad d \Omega^0 = 4\,\im\,\eta\wedge\Omega^0  \,.
\end{equation}
More generally, upon turning on a general deformation based on the unfolding \eqref{eq:unfolding}, the holomorphic form $\Omega_4$ on $M_8$ still induces an integrable CR structure with a three-form $\Omega$. However, now this is not a form with a well-defined charge under the vector $\xi$, but rather a linear combination of such forms, similar to what was seen for the twisted cohomology on the base $M_6$ in \eqref{eq:Omu}.

We can describe the properties of the relevant forms on $M_7$ in terms of the forms $\Sigma_4$ introduced in section \ref{sub:fixed}. A priori, each form gives rise to two different forms on $M_7$, by taking the Neumann and Dirichlet parts. However, due to the twisted self-duality property (\ref{eq:Hodge-4}), half of these forms are related to the other half by Hodge duality, so that in the end we still end up with $2 \mu + 2$ forms. In fact, recall from \eqref{eq:aDbN} that one may choose the $\alpha_\alpha$ to have only Dirichlet part, while the $\beta^\alpha$ to have only Neumann part. Hence we can define three-forms on $M_7$ as
\begin{equation}\label{eq:abiota}
 \alpha_{\alpha}{}_7 = \iota_\rho \alpha_\alpha\,, \qquad \beta^\alpha = \eta\wedge \beta^\alpha{}_7\,,
\end{equation}
in terms of the four-forms $\alpha_{\alpha}$, $\beta^\alpha$ on $M_8$. We can similarly define $\alpha_{0,7}$ and $\beta^0{}_7$ by restricting the forms in \eqref{eq:ab0} according to \eqref{eq:CR-form}, as 
\begin{equation}
\eta\wedge\left( \alpha_{0,7} + i \beta^0{}_7 \right) = \Omega_4^0|_{M_7} \,.
\end{equation}
Since from now on we will work on $M_7$, we will drop the label ${}_7$ on all forms, hoping no confusion is generated. 
We again collect the $\alpha_\Lambda$ and $\beta^\Lambda$ into a single vector of forms 
\begin{equation}\label{eq:Sig-M7}
\Sigma=\binom{\beta^\La}{\alpha_\La} \,,
\end{equation} 
just like we did on $M_8$ in \eqref{eq:SigmaDef1}.

With these definitions, using (\ref{eq:Omega4XaFb}), (\ref{eq:Ofour-dimensionalre}) and the fact that $\Omega_4$ is self-dual and primitive, one can see that the CR three-form $\Omega$ takes the form
\begin{equation}\label{eq:CR-struct-1}
 \Omega = X^0 \alpha_0 - F_\Lambda \beta^\Lambda + \tgd(\ldots) \,.
\end{equation} 
We have also used the fact that the Dolbeault differential in \eqref{eq:Omega4XaFb} reduces to the KR differential $\tgd$, as reviewed in section \ref{sub:CR}. The intersection pairing for the components of $\Sigma$, given by $\int_{M_7} \eta \wedge \Sigma_A \wedge \Sigma_B$, is just any antisymmetric matrix at this point, which is guaranteed to be of maximal rank due to the non-orthogonality of the Dirichlet and Neumann forms, as mentioned above \eqref{eq:H-dec-M8}, see also \cite{Shonkwiler}. It is convenient to go to a basis where this reads like a canonical intersection pairing, so that the only non-zero pairings are\footnote{There might be something interesting hidden in this change of basis. The intersection pairing of four-cycles in $M_8$ is related to a linking number $l$ of three-cycles in $M_7$. The relation works as follows \cite[Section 2.3]{Arnold-Gusein-Varchenko}: $\Iprod{A_\alpha}{B^\beta} = \mathrm{l}({\mathrm Var}A_\alpha, \Gamma_{1/2} B^\beta)$, 
where $A$ and $B$ are a compact and a relative cycle respectively, ${\mathrm Var}$ is the variation operator introduced in \eqref{eq:var}, and $\Gamma_{1/2}$ is a half-monodromy around the singularity. The linking number $\mathrm{l}(a, b)$ for two non-intersecting three-cycles $a$ and $b$ on the link is defined simply as 
$\mathrm{l}(a, b)=\Iprod{A}{b}$, with $\partial A =a$ for an appropriate a four-chain $A$. It would be interesting to work out the cohomology counterpart of this: it would presumably involve an integration by parts from $M_8$ to $M_7$.} 
\begin{equation}\label{eq:ab-M7}
 \int_{M_7} \eta\wedge\alpha_\Lambda \wedge \beta^\Sigma =  \delta_\Lambda^\Sigma\,.
\end{equation}
After doing so, the expression for $\Omega$ also gets more complicated, but it remains a linear combination of the $\Sigma$'s:
\begin{equation}\label{eq:Omega7XaFb}
\Omega = X^\Lambda \alpha_\Lambda - F_\Lambda \beta^\Lambda + \tgd(\ldots)\ ,
\end{equation}
where $X^\La$ are complex variables, while the $F_\La$ are holomorphic functions of the $X^\La$. 
This is similar to the expansion along the space \eqref{eq:Htw}, except that now the cohomology groups are defined in terms of the ``moving'' KR differential $\tgd$, rather than the differential defined at the Sasaki--Einstein point. 

It then follows that the forms $\Sigma$ are seen as belonging to KR cohomology. If in addition we choose $\Sigma$ that are harmonic in the KR sense, they remain so on $M_8-Z$, where $Z$ denotes the singular locus, as in \eqref{eq:KRdol}. On the other hand, we can take harmonic representatives in the Dolbeault cohomology $H^{3,1}(M_8-Z)$, and in the holomorphic de Rham cohomology $H^4_h(M_8-Z)$. As we saw in \eqref{eq:KRdol}, these are respectively equal to $H^{3,1}_\stgdb$ and $H^4_\stgd$. By \eqref{eq:31hol}, there will be $\mu$ (complex) harmonic representatives in these two groups, which will be mapped to $H^{2,1}_\stgdb$ and $H^3_\stgd$ respectively, through \eqref{eq:abiota}. These representatives can also be described in terms of the $\Sigma$, upon varying \eqref{eq:Omega7XaFb} to obtain a basis of complex forms
\begin{equation}\label{eq:DOmega7XaFb}
\del_\Lambda \Omega = \alpha_\Lambda - \del_\Lambda F_\Sigma \beta^\Sigma \,,
\end{equation} 
which are of type $(3,0)+(2,1)$ around the conical point. In the generic case, \eqref{eq:DOmega7XaFb} describes the set of $(3,0)$ forms spanning the space of allowed CR structures that descend from the holomorphic forms in $H^4_\stgd$. Note that we have dropped the $\tgd$-exact terms, following the suggestion in \cite[Sec.~3.2.2]{kashanipoor-minasian}, because the projector on the space of KR-harmonic forms commutes with the projector on the space of $(p,q)$-forms. 

We now describe the properties of the forms $\Sigma$ under Hodge duality and under the U(1) symmetry. Starting from the former, we note that the property \eqref{eq:Hodge-4} implies that
\begin{equation}\label{eq:Hodge-M7}
*_7 \Sigma = \eta\wedge\mathcal{M}\?\Sigma\,,
\end{equation}
which again expresses the duality between the Dirichlet and Neumann forms, consistent with the selfduality of the three-form, as in $*_7 \Omega = i \eta \wedge \Omega$. We once again abuse notation by using the same symbol for the matrices in \eqref{eq:Hodge-4} and \eqref{eq:Hodge-M7}, despite the fact that they are not equal. The matrix ${\cal M}$ in \eqref{eq:Hodge-M7} is required to be symplectic by consistency of the Hodge star, and is related to the one in \eqref{eq:Hodge-4} by the involution induced by the decomposition of forms according to \eqref{eq:Ofour-dimensionalre}, followed by the redefinition required to arrange for \eqref{eq:ab-M7}.  
Combining \eqref{eq:DOmega7XaFb} and \eqref{eq:Hodge-M7}, one can determine the form of ${\cal M}$ in terms of the $F_\Sigma$ in the usual way, as we will review shortly.

We now turn to the properties of the forms $\Sigma$ under the action of the Lie derivative along the vector $\xi$, dual to the contact one-form $\eta$. We expect the $\Sigma$ to obey a relation of the form
\begin{equation}\label{eq:dab-M7}
d\Sigma= \eta\wedge\Um\?\Sigma\,,
\end{equation}
with $\Um$ in block-diagonal form with blocks of the type $\left(\begin{smallmatrix} 0 & k \\ -k & 0 \end{smallmatrix}\right)$. At the conical point this is a way of summarizing the results of section \ref{sub:derham}, while in the more general deformed case, this is analogous to \eqref{eq:Lie-defs-fin} on $M_8$.
This implies that it might be possible to obtain a proof in a similar way as in section \ref{sub:derham}, but we have not done so. One can also attempt to derive it through a change of basis: both the twisted cohomology forms and the $\Sigma$ correspond to forms in de Rham cohomology of $M_8$; they differ by the fact that they are coclosed with respect to different metrics (the conical and deformed ones). Hence, they are related by exact pieces which might be proven to be immaterial.\footnote{One might also think of using directly the twisted cohomology forms, as we tried to do at the end of section \ref{sec:cmplx-defs}. The flaw with this strategy is that it is not clear why $*$ would close on them.} 

Here, we use the existence of a set of forms satisfying \eqref{eq:Lie-defs-fin} in $M_8$ to argue towards \eqref{eq:dab-M7}, through a reduction of this basis of complex forms, on the link $M_7$. 
We start by expanding each form around the location of the link at $\rho=0$ in the local coordinates discussed in \eqref{eq:metr-near-link}, as
\begin{equation}\label{eq:form-near-M7}
 \Omega_\alpha = d\rho\wedge\Omega_\alpha^\rho + \eta\wedge\Omega_\alpha^7 + (\rho-\rho_0)\,  \eta\wedge\Omega^{(1)}_\alpha + {\cal O}(\rho-\rho_0)^2 \,,
\end{equation}
where $\Omega_\alpha^\rho$, $\Omega_\alpha^7$ and $\Omega^{(1)}_\alpha$ are three-forms on $M_7$, and we have discarded some possible additional terms at order ${\cal O}(\rho-\rho_0)$ that can be seen to belong in ${\rm cE}^4_{\rm N}$ in \eqref{eq:hodge-dec}. Holomorphicity and primitivity of the $\Omega_\alpha$ imply that $\iota_\xi \Omega_\alpha^\rho=0$ and that all three forms in the r.h.s.~of \eqref{eq:form-near-M7} are primitive with respect to $d\eta$. One can insert this expansion in \eqref{eq:Lie-defs-fin} to obtain
\begin{equation} \label{eq:Lie-on-M7}
   d\Omega_\alpha^7  = \im\?\lambda_\alpha\,  \eta\wedge\Omega_\alpha^7 \,, \qquad d\Omega_\alpha^\rho = \im\?\lambda_\alpha\,  \eta\wedge\Omega_\alpha^\rho  \,.
\end{equation}
Since $\Omega_\alpha^\rho$ and $\Omega_\alpha^7$ in the expansion \eqref{eq:form-near-M7} correspond to Dirichlet and Neumann forms respectively, \eqref{eq:Lie-on-M7} implies that there exist appropriate complex combinations of each type transforming nicely under the Lie derivative.
One can follow exactly the same steps for the complex structure four-form $\Omega_4$, leading to two corresponding components $\Omega^\rho$ and $\Omega^7$, to conclude that the $\mu+1$ complex forms lead to $\mu+1$ pairs of forms as in \eqref{eq:Lie-on-M7}.
Now, since the original $\Omega_4$, $\Omega_\alpha$ can be expanded on the basis $\Sigma$ in \eqref{eq:Sig-M7}, one can do the same on both sides of \eqref{eq:Lie-on-M7}, as in 
\begin{equation}\label{eq:Om-U-1}
\binom{\Omega^7}{\Omega_\alpha^7}
 =
 \begin{pmatrix}  X^\La & - F_\La \\   X^\La{}_\alpha & -  F_\La{}_\alpha \end{pmatrix}\, \binom{\beta^\La}{\alpha_\La}\,,
\end{equation}
in terms of appropriate expansion coefficients $X^\La{}_\alpha$, $ F_\La{}_\alpha$. One can consider a similar expansion for $(\Omega^\rho\?,\, \Omega_\alpha^\rho)^T$, but the corresponding expansion coefficients are fixed in terms of the ones in \eqref{eq:Om-U-1}, since the Hodge duality in \eqref{eq:Hodge-M7} exchanges the Dirichlet and Neumann parts.
One expects that the scalar-dependent matrix in \eqref{eq:Om-U-1} is invertible at a generic point, since the forms $\Omega_\alpha$ on $M_8$ are in one to one correspondence to the deformation parameters. It then follows that \eqref{eq:Lie-on-M7} can be represented on the forms $\Sigma$ as
\begin{equation}\label{eq:Lie-alpha-beta}
 d \binom{\beta^\La}{\alpha_\La} = 
 \begin{pmatrix}  \,\Um^\La{}_\Si \,&\, \Um^{\La \Si}\, \rule{0pt}{2ex}\\ \, \Um_{\La \Si} \, & \, \Um_\La{}^\Si \,\rule{0pt}{2ex} \end{pmatrix}\, \eta\wedge\binom{\beta^\Si}{\alpha_\Si}\,.
\end{equation}
Here,  $\Um$ is an appropriate symplectic matrix standing for a real form of the action \eqref{eq:Lie-on-M7}, that can be brought in the block-diagonal form mentioned below \eqref{eq:dab-M7} with eigenvalues $(4\?\im , \im\?\lambda_\alpha)$.  
Note that this argument relies on reducing the complex forms of \eqref{eq:Lie-defs-fin} on the link $M_7$ before changing to the real basis, so that we have no direct way to realise the action of $\Um$ on the real forms in $M_8$, which would mix the forms in $H^4(\bar M)$ and $H^4(\bar M, \partial M)$. Such an action may exist, presumably involving the variation operator.

\subsubsection{Special \Kah geometry}
The relation between the complex and real forms is expressed through the expansion \eqref{eq:Omega7XaFb}, which provides a natural parametrisation of the CR structure in terms of the relevant basis of forms $\Sigma$, and \eqref{eq:DOmega7XaFb}, which provides a parametrisation of the $\mu$ forms that control the deformations (plus a linear combination equal to the original CR structure in \eqref{eq:Omega7XaFb}). 
In view of the holomorphicity properties of these forms, the integral $\int\eta\wedge\Omega \wedge \partial_\La\Omega$ must always vanish, since there exist at most four holomorphic directions in the sense of \eqref{eq:CR-Om-decomp}. Using this, a standard argument shows that the $F_\La$ are given by derivatives of a prepotential function $F(X)$, as
\begin{equation}
\int\eta\wedge\Omega \wedge \partial_\La\Omega=  F_\Lambda - X^\Sigma \del_\Lambda F_\Sigma=0 \qquad\Rightarrow\qquad F_\La = \frac{\partial{F}}{\partial{X^\La}}\,,
\end{equation}
which must be homogeneous of degree two in the $X^\La$.
One can then choose special coordinates $z^\alpha = X^\alpha/X^0$, $\alpha=1,\ldots,\mu$ (away from the $X^0=0$ locus) and define the forms
\begin{equation}\label{eq:chi-kap}
\chi_\alpha = \frac{\partial}{\partial z^\alpha} \Omega - \kappa_\alpha\?\Omega\,,
\end{equation}
where the $\kappa_\alpha$ are defined by the requirement that $\int\eta\wedge\bar\Omega \wedge \chi_\alpha=0$, i.e.~that the direction along $\Omega$ is projected out. 
At the conical (Sasaki--Einstein) point, the $\chi_\alpha$ are KR representatives of type (2,1) and the terms proportional to $\kappa_\alpha$ remove the (3,0) part generically present in the derivative, while in the general case the $\chi_\alpha$ represent the variation away from the chosen CR structure, where one again has to remove the linear combination \eqref{eq:Omega7XaFb}. Note that the $\kappa_\alpha$ arising from a variation of a three-form as in \eqref{eq:chi-kap} may a priori be holomorphic functions\footnote{While holomorphic functions on compact complex manifolds are necessarily constants, we remind the reader there do exist nontrivial holomorphic functions on a CR manifold, satisfying \eqref{eq:hol-CR-fun}.} on $M_7$. However, this is not the case here, since \eqref{eq:chi-kap} is a rewriting of \eqref{eq:DOmega7XaFb}, which describes the finite-dimensional $H^{4}_\stgd$ in terms of parameters $X^\La$, $F_\La$, that are constant on $M_7$. 
Alternatively, one may use the fact that $\mathcal{M}$ in \eqref{eq:Hodge-M7} is constant to show that the $\kappa_\alpha$ are constant \cite{kashanipoor-minasian}.

Using this parametrisation, one can define a \Kah potential, $\KahO$, as
\begin{equation}\label{eq:Kah-complx}
\ee^{-\KahO}=\,  -i \,\int_{M_7}\eta \wedge \bar{\Omega} \wedge\Omega
\,=\,  i \left( \bar{X}^\La F_\La -X^\La \bar{F}_\La \right)\,, 
\end{equation}
on the space of deformations. This \Kah potential has been shown to describe the CR deformations of the link around the quadratic hypersurface singularity (the conifold/Stenzel space) in \cite{Akahori-Kahler}, where it was conjectured to hold more generally. In the physics literature, this is standard for SU(3) structure compactifications, see e.g. \cite{Micu:2006ey, cassani-koerber-varela}. Moreover, we expect the recent developments of \cite{Aleshkin:2017fuz} to be applicable to the computation of this \Kah potential.
Given \eqref{eq:Kah-complx}, one can compute the K\"ahler metric, 
\begin{equation}\label{eq:Kah-metr-complx}
g_{\alpha\bar \beta}=\partial_\alpha\partial_{\bar \beta}\KahO\,,
\end{equation}
on the space of deformations, in special coordinates.
We finally return to the scalar-dependent matrix $\mathcal{M}$ in \eqref{eq:Hodge-M7}, and note that it can be parametrised in terms of the prepotential, as in \cite[Sec. 3.2.2]{kashanipoor-minasian}. The result is the familiar expression from $\cN=2$ supergravity theories, explicitly 
\begin{equation}\label{eq:M-matrix}
\mathcal{M}=
\begin{pmatrix}
\Im\cN + \Re\cN\? (\Im\cN)^{-1}\Re\cN  & -\Re\cN\?(\Im\cN)^{-1}\\
-(\Im\cN)^{-1}\Re\cN & (\Im\cN)^{-1} 
\end{pmatrix}\,,
\end{equation}
where the period matrix $\cN_{\La\Sigma}$ is given by
\begin{equation}\label{eq:per-matr}
\cN_{\La\Sigma}= \overline{F}_{\La\Sigma} + 2i \frac{(\Im F)_{\La P} X^P (\Im F)_{\Sigma T} X^T}{(\Im F)_{P T} X^P X^T }\,,
\end{equation}
and matrix multiplication is implicitly assumed in \eqref{eq:M-matrix}. Given that the matrix $\mathcal{M}$ originally arises on the ambient complex manifold $M_8$, it would be interesting to obtain a direct derivation using the properties of the deformed singularity.

\subsection{\Kah deformations}
\label{sec:Kah-deformations}

We now turn to \Kah deformations, whose description is simpler, since all relevant forms are closed. However, one still needs to refer to CR geometry: the analysis in section \ref{sec:Kah-defs}, where it was shown that there exists a metric leading to a fixed Ricci form $\rho$ in every \Kah class, made use of Yau's theorem, which implicitly employs a complex structure, that is a priori only available at the Sasaki--Einstein point.\footnote{We thank D. Cassani for comments on this point.} One may extend this result to the case of CR structure, defined by an appropriate three-form on the seven-dimensional manifold, by the version of Yau's theorem pertaining to CR structures \cite{CaoChang2006,chang2008,chang2014}. 

In this paper, we have focused on Sasaki--Einstein manifolds enclosing hypersurface singularities, i.e.~defined in terms of a single holomorphic function, so it follows that there is a unique \Kah class descending from the one in $\mathbb{C}^5$ by restriction. Hence, there are in fact no non-universal \Kah deformations in this case --- there is only the freedom of rescaling the \Kah class, giving rise to a single \Kah modulus; this is exactly the same as for compact hypersurface CY's, which also only have universal \Kah deformations. Nevertheless, we provide this discussion for completeness, since we have implicitly used the CR version of Yau's theorem in the previous section, and anticipating applications to deformations of Sasaki--Einstein manifolds arising as links of complete intersection singularities.

In view of the decomposition of the exterior form bundle according to the CR structure as in \eqref{eq:CR-Om-decomp}, one can introduce an adapted vielbein $\{\eta,\? \theta^m, \theta^{\bar m} \}$, where $\eta$ is the one-form defining the CR-structure, while the $\theta^m$ and the $\theta^{\bar m}$ span the holomorphic and anti-holomorphic directions respectively. One can then define the connection $\omega^m{}_n$ and the torsion, $A_{m n}=A_{n m}$, by the relations
\begin{equation}
\begin{split}	\label{eq:conn-def}	
	d \eta= h_m{}_{\bar n}\?\theta^m\wedge &\theta^{\bar n}  \,,
	\quad
	d \theta^m = \omega^m{}_n\wedge \theta^n + A^m{}_{\bar n}\? \theta^{\bar n}\wedge\eta\,,
	\\
	&\omega_{\bar m}{}_n + \omega_n{}_{\bar m} = d h_n{}_{\bar m}\,,
\end{split}	
\end{equation}
where $h_m{}_{\bar n}$ denotes the Levi metric.
Here and in the following, we use the Levi metric to raise and lower indices.
One can define the associated curvature $\Pi^m{}_n$, which can be decomposed as follows
\begin{equation}\label{eq:ps-Riem}
\begin{split}
	\Pi^m{}_n \equiv &\, d \omega^m{}_n + \omega^m{}_p\wedge \omega^p{}_n \\
	=&\,  {\cal R}^m{}_{n p \bar{q}}\?\theta^p\wedge \theta^{\bar q} 
	+ \left( A_{n p}{}^{;m}\?\theta^p -A^m{}_{\bar q;n}\?\theta^{\bar q} \right)\wedge\eta
	+ \im \left( A_{np}\theta^p\wedge \theta^m - A^{m p}\theta_n\wedge \theta_p \right)\,,	
\end{split}	
\end{equation}
where $;m$ denotes the covariant derivative. Here, ${\cal R}^m{}_{n p \bar{q}}$ is the so called pseudo-hermitian Riemann tensor, which satisfies
\begin{equation}
{\cal R}_{n {\bar m} p \bar{q} } = {\cal R}_{{\bar m} n \bar{q} p } = {\cal R}_{p {\bar m} n \bar{q}}\,.
\end{equation}
Contraction of \eqref{eq:ps-Riem} yields the analogous definition of the pseudo-hermitian Ricci tensor ${\cal R}_{p \bar{q}}$, 
\begin{align}\label{eq:ps-Ricci}
\Pi^m{}_m = &\, d \omega^m{}_m
={\cal R}_{p \bar{q}}\?\theta^p\wedge \theta^{\bar q} \,,
\end{align}
which plays a central role in the following.

A contact structure $\eta$ on a seven-dimensional CR manifold is defined to be pseudo-Einstein if the associated Levi metric and pseudo-hermitian Ricci tensor are proportional, as
\begin{equation}\label{eq:ps-Einst}
{\cal R}_{m \bar{n}} = \frac{R}{3}\?h_m{}_{\bar n}\,.
\end{equation}
A Sasaki--Einstein manifold is clearly a pseudo-Einstein manifold, since in this case \eqref{eq:conn-def}--\eqref{eq:ps-Ricci} reduce to the standard definitions of the geometrical quantities on the \KE base. Then, the pseudo-Einstein condition \eqref{eq:ps-Einst} simply expresses the Einstein condition on the base, upon identifying the Levi metric with the \Kah form in \eqref{eq:SE-point}, as $h_m{}_{\bar n}=2\? (J_\se)_m{}_{\bar n}$.

More generally, the pseudo-Einstein condition \eqref{eq:ps-Einst} can also be imposed away from the Sasaki--Einstein point; as it turns out, a CR manifold with a non-degenerate $h_m{}_{\bar n}$ that can be realised as the boundary of a complex manifold admits a pseudo-Einstein structure. The latter is constructed as follows: starting from the one-form $\eta$, one can pass to a rescaled form, defined as $\tilde\eta=\ee^u \eta$ for a function $u$, which is determined by demanding that \eqref{eq:ps-Einst} holds for the pseudo-Ricci form ${\cal \tilde R}$ associated to $\tilde \eta$. This results in a differential condition on $u$, through a construction that can be thought of a CR analogue of Yau's theorem \cite[Th. 4.2]{chang2014}. 
However, the function $u$ is not fixed uniquely, as it is only given up to the trace of a harmonic (1,1)-form $J$, which is a zero mode of the differential condition that ensures \eqref{eq:ps-Einst} is true \cite[Lem. 4.2]{chang2014}. We can parametrise the possible pseudo-Einstein structures starting from a solution $u_0$, as 
\begin{equation}\label{eq:u-psEinst}
 u = u_0 + h^{m\?\bar n} J_{m\?\bar n}\,,
\end{equation}
by introducing an expansion for $J$ over the basis $\omega_a=\{ d\tilde\eta ,\, \omega^{\scriptscriptstyle KR}_{\hat{a}} \}$ where the $\omega^{\scriptscriptstyle KR}_{\hat{a}}$ for $\hat{a}=1,\dots \dim\! H^{1,1}_\stgdb$ are a basis of harmonic elements of $H^{1,1}_\stgdb$ as
\begin{equation}\label{eq:Kah-form}
 J = t^a \omega_a \,,
\end{equation}
for real parameters $t^a$, with $a=1,\dots \dim\! H^{1,1}_\stgdb+1$.
Here, we extend the basis of the $\omega^{\scriptscriptstyle KR}_{\hat{a}}$ by addition of the Levi form for later convenience, in order to be able to use \eqref{eq:Kah-form} as a metric along the directions complementary to $\tilde\eta$.
Note that this additional component of \eqref{eq:Kah-form} along $d\tilde\eta$ can be viewed as parametrising shifts of \eqref{eq:u-psEinst} by a constant.
It then follows that any $u(t^a)$ given by \eqref{eq:u-psEinst}--\eqref{eq:Kah-form} provides a pseudo-Einstein structure, so that \eqref{eq:ps-Einst} is satisfied.\footnote{A corresponding picture from the point of view of deformations of the singular cone over the Sasaki--Einstein manifold arises by the results of \cite{Conlon-Hein-1,Conlon-Hein-2,Conlon-Hein-3}, realising the predictions in \cite{Tian1991}, to infer the existence of asymptotically conical metrics on $M_8$ in each \Kah class $[J_8]$, which is of the form $J_8=d\left(\ee^{u_8}\eta\right) $, for an appropriate function $u_8$.} 
For any choice of the $t^a$, this also implies that the first Chern class $c_1(T_0)$ is trivial, since the pseudo-Ricci form is proportional to the exact Levi form, which is exact \cite[Prop. D]{lee-pseudo}. 

The above parametrisation of the space of pseudo-Einstein conditions for a given CR structure in terms of harmonic KR (1,1)-forms is analogous to the parametrisation of the space of Ricci-flat metrics on a Calabi--Yau manifold for a given complex structure by expanding the \Kah form on a harmonic basis. In addition, it reduces to the discussion in section \ref{sec:Kah-defs} when the CR structure is restricted to the Sasaki--Einstein complex structure, where the expansion \eqref{eq:Kah-form} can be identified with the expansion of the \Kah form.\footnote{The different description compared to the standard treatment of \Kah moduli is due to the fact that the $\del \bar{\del}$-lemma does not hold on a CR manifold. One therefore has to rely on fixed contact class rather than fixed \Kah class in general, away from special situations as in section \ref{sec:Kah-defs}} We will therefore view \eqref{eq:Kah-form} as the hermitian form of a metric along the complement of $\tilde{\eta}$, generalising the \Kah form at the Sasaki--Einstein point. Similarly, we refer to the deformations in \eqref{eq:u-psEinst} as \Kah deformations and to the parameters $t^a$ as \Kah moduli in the general case as well, for simplicity.
In particular, comparing with section \ref{sec:Kah-defs}, one finds that the basis $\omega_a$ can be identified with the basis of (1,1)-cohomology in \eqref{eq:J-rho}, in that limit. Since we keep the contact class fixed, we continue to use the expansion coefficients $m^a$ to parametrise the pseudo-Ricci form, in a general basis, as
\begin{equation}\label{eq:d-eta-gen}
 d\tilde{\eta} = m^a \omega_a \,,
\end{equation}
keeping in mind that in the preferred basis defined above \eqref{eq:Kah-form}, this expansion has only a single term. 
We henceforth drop the tilde on the contact form $\eta$, assuming that it always corresponds to a pseudo-Einstein structure.

We now turn to some properties of the \Kah moduli space that will be useful in the dimensional reduction in the next section. We first address the issue of the moduli dependence of the harmonic forms in \eqref{eq:Kah-form}, since the harmonic representatives $\omega_a$ may in general depend on the $t^a$. This is again parallel to the situation in section \ref{sec:Kah-defs}, since the variation of \eqref{eq:Kah-form} leads to 
\begin{equation}\label{metricvariation-2}
\frac{\partial J_{m \bar{n}}}{\partial t^a} = \omega_{a\?m\bar{n}} +  t^b \frac{\partial}{\partial t^a}\omega_{b\?m\bar{n}}  \,,
\end{equation}
which is identical to \eqref{metricvariation} and the last term is similarly required to be $\tgdb$-exact. Now, since a variation $\delta u$ of \eqref{eq:u-psEinst} must be the trace of a harmonic form, it follows that $\delta J_{m \bar{n}}$ must be harmonic, so that one can repeat the argument above \eqref{eq:consistent-Kah} to derive the same constraint on the $\del_a \omega_b$. Note that the component along the Levi form in the preferred basis above \eqref{eq:Kah-form} is by definition independent of the $t^a$, since all deformations are for fixed contact class, and therefore does not appear in the last term of \eqref{metricvariation-2}. 

Given the basis of $(1,1)$-forms $\omega_a$, it follows by Poincar\'e duality that the $(3,2)$ cohomology in the diamond \eqref{eq:hodge} is also of dimension ${\rm dim} H^{1,1}_\stgdb$. Denoting the relevant set of harmonic $(3,2)$ forms as $\eta\wedge\omega^{\hat{a}}_{\scriptscriptstyle KR}$, and extending by one more element along the Levi form as in \eqref{eq:Kah-form}, the relevant basis is $\eta\wedge\omega^a = \{ \eta\wedge d\eta, \, \eta\wedge\omega^{\hat{a}}_{\scriptscriptstyle KR} \}$. We then find the following conditions
\begin{equation}\label{eq:K-tensor}
 \int_{M_7} \eta\wedge\omega_a \wedge \omega^b = \delta_a^b\,, \qquad 
 \int_{M_7} \eta\wedge\omega_a \wedge \omega_b \wedge \omega_c = \cK_{abc} \,,
\end{equation}
where the $\cK_{abc}$ are constants.
Using our above assumption of taking $J$ in \eqref{eq:Kah-form} as a metric along the six directions complementary to $\eta$, we can apply the standard formula for the Hodge dual on an SU(3) structure manifold, to obtain the following expression for the metric on the moduli space of the \Kah form
\begin{align}\label{eq:Kah-metr-4}
 g_{ab} \,=&\, \frac{1}{4\,\cK} \,\int_{M_7} *\,\omega_a \wedge \omega_b 
  \CR
 = &\, -\frac{1}{4\,\cK} \,\int_{M_7} \eta\wedge\left( J \wedge \omega_a - \left(J \llcorner \omega_a \right)\, \frac{1}{2}\, J \wedge J \right) \,\wedge \omega_b
 \CR
 = &\, -\frac{1}{4\,\cK} \,\left( \cK_{abc}\,t^c - \frac1{4\,\cK}\,\cK_{acd}\,t^c t^d \? \cK_{bef}\,t^e t^f \right)  \,,
\end{align}
on the seven-dimensional manifold. Here, $J \llcorner \omega_a = J^{m\bar{n}} \omega_{a\?m\bar{n}}$ and we define the shorthand
\begin{equation}\label{eq:K-fun}
 \cK = \frac16\,\cK_{abc}\,t^a t^b t^c\,,
\end{equation}
that will be extensively used in the following.

Exactly as for Calabi--Yau compactifications, the \Kah moduli are complexified in the process of the reduction, once paired with the components of the B-field along the $(1,1)$ directions, denoted as $b^a$ in the reduction Ansatz (see \eqref{eq:A3-ansatz} below). We then define the complex fields
\begin{equation}\label{eq:compl-Kah}
 v^a = b^a + \im\,t^a\,,
\end{equation}
and the \Kah potential for the metric \eqref{eq:Kah-metr-4}
\begin{equation}\label{eq:K-Kah}
 \ee^{-\KahJ} \,=\, -\frac{\im}{6}\,\cK_{abc} (v^a-\bar v^a)(v^b-\bar v^b)(v^c-\bar v^c) \,=\, 8\,\cK\,,
\end{equation}
for which it is straightforward to check that $g_{ab} = \frac{\partial}{\partial v^a}\frac{\partial}{\partial\bar v^b} \KahJ$. The associated prepotential that determines the supergravity action is given by
\begin{equation}\label{eq:Kah-prepotential}
 G(Z) = -\frac{1}{6} \cK_{abc} \frac{Z^a Z^b Z^c}{Z^0}\,,
\end{equation}
where the projective coordinates $Z^I \equiv (Z^0,Z^a)$ are identified as $Z^I= Z^0\?(\?1\,, v^a)$. The relation to the \Kah potential is given by $\ee^{-\KahJ} =  \im\,(\,\bar Z^I G_I - Z^I\bar{G}_I\,)$, with $G_I = \tfrac{\del G}{\del Z^I}$. 

Finally, we comment on the compatibility between the \Kah and complex structure deformations, which is enforced by the condition 
\begin{equation}\label{eq:compatibility}
 \eta \wedge \Sigma \wedge \omega_a =0 \,,
\end{equation}
ensuring that the forms in the basis $\omega_a$ remain of type $(1,1)$ for all possible choices of complex structure. The triviality of the product in \eqref{eq:compatibility} in cohomology relies on the vanishing of the group $H^{4,2}_\stgdb \?\!\cong\?\! H^{0,1}_\stgdb$, which holds true at the Sasaki--Einstein point and we take as a working assumption for deformations away from it. Around a generic point, the three-forms $\Sigma$ are in fact holomorphic, as explained around \eqref{eq:Omega7XaFb}--\eqref{eq:DOmega7XaFb}, so that the triviality of \eqref{eq:compatibility} in cohomology is again guaranteed by the fact that there do not exist any (5,1)-forms. The strict vanishing of the quantity \eqref{eq:compatibility} is not necessary for the various considerations of this paper, but we make this choice for simplicity.

\section{Reduction of M-theory to four-dimensional supergravity}
\label{sec:reduction}

We now turn to the reduction of M-theory on our class of seven-dimensional manifolds admitting SU(3) structures parametrised by sets of two- and three-forms, as summarised in section \ref{sec:CR-struct}. In this task, we make use of the literature on SU(3) structure compactifications, by connecting to the structures postulated in \cite{Micu:2006ey, cassani-koerber-varela} to obtain a four-dimensional $\cN=2$ supergravity. Since the reduction is described in detail in these references, we give a concise overview of this procedure and the results for the sake of completeness in sections \ref{sec:ansatz} and \ref{sec:4d-sugra}. In section \ref{sec:vac-moduli} we discuss the supergravity moduli space at the AdS$_4$ vacuum and check that it agrees with the moduli of the Einstein metric along the internal directions, for the examples considered sections \ref{sec:cmplx-defs} and \ref{sec:CR-cone}. Section \ref{sec:physical} concludes with some comments on the consistency of the reduction based on our SU(3) structures.

\subsection{The reduction Ansatz}
\label{sec:ansatz}

In terms of the objects introduced in section \ref{sec:CR-struct}, it is straightforward to define a family of SU(3)-structures on $M_7$ as
\begin{gather}
\theta = \ee^V \eta , \;\qquad J= \ee^{-V}  t^a \omega_a \,, \CR
\Omega = \ee^{-\frac{3}{2}\,V}\Iprod{\cV}{\Si}
= \ee^{-\frac{3}{2}\,V}( X^\La \alpha_\La - F_\La \beta^\La)\,, 
\label{eq:SU3-str}
\end{gather}
where $V$ is a real parameter, $t^a$ is a set of real parameters, $\cV=(X^\La$\,, $F_\La)^T$ is a vector of complex parameters and we defined the symplectic inner product for convenience. The parameter $\ee^V$ will eventually correspond to the dilaton, up to a redefinition, while the $t^a$ and $\cV$ parametrise the \Kah form and the CR structure respectively, as discussed in sections \ref{sec:CR-deformations} and \ref{sec:Kah-deformations}. As a result, the manifold $M_7$ admits an SU(3) structure in the class of \eqref{eq:su3-struct}, parametrised by the forms \eqref{eq:SU3-str} above. 

We now specify the reduction Ansatz for the bosonic fields of M-theory on $M_7$, following the treatment of \cite{Micu:2006ey, cassani-koerber-varela}.
The bosonic content of 11D supergravity \cite{cremmer-julia-scherk} includes the metric and a three-form potential $A_3$, with field strength $F_4 = dA_3$. The bosonic part of the action is
\begin{equation}\label{eq:11daction}
S_{11} = \frac{1}{2\?\kappa_{11}^2} \int \left( R \?*_{\scriptscriptstyle 11}\!1 -\frac{1}{2}\? F_4 \wedge *_{\scriptscriptstyle 11}F_4 -\frac{1}{6}\? A_3 \wedge F_4 \wedge F_4  \right)\,,
\end{equation}
where $*_{\scriptscriptstyle 11}\!$ and $\kappa_{11}$ are the Hodge star and the gravitational constant in 11D, respectively. 
The Ansatz for the reduction of the metric takes the form
\begin{equation}\label{eq:metric-fin}
ds^2 = \ee^{2\?V} \cK^{-1} ds^2_4 +  \ee^{-V} ds^2(M_6) + \ee^{2\?V} \big(\eta + A^0 \big)^2 \,,
\end{equation}
where $ds^2_4$ is the four-dimensional spacetime metric in the Einstein frame and the internal seven-dimensional part is the metric induced by the SU(3) structure \eqref{eq:SU3-str}, which belongs to the class defined in \eqref{eq:su3-struct}. The one-form $A^0$, corresponding to reparametrisation invariance of the vector dual to $\eta$, is accordingly viewed as a gauge field in four dimensions. Similarly, the parameters $V$, $t^a$, $X^\La$ are now viewed as scalar fields on the four-dimensional space-time. We will refer to the $t^a$ as \Kah moduli and to the $X^\La$ as CR structure moduli, as has become customary in the literature on related reductions for parameters of continuous families of SU(3) structures, as the ones described in section \ref{sec:CR-deformations}. However, we point out that these are not moduli from the four-dimensional point of view, i.e.~these scalar fields appear in the supergravity potential below.

The reduction Ansatz for the three-form potential $A_3$ is defined by expanding over the forms in \eqref{eq:SU3-str}, as:
\begin{equation}\label{eq:A3-ansatz}
A_3 = C_3 + (B + b^a\, \omega_a ) \wedge (\eta + A^0) - A^a \wedge \omega_a + \xi^\La \alpha_\La - \xi_\La \beta^\La \,.
\end{equation}
The quantities on the right hand side are fields on the four-dimensional spacetime, namely a (non-dynamical) three-form,  $C_3$, a two-form $B$, a set of gauge vector fields $A^a$, the real scalars $b^a$, and the symplectic vector of real scalars $(\xi^\La\,, \xi_\La)$.
For the field strength we take
\begin{equation}\label{eq:F4-ansatz}
F_4 \,=\, dA_3 + (p^\La \alpha_\La - q_\La \beta^\La) \wedge \theta + e_a\, \omega^a\,,
\end{equation}
where $e_a$ and $(p^\La\,, q_\La)\equiv Q_0$ are constants parametrizing the flux of the four-form on the internal manifold. 
Note that these fluxes may or may not be physical on a given manifold; for example, when the base of the Sasaki--Einstein manifold admits a single $(2,2)$-form, as in the examples of sections \ref{sub:hypersurf} and \ref{sec:singularities}, this is necessarily exact on $M_7$ and the corresponding term in \eqref{eq:F4-ansatz} is a total derivative that can be absorbed by a redefinition of the $b^a$ and $A^a$. Exactly the same holds for the fluxes $Q_0$, which can be absorbed by a redefinition of the axions $(\xi^\La\,, \xi_\La)$, if the matrix $\Um$ in \eqref{eq:Lie-alpha-beta} is non-degenerate, while in the general case a similar redefinition allows to reduce to a restricted $Q_0$, spanning the null eigenspace of $\Um$.

As mentioned above, the three-form $C_3$ is non-dynamical in four dimensions and therefore it can be dualised to a constant, denoted $e_0$, identified as the Freund--Rubin parameter of the compactification, see \cite{louis-micu, cassani-koerber-varela} for more details. Combining this constant with the additional flux parameters $e_a$ in \eqref{eq:F4-ansatz} one can define the vector
\begin{equation}
e_I = (e_0\,,\, e_a)\,,
\end{equation}
while a similar convenient vector can be defined using the components of the Ricci form, defined in \eqref{eq:J-rho} and \eqref{eq:d-eta-gen}, as
\begin{equation}
m^I = (0\,,\, m^a)\,.
\end{equation}
Note that from a four-dimensional perspective, the first of the magnetic components, $m^0$, can also be a nonzero constant. This corresponds to a Romans mass in a Type IIA setting and does not arise in this paper.

\subsection{The four-dimensional gauged supergravity}
\label{sec:4d-sugra}

The four-dimensional Lagrangian obtained by dimensional reduction reads \cite{cassani-koerber-varela}
\begin{align}\label{eq:four-dimensional-action}
S = \frac{1}{\kappa_4^2} \int\!&\left[\, \Bigr. 
\tfrac{1}{2}R_4*\!1  
+ g_{ab}\? dv^a\wedge * d\bar v^{b} + g_{\alpha \bar \beta}\? Dz^\alpha \wedge *D{\bar z}^{\beta} 
\right. 
\CR
&\,\left. 
+  d\phi\wedge* d\phi  -\tfrac{1}{4}\? \ee^{2\phi} \Iprod{D\xi}{\mathcal{M} * D\xi} + \tfrac{1}{4}\? \ee^{-4\phi} dB \wedge *dB  
\right. 
\CR
&\,\left. 
+ \tfrac{1}{4}\? \Im\,\mathcal N_{IJ} F^I \wedge *F^J + \tfrac{1}{4}\? \Re\mathcal N_{IJ} F^I \wedge F^J \right. \rule{0pt}{3ex}
\CR
&\,\left. -\tfrac{1}{4}\? dB \wedge \left(2\,e_I \,A^I  + \Iprod{Q_0}{\xi}\,A^0  -\Iprod{\xi}{D\xi}\right)  \right. \rule{0pt}{3ex}
\CR
&\,\left.  -\tfrac{1}{4}\? e_I m^I B \wedge B -V *\!1\Bigr. \right]\,,\rule{0pt}{3ex}
\end{align}
where we defined the gravitational coupling constant $\kappa^{-2}_4 = \kappa_{11}^{-2} \ee^{2V}\,\cK^{-1} $, and the Hodge star $*$ with respect to the four-dimensional metric $ds^2_4$. We now list the various fields and quantities appearing in \eqref{eq:four-dimensional-action}, starting with $g_{ab}$, the special \Kah metric in \eqref{eq:Kah-metr-4} for the complexified \Kah moduli $v^a$ in \eqref{eq:compl-Kah} and the corresponding special K\"ahler metric $g_{\alpha \bar \beta}$ in \eqref{eq:Kah-metr-complx} for the CR structure moduli $z^\alpha$ in \eqref{eq:chi-kap}.  Furthermore, the symplectic matrix $\mathcal{M}$ is the one in \eqref{eq:M-matrix}. The gauge field strengths in four dimensions are defined in terms of the vectors $A^I = (A^0, A^a)$, as
\begin{equation} \label{eq:four-dimensional-Fs}
F^I = dA^I - m^I\, B\,,
\end{equation}
and the gauge kinetic matrix $\mathcal N_{IJ}$ is given by
\begin{align}
\Re \,\mathcal N_{00} &=   -\frac{1}{3}\? \mathcal K_{abc} b^a b^b b^c\,,\qquad\quad
\Re \,\mathcal N_{0a} = \frac{1}{2}\?\mathcal K_{abc} b^b b^c\,,\qquad
\Re \,\mathcal N_{ab} = -\mathcal K_{abc} b^c 
,\CR
\Im \,\mathcal N_{00} &=  - \mathcal{K}\?(1+ 4\?g_{ab} b^a b^b)\,,\qquad
\Im \,\mathcal N_{0a} = 4\? \mathcal K\?g_{ab} b^b\,,\qquad
\Im \,\mathcal N_{ab} = -4\? \mathcal K\? g_{ab} \,,
\end{align}
where we used the quantities in \eqref{eq:K-tensor} and \eqref{eq:K-fun}.
The real scalar $\phi$, sometimes called the four-dimensional dilaton, is defined as
\begin{equation} \label{eq:four-dimensionaldilaton}
\ee^{2\phi} = \ee^{3V} \cK^{-1}\,,
\end{equation}
and, together with the two-form $B$, is part of a tensor multiplet.
Finally, the scalar potential $V$ is given by 
\begin{align} \label{eq:four-dimensional-V}
V  \,= &\,\;  -2\,\ee^{\KahJ+2\phi} \Iprod{Q_0 + \Um \xi}{\mathcal{M}(Q_0 + \Um \xi)}
+ 8\,\im\,\ee^{\KahJ + \KahO}  \, \Iprod{ \Um \cV}{\Um \bar\cV} 
\CR
&\, +8\, \ee^{\KahJ + 2\KahO} \Iprod{\cV}{\Um \bar\cV}^2 + 4\,\ee^{\KahJ + \KahO + 2\,\phi} \cK_{abc} m^a t^b t^c\,\Iprod{\cV}{\Um \bar\cV} 
\CR
& \,- \tfrac{1}{4}\ee^{4\phi} \left[ \Im \cN_{IJ} m^I  m^J 
+ \big({\cal E}_I -\Re{\cal N}_{IK} m^K \big)\,(\Im{\cal N})^{-1 IJ} \big({\cal E}_J - \Re{\cal N}_{JL} m^L \big) \right],
\end{align}
where the \Kah potentials $\KahJ$, $\KahO$ for the \Kah and CR deformations were defined in \eqref{eq:K-Kah} and \eqref{eq:Kah-complx} respectively, while the quantities ${\cal E}_I$ are defined as
\begin{equation}\label{defE_I}
{\cal E}_I \,=\, e_I  + \delta_I^0\,\left( \Iprod{Q_0}{\xi} - \tfrac 12\,\Iprod{\xi}{\Um \xi} \right)\,.
\end{equation}

The action \eqref{eq:four-dimensional-action} is consistent with the bosonic sector of gauged $N=2$ supergravity coupled to $\nv$ vector multiplets, $\mu$ hypermultiplets and one tensor multiplet, where the various fields are grouped as follows
\begin{align}\label{eq:matt-cont}
\text{gravity multiplet:}& \qquad\qquad \{\,\, g_{\mu\nu}\,,\hspace{4pt}  A^0_\mu \,\, \}
\CR 
\nv \text{ vector multiplets:}& \qquad\qquad \{ \,\, A^a_\mu\,,  \hspace{4pt} v^a  \,\, \}   
\CR 
\text{ tensor multiplet:}& \qquad\qquad \{ \,\, B\,,  \hspace{4pt} \phi \,,  \hspace{4pt} \xi^{\textsf t} \,,  \hspace{4pt} \xi_{\textsf t}  \,\, \}  
\\
\mu \text{ hypermultiplets:}& \qquad\qquad \{ \,\, z^\alpha\,,  \hspace{4pt} \xi^\alpha \,,  \hspace{4pt} \xi_\alpha \,\, \} \,.\nonumber
\end{align}
Here, $\xi^{\textsf t}$ and $\xi_{\textsf t}$ are the linear combinations of the $(\xi^\La\,, \xi_\La)$ that belong to the tensor multiplet, whose expressions we will not need here. 

In four spacetime dimensions, one can dualise the two-form $B$ to a scalar $a$, thus transforming the tensor multiplet to a hypermultiplet, so that one obtains a system of $\mu+1$ hypermultiplets, with the $\phi$, $a$ and $z^\alpha$ parametrizing a special \Kah manifold and all the $\xi^\La$, $\xi_\La$ on the same footing. In this setting, the hypermultiplets are described by coordinates $q^u = \{\phi, a, z^\alpha, \xi^\Lambda, \xi_\Lambda\}$, with $u=1,\ldots,4\,(\mu+1)$. Since the tensor to be dualised interacts with the vector fields, this requires an electric/magnetic duality operation that also leads to magnetic vector fields \cite{deWit:2005ub,deWit:2011gk}. In the case at hand, this is straightforward and leads to a dual theory where the hypermultiplet dual to the tensor multiplet in \eqref{eq:matt-cont}, along with the $\mu$ other hypermultiplets arising from the three-form, are all charged under the electric and magnetic gauge fields. For the details of this operation, we again refer the reader to \cite{cassani-koerber-varela}.

For the remainder of the section, we concentrate on the theory one obtains after performing this dualisation, so that only uncharged vector multiplets and charged hypermultiplets are present. The resulting kinetic terms of the hypermultiplet scalars are described by the $\sigma$-model quaternionic-K\"ahler metric
\begin{equation} \label{eq:cmapmetric}
h_{uv} dq^u dq^v \,=\,  d\phi^2 + g_{\alpha \bar \beta} dz^\alpha d{\bar z}^{\beta}
+ \tfrac{1}{4} \ee^{4\phi} \left(da + \tfrac{1}{2}\,\Iprod{\xi}{d\xi} \right)^2 -\tfrac{1}{4} \ee^{2\phi} \Iprod{d\xi}{\mathcal{M} d\xi}\,.
\end{equation}
This metric is in the image of the c-map \cite{Ferrara:1989ik}, so that it may be obtained from a set of vector multiplets described by the special \Kah manifold spanned by the $z^\alpha$. This situation is generic and arises naturally in Calabi--Yau compactifications of Type II theories.

The potential \eqref{eq:four-dimensional-V} obtained by the dimensional reduction can be seen as arising from the gauging of isometries in the hypermultiplet sector of $\cN=2$ supergravity, in the presence of the magnetic vector fields mentioned above \cite{deWit:2011gk}. Given that the hypermultiplet target space \eqref{eq:cmapmetric} is in the image of the c-map, there is a variety of isometries that could be possibly gauged from the four-dimensional point of view. In terms of the reduction, we will only require those corresponding to the Killing vectors
\begin{align}\label{eq:Killing-vecs}
k_I^u\frac{\partial}{\partial q^u}  \,
&=\,\delta_I^0 \, \left( k_{\Ums} + Q_0{}^\Lambda h_\Lambda  +  Q_{0\,\Lambda}\, h^\Lambda \right) - e_I\, h \,,
\CR
k^{I\,u}\frac{\partial}{\partial q^u}\, &=\, - \, m^I h\,.
\end{align}
Here, the vectors $k_{\Ums}$, $h_\Lambda$, $h^\Lambda$ and $h$ are defined as follows. We first define 
\begin{equation}
k_{\Ums} \,=\, (\Um \cV)^\Lambda \frac{\partial}{\partial X^\Lambda} + (\Um \bar \cV)^\Lambda \frac{\partial}{\partial \bar X^\Lambda} + (\Um \xi)^\Lambda \frac{\partial}{\partial\xi^\Lambda} + (\Um \xi)_\Lambda \frac{\partial}{\partial \xi_\Lambda}\,,
\end{equation}
as the Killing vector corresponding to the isometry
\begin{equation}
\delta \cV \,=\,  \Um \cV \,,\qquad\qquad \delta \xi \,=\,  \Um \xi\,,
\end{equation}
where we remind the reader that $\cV=(X^\La$\,, $F_\La)^T$. The matrix $\Um$ parametrising this isometry captures the information on the charges of the $\mu$ three-forms, that describe the deformations away from the Sasaki--Einstein internal metric. The remaining Killing vectors are given by
\begin{equation} \label{eq:Heisenberg}
h^\La = \frac{\partial}{\partial\xi_\La} + \frac12\,\xi^\La \frac{\partial}{\partial a} \,, 
\qquad 
h_\La = \frac{\partial}{\partial\xi^\La} - \frac12\,\xi_\La \frac{\partial}{\partial a} \,,
\qquad 
h = \frac{\partial}{\partial a}\,,
\end{equation}
and satisfy the commutation relations of the Heisenberg algebra 
\begin{equation}
[h_A,h^B] \,=\, \delta_A^B\, h\,.
\end{equation}
Finally, we record the triplet of moment maps $( \mathcal  P^x{}^I,\, \mathcal  P^x{}_I)$, for $x=1,2,3$, corresponding to this particular gauging, 
\begin{align}\label{eq:mom-maps}
\mathcal P^1_I + i \mathcal P^2_I \,&=\, \sqrt 2 \, \ee^{\frac{\KahO}{2} + \phi}\, \Iprod{\cV}{Q_0 +  \Um \xi}\,\delta_I^0 ,
\CR
\mathcal P^3_I \,&=\, -\tfrac{1}{2} \,\ee^{2\phi}\left( e_I + \Iprod{Q_0}{\xi}\,\delta_I^0  - \tfrac 12 \, \Iprod{\xi}{\Um\xi}\,\delta_I^0 \right) -\ee^{\KahO} \Iprod{\cV}{\Um\bar \cV}\,\delta_I^0  \,,
\\
\mathcal  P^{3I} \,&=\, - \tfrac{1}{2}\,  \ee^{2\phi}\, m^I \,.\nonumber
\end{align}

\subsection{Moduli space of the \texorpdfstring{AdS$_4$}{AdS4} vacuum and \KE space metrics}
\label{sec:vac-moduli}

We now turn to a discussion of the possible moduli spaces arising for the universal $\cN=2$ AdS$_4$ vacuum associated to the class of regular Sasaki--Einstein manifolds analysed in the previous sections. In the context of four-dimensional $\cN=2$ supergravity, the structure of the possible AdS$4$ vacua has been explored in a number of papers, see e.g. \cite{Hristov:2009uj, louis-smyth-triendl, deAlwis:2013jaa, Erbin:2014hsa}. We refer to these works for more details on the derivation of the BPS vacuum conditions and the conditions on the possible moduli spaces, restricting attention to the implications of these results for the compactifications considered in this paper.

In terms of the quantities defined in the previous subsection, the relevant BPS conditions for an AdS$4$ vacuum are given by
\begin{gather}
Z^I \? k^u_{\? I}  - G_I\? k^{u\?I} =0\,,  
\label{eq:BPS-AdS4-1}
\\
2\,R^{-1}\?\Im\!\left( \ee^{-\im\?\alpha}\binom{Z^I}{G_I} \right) 
= \binom{\mathcal{P}^{x\?I}}{\mathcal P^x_{\? I}}\? \ee^x\,, \rule{0pt}{5ex}
\label{eq:BPS-AdS4-2}
\end{gather}
where $R$ stands for the radius of AdS$_4$ and $\ee^x$ is an arbitrary unit vector on S$^2$. The $k^u$ are the Killing vectors of the hypermultiplet target space in \eqref{eq:Killing-vecs} that are being gauged, while the $Z^I$, $G_I$ are the standard projective variables for the \Kah moduli defined below \eqref{eq:Kah-prepotential}. We restrict the gaugings in \eqref{eq:Killing-vecs} by setting both $e_a=0$ and $Q_0=0$. The former holds for the examples considered in this paper, since they admit a single  $(1,1)$-form, while we disregard the possibility of a $Q_0$ along the null directions of $\Um$ for simplicity (see the comments below \eqref{eq:F4-ansatz}).

In this setting, we can assume that the prepotentials have been aligned along the third direction, i.e.~$\mathcal P^1_{\? I} = \mathcal P^2_{\? I}=0$ and we set the the unit vector $\ee^x=\delta^x_3$ in \eqref{eq:BPS-AdS4-2} above\footnote{This is not feasible in general, but can be done for the examples considered here \cite{Erbin:2014hsa}.}. This allows to solve half of the conditions in \eqref{eq:BPS-AdS4-1} through the restriction $\xi^\La = \xi_\La =0$, implying that only the scalars in the special \Kah base of the hypermultiplet sector are relevant in the discussion.

It is useful to recognise that \eqref{eq:BPS-AdS4-2} is identical to the attractor equation for static asymptotically flat BPS black holes, with the $({\mathcal{P}^{3\?I}}\,, {\mathcal P^3_{\? I}}\? )$ in place of the charges. One can then readily write down its solution \cite{shmakova} for the $Z^I$, $G_I$, whose explicit form is not needed here. We only record the resulting expression for the radius of AdS$_4$ in terms of the remaining hypermultiplet scalars
\begin{equation}\label{eq:R-AdS4-n}
R^{-4} 
= -\tfrac23\? \cK_{abc}\mathcal  P^{3\?a}\?\mathcal  P^{3\?b}\?\mathcal  P^{3\?c} \? \mathcal  P^3_{\? 0} 
= -\tfrac1{24}\,\ee^{6\phi}\, \cK_{abc} m^a\?m^b\?m^c \,\left( \ee^{2\phi} e_0 + 2\?\ee^{\KahO} \Iprod{\cV}{\Um\bar \cV} \right)\,,
\end{equation}
where we used the explicit expressions in \eqref{eq:Killing-vecs} and \eqref{eq:mom-maps}.  
We then return to \eqref{eq:BPS-AdS4-1}, which simplifies to
\begin{gather}\label{eq:BPS-AdS4-n}
3\,\left( \ee^{2\phi} e_0 + 2\?\ee^{\KahO} \Iprod{\cV}{\Um\bar \cV} \right)\, h
+\,\ee^{2\phi}\, \left( e_0\, h- k_{\Ums} \right)
=0\,.
\end{gather}
To obtain a vacuum, one has to set to zero all linearly independent components of this equation. Assuming a finite dilaton, the component along $k_{\Ums}$ can only vanish if one sets $k_{\Ums}=0$, which specifies the scalars contained in $\cV$. One can then solve for the vacuum expectation value of the dilaton $\phi$ from the component of \eqref{eq:BPS-AdS4-n} along $h$. 

The general solution for a vanishing $k_{\Ums}$ is a priori complicated, since it would depend on the details and chosen parametrisation of the matrix $\Um$ and the holomorphic prepotential for the special \Kah base of the hypermultiplet sector. However, in this discussion we are only interested in the maximally supersymmetric AdS$_4$ vacuum, which is described by tuning moduli so that the eigenvalue condition
\begin{equation}\label{eq:AdS4-vac}
\Um \cV = -4\?\im\?\cV\,,
\end{equation}
is met. This corresponds to the Sasaki--Einstein point for the internal manifold, as one can compute that the derivative of the three-form in \eqref{eq:SU3-str} reduces to the undeformed result in \eqref{eq:SE-point}, as 
\begin{equation}
 d\Omega = \ee^{-\frac{3}{2}\,V}\Iprod{\cV}{d\Si} = \ee^{-\frac{3}{2}\,V}\eta\wedge\Iprod{\cV}{\Um\?\Si} = 4\?\im\?\eta\wedge\Omega \,,
\end{equation}
where we used \eqref{eq:dab-M7} and \eqref{eq:AdS4-vac}. Inserting \eqref{eq:AdS4-vac} in \eqref{eq:R-AdS4-n}--\eqref{eq:BPS-AdS4-n} leads to the following values for the dilaton and the AdS$_4$ radius
\begin{equation}\label{eq:dil-R-AdS4-vac}
\ee^{2\phi} = \frac{6}{e_0}\,, \qquad R^{-4} = \tfrac1{12}\,\cK_{abc} m^a\?m^b\?m^c \,\left( \frac{6}{e_0} \right)^{3}\,.
\end{equation}

The fact that \eqref{eq:AdS4-vac} leads to an AdS$_4$ solution while keeping the parametrisation for the scalars implicit, allows to easily characterise the moduli of this vacuum. Indeed, we are guaranteed to have at least one solution to \eqref{eq:AdS4-vac} by construction, since there is at least one pair of eigenvalues with this value, corresponding to the indices labelled by '0' in \eqref{eq:Lie-alpha-beta}, coming from \eqref{eq:Lie-omega-se}. However, the multiplicity of this eigenvalue can be higher, with the corresponding additional directions being unconstrained by \eqref{eq:AdS4-vac}, thus corresponding to exact moduli of the vacuum. The expressions \eqref{eq:dil-R-AdS4-vac} remain true for any choice of such moduli, since only \eqref{eq:AdS4-vac} was used in their derivation.

Referring to the examples discussed in sections \ref{sub:hypersurf} and \ref{sec:singularities}, one can see from Table \ref{tab:jacobi} on page~\pageref{tab:jacobi} that there exist twisted Beltrami's with $k=0$ for the hypersurfaces with $d=3,4$, which arise in Figure \ref{fig:fermat} as three-forms carrying exactly the same charge as the corresponding $\Omega_0$ in each case. These are characterised as elements of the Jacobi ring which are of the same degree under the $\mathbb{C}^*$ action as the defining function.\footnote{In the case of the quadric, for $d=2$, there are clearly no such elements, since the only element of the Jacobi ring is the constant.} Explicitly, for $d=3$ one may write down the deformation
\begin{equation}
f(z) + \sum_{i\neq j\neq k}t^{ijk}\?z_i\? z_j\? z_k \,,
\end{equation}
for any 10 constants $t^{ijk}$, which is the part of the universal unfolding that preserves the degree of the function. Similarly, Table \ref{tab:jacobi} shows that for the quartic hypersurface there are 45 elements of the Jacobi ring that are quartic in the coordinates, corresponding to an equal number of three-forms carrying the same charge as $\Omega_0$, in Figure \ref{fig:fermat}. The same comments apply for the deformations of the cone discussed in section \ref{sec:singularities}, implying that the same number of eigenvalues as in \eqref{eq:Lie-defs-fin} are $\lambda_\alpha=4$, matching with the one for $\Omega_4^0$ in \eqref{eq:Lie-omega}. We therefore conclude that the eigenspace of the matrix $\Um$ in \eqref{eq:AdS4-vac} is 11- and 46-dimensional for the cubic and quartic hypersurfaces, respectively. It follows that the AdS$_4$ vacuum has 10 and 45 moduli in each case, which can be viewed as moduli of the \KE base of the internal Sasaki--Einstein manifold. This situation is exactly the same as for hypersurface Calabi--Yau manifolds, whose complex structure deformations are precisely of the type described here, since they correspond to deformations of the defining polynomial by homogeneous terms of the same degree as the defining polynomial, leading to a three-form of the same charge and the same Ricci form (both vanishing in the CY case).

A priori, from a geometrical point of view, it might not be obvious that a complex deformation induced by the zero-charge Beltrami differentials discussed above preserves the K\"ahler--Einstein condition on $M_6$. Indeed, after deforming $\Omega$, the metric changes, and it might no longer satisfy the Einstein condition. However, results in \cite{szekelyhidi} can be used to show that first-order deformations preserve the K\"ahler--Einstein condition if and only if they are polystable with respect to the action of the automorphisms of $M_6$; since in our case this is a finite group, all deformations are polystable, and the K\"ahler--Einstein condition is preserved.\footnote{\label{foot:szek}We thank J.~Stoppa for this argument.} This is in agreement with our finding that there are as many moduli as uncharged Beltrami differentials.

\subsection{Comments on the consistency of the reduction}
\label{sec:physical}

In this section, we comment on the consistency of our M-theory compactification on deformed regular Sasaki--Einstein manifolds, that only retains the modes described in section \ref{sec:CR-struct}. These modes are uniquely distinguished as arising by reduction of the harmonic forms on an asymptotically conical manifold, described as a deformation of the cone over the Sasaki--Einstein manifold. The deformation space of these non-compact manifolds has well-studied and rich properties, and is known to admit a flat structure on its tangent space, similar to the known structures for compact Calabi--Yau manifolds. While these properties are essential in defining a sensible compactification Ansatz as discussed in section \ref{sub:gen}, to the best of our understanding, they are not strong enough to establish consistency of the truncation to these modes, unless extra symmetry is present (e.g. in the case of the quadric hypersurface, for $d=2$ in \eqref{eq:fermat}, which is a coset space).

Our compactification Ansatz is explicitly based on a \emph{fixed contact class}, i.e.~a one-form on $M_7$ that is identified up to rescaling with the Reeb vector of the undeformed Sasaki--Einstein manifold. This is natural in the context of deformations of CR structures, and allows to treat this family of deformations similar to the compact Calabi--Yau case. The implications of such a structure have been explored around the Sasaki--Einstein point, in which case one can explicitly show that the four-dimensional spectrum is organised in terms of forms on $M_7$ of definite charge, along the base and the contact structure \cite{eager-schmude-tachikawa, eager-schmude, schmude}. More concretely, restricting to primitive forms, the Laplacian operator acting on a $p$-form on a Sasaki--Einstein space is given by \eqref{eq:Lapl-SE}, where $\Delta_\stgdb$ stands for the Kohn--Rossi Laplacian. This decomposition allows to describe the spectrum of the Laplacian in terms of the Kohn--Rossi complex. The forms used in the reduction Ansatz in this section are precisely the modes for which $\Delta_\stgdb = 0$, extended to a general CR structure away from the Sasaki--Einstein point. While we have not constructed an explicit analogue of \eqref{eq:Lapl-SE} in the general case, similar techniques can be applied, see \cite[Th. 1.19]{dragomir-tomassini} for a partial result.

These observations lead to an intuitive picture for the spectrum of the Laplacian on $M_7$, with eigenmodes labelled by their eigenvalues under $\Delta_\stgdb$ and $\pounds_{\xi}$. We have shown that there is a finite number of $\pounds_{\xi}$ eigenmodes for the lowest eigenvalue along the base, i.e.~for $\Delta_\stgdb=0$, for any point in our family of CR structures, while we expect the same to hold for higher $\Delta_\stgdb$ eigenvalues\footnote{The eigenvalues of $\Delta_\stgdb$ are discrete by similar arguments as for compact complex manifolds, while the gap to the first nonzero eigenvalue is known to depend only on the contact class \cite[Ch. 9]{dragomir-tomassini}} as well, by similar arguments.
This is again parallel to the situation for M-theory compactifications on CY$_6\times S^1$, where one may consider compactifications using forms with vanishing eigenvalue of the Laplacian along the Calabi--Yau space CY$_6$, but allow for a twist along the circle \cite{Aharony:2008rx, Looyestijn:2010pb}. In this case, one finds a finite range of eigenvalues for the forms in each eigenspace of the CY$_6$ Laplacian.

In addition, we note that the modes arising for the vanishing eigenvalue of $\Delta_\stgdb$ naturally combine in short multiplets from the point of view of four-dimensional supergravity, while higher eigenvalues lead to long multiplets. This property was crucial for matching the computation of the index to the dual CFT in \cite{eager-schmude} at the Sasaki--Einstein point. The extension of these structures away from the vacuum is a nontrivial property, that is not a priori obvious for a general AdS$_4$ supergravity. One could then hope that a rearrangement of the supergravity modes based on the KR Laplacian, rather than the full internal Laplacian, may be more natural both from the point of view of supergravity and the dual CFT, despite the lack of a clear separation of scales. We hope to return to some of these points in the future.

\section*{Acknowledgements}
We would like to thank A.~Belavin, G.~Bossard, D.~Cassani, M.~Gra\~na, A.~Grassi, A.~Kashani-Poor, D.~Morrison, C.~Strickland-Constable, D.~Waldram and A.~Zaffaroni for interesting discussions; we especially thank A.~Dimca, C.~Sabbah and J.~Stoppa for invaluable help with some mathematical aspects of this work. S.K.~and A.T.~were supported in part by INFN and by the European Research Council under the European Union's Seventh Framework Program (FP/2007-2013) -- ERC Grant Agreement n. 307286 (XD-STRING). The work of S.K.~was further supported in part by the John Templeton Foundation Grant 48222, by the KU Leuven C1 grant ZKD1118 C16/16/005, by the Belgian Federal Science Policy Office through the Inter-University Attraction Pole P7/37, and by the COST Action MP1210 The String Theory Universe.  The research of A.T.~was also supported by the MIUR-FIRB grant RBFR10QS5J ``String Theory and Fundamental Interactions''.

\appendix

\section{Sasaki--Einstein manifolds arising from complete intersections} 
\label{app:ex}

In this appendix, we make a few comments on the extension of the results discussed in section \ref{sec:CR-cone} for hypersurfaces in $\mathbb{C}^5$, to the more general case of complete intersections of quasihomogeneous polynomials in some higher-dimensional $\mathbb{C}^N$. The various considerations in section \ref{sec:CR-cone} may be followed in an analogous but more involved manner, leading to similar results, see \cite{Ebeling} for an extended list of references. In particular, the monodromy operator acting on the $\mu$-dimensional cohomology is again semisimple with eigenvalues given by the exponents of the Poincar\'e polynomial generalising the spectral polynomial \eqref{eq:spec-pol}. More concretely, for $s$ quasihomogeneous polynomials $f_r$ of degree $d_r$, with $r=1,\,\dots s$ and weights $w_i$ for the coordinates $z_i$ of $\mathbb{C}^N$, the Poincar\'e polynomial $P(T)$ is computed as \cite{Hamm1978}
\begin{equation}\label{eq:poin-pol}
 P(T)
 = \frac{1}{T^2}\,\mathsf{res}_{q=0}\frac{q^{s-N-1}}{q+1}\,\left[ 
 \prod_{i=1}^{N}\frac{1 + q\,T^{w_i}}{1-T^{w_i}}\,
 \prod_{r=1}^s\frac{1-T^{d_r}}{1 + q\,T^{d_r}} + q \right]
 = \sum_{[\alpha]}\, b_{[\alpha]}\, T^{[\nu_\alpha]}\,,
\end{equation}
which reduces to \eqref{eq:spec-pol} for $s=1$ and $w_i=1/d_i$. This explicit formula doubles as a convenient way of computing the Milnor number, given by $\mu=P(1)$, which is a slightly more involved task in the case of complete intersections; see \cite[Thm.~1]{randell} for a direct method. The exponents $\nu_\alpha$ in the last expression in \eqref{eq:poin-pol} again correspond to the eigenvalues of the monodromy operator and specify the action of the Reeb vector on the cohomology elements as in \eqref{eq:Lie-defs}. The residue construction of representatives in \eqref{eq:top-hol-res}--\eqref{eq:Oa} can be followed for multiple polynomials $f_r$, but we refrain from giving any details for the general case.

Of course, not all quasi-homogeneous complete intersections are conical Calabi--Yau's, and thus they are not all cones over Sasaki--Einstein manifolds. However, there are several examples where this is the case. Some very old examples are the ones in \cite{nadel}: three Fermat quadrics in $\cc^7$, two quadrics in $\cc^6$, a quadric and a cubic in $\cc^6$ (the latter two with certain specially chosen coefficients). Some examples of \emph{finite} deformations of the quartic in $\cc^5$ are given in \cite{arezzo-ghigi-pirola}; see also \cite{dervan}. Some more modern examples, e.g.~the ones in \cite{suess}, could also admit a description as complete intersections.

\section{Gibbons--Hawking spaces} 
\label{app:GH}

In this appendix, we collect some useful facts for Gibbons--Hawking metrics for four-dimensional $A_k$ singularities, which represent a simple example class of hypersurface singularities that can be treated exactly. The relevant function defining these hypersurface singularities is given by setting $n=2$ in \eqref{eq:Ak-def} as
\begin{equation} \label{eq:Ak-sing}
f_{\scriptscriptstyle\sf A_\mu} = z_1^{\mu+1} + z_2^{2} + z_3^{2} \,,
\end{equation}
where we have traded the parameter $k$ for the Milnor number $\mu$. The resulting cone
\begin{equation}
 M_0 = \{\, z \in \mathbb{C}^{3}\quad |\,\, f_{\scriptscriptstyle\sf A_\mu}= 0 \, \}\,,
\end{equation}
has an isolated conical singularity at $z_i=0$. It is well-known that $M_0 \cong \mathbb{C}^{2}/\mathbb{Z}_{\mu+1}$, which is made manifest by the relevant metric, that can be written as
\begin{equation}\label{eq:con-GH}
ds^2 = \frac{r}{\mu+1}\, (d\psi + \mu\,\cos\theta\, d\phi)^2 + \frac{\mu+1}{r}\,\left( dr^2 + r^2 d\theta^2 + r^2 \sin^2\theta \,d\phi^2 \right)\,.
\end{equation}
This is exactly of the conical type, with base given by the quotient $S^{3}/\mathbb{Z}_{\mu+1}$. The deformations of the singularity \eqref{eq:Ak-sing} are given by the unfolding \eqref{eq:unfolding}, in terms of the Jacobi ring basis
\begin{equation}\label{eq:GH-Milnor}
 \jac_{\scriptscriptstyle\sf A_\mu} = \{1\,, z_1\,, z_{1}^2\,, \dots \,, z_1^{\mu-1} \}\,,
\end{equation}
leading to a deformed manifold
\begin{equation}
 M = \{\, z \in \mathbb{C}^{3}\quad |\,\,\, F_{\scriptscriptstyle\sf A_\mu}= 0 \, \}\,,
\end{equation}
which features $\mu$ topological $S^2$'s. The associated metrics can be written down explicitly for this special class, owing to its hyper-\Kah property, and are given as a special case of the  Gibbons--Hawking multi-instantons \cite{Gibbons:1979zt}, as
\begin{equation}\label{eq:def-GH}
ds^2 = V^{-1}\, (d\psi + \chi)^2 + V\,\left( dr^2 + r^2 d\theta^2 + r^2 \sin^2\theta \,d\phi^2 \right)\,,
\end{equation}
where 
\begin{equation}
V = \sum_{I=1}^{\mu+1} \frac{1}{r_I}\,, \qquad d\chi = *_3 dV\,,
\end{equation}
Here, $r_I\equiv || \vec x- \vec x_I ||$ denotes the distance in $\mathbb{R}^3$ from the point at $\vec x = \vec x_I$, labelled by the index $I$, running over $\mu+1$ such points. We refer to these points as centres, while we note that the $\mu$ topological spheres are defined by the fibration of the coordinate $\psi$ over the collection of lines between the centres. The metric \eqref{eq:def-GH} is only asymptotically conical, asymptoting to \eqref{eq:con-GH} for large distances and reduces to it if all the centres coincide.\footnote{Note that the relative positions of the centres correspond to $3 \mu$ parameters, while the unfolding of the singularity along the Milnor ring in \eqref{eq:GH-Milnor} is parametrised by $\mu$ complex parameters. The additional $\mu$ parameters correspond to \Kah deformations, as the $3 \mu$ position parameters altogether describe deformations of the complete hyper-\Kah structure triplet $J^x$ for $x=1,2,3$. Taking $J^3$ as the \Kah form, one can think of the parameters of deformations in \eqref{eq:GH-Milnor} as the positions of the centres on the plane along the other two directions.} 

Following e.g. \cite{Franchetti:2014lza}, one may write down the $\mu$ cohomology representatives, which can be chosen to definite selfduality in this case. We define the following sets of two-forms
\begin{align}\label{eq:GH-ab-forms}
\alpha_I =&\, V\,*_3 d\left( \frac{1}{r_I} \right) - (d\psi + \chi)\wedge d \left( \frac{1}{r_I} \right) \, ,
\CR
\beta^I =&\, V\,*_3 d\left( \frac{1/r_I}{V} \right) + (d\psi + \chi)\wedge d\left( \frac{1/r_I}{V} \right) \,,
\end{align}
which are explicitly anti-selfdual and selfdual respectively with respect to the metric \eqref{eq:def-GH}, as
\begin{equation}\label{eq:GH-dual}
* \alpha_I = - \alpha_I \,, \qquad  * \beta^I = \beta^I \,,
\end{equation}
for each index $I$ separately. Of these pairs of forms, only $\mu$ are linearly independent, since
\begin{equation}
\sum_{I=1}^{\mu+1} \alpha_I =\sum_{I=1}^{\mu+1} \beta^I = 0 \,.
\end{equation}
while one may relate one set to the other by exact pieces, so that only $\mu$ forms are nontrivial in cohomology. In principle, one may choose any $\mu$ forms out of the $2\?(\mu+1)$ forms $\alpha_I$, $\beta^I$ to represent the cohomology group $H^4(M)$. 
However, there is a clear distinction between the two sets of forms in \eqref{eq:GH-ab-forms}, in terms of the de Rham and relative cohomology we reviewed in section \ref{sub:fixed}. 

To see this, let us first define some notable cycles. As mentioned above, one can obtain compact $S^2$ cycles by considering the total space the U(1) fibration (spanned by $\psi$) over a segment joining two centers. For example we can obtain compact cycles $A_\alpha$, $\alpha=1,\ldots,\mu$, by considering the segment from the $\alpha$-th to the $(\alpha+1)$-th center. One can also obtain ``co-compact'' cycles $B^\alpha$ by considering a plane $\rr^2\subset\rr^3$ that meets the segment from the $\alpha$-th to the $(\alpha+1)$-th center exactly once, and choosing a lift of this plane from $\rr^3$ to $M_4$.\footnote{\label{foot:lift}For example we can choose this lift so that the boundary of the $\alpha$-th plane lifts to an $S^1$ that winds $\alpha$ times around the U(1) fibre.} By construction we have that (\ref{eq:ABd}) holds. 

Just as described in section \ref{sub:fixed}, the $A_\alpha$ are both in $H_2(M)$ and in $H_2(M,\del M)$, while the $B^\alpha$ are in $H_2(M,\del M)$, but not in $H_2(M)$ (since they have a boundary on $\del M$). Among the forms in (\ref{eq:GH-ab-forms}), the $\beta^I$ span $H^2(M)$ but do not belong to $H^2(M,\del M)$, since they reduce to forms describing Dirac monopoles on the boundary, as one can see be rewriting them as
\begin{equation}
\beta^I = *_3 d\left( \frac{1}{r_I} \right) - d\left(  \frac{1/r_I}{V}\, (d\psi + \chi) \right)\,.
\end{equation}
We can take the $\mu$ linearly independent $\beta^I$, to be a basis for $H^2(M)$, as is standard in the literature. On the other hand, one can show that appropriate linear combinations of the $\alpha_I$ with the $\beta^I$ reduce to exact forms $d\left( \lambda_I(d\psi + \chi) \right)$ for appropriate functions $ \lambda_I$, so that they are both in $H^2(M)$ and $H^2(M,\del M)$. We can take $\mu$ of these forms to be a basis for either of the two cohomologies.

Thus in $H_2(M,\del M)$ we have two competing bases.
One can find a topological relation between them as follows. First consider the sphere $S^\alpha\subset \rr^3$ surrounding the $\alpha$-th center. Since the U(1) fibration is topologically non-trivial, $S^\alpha$ cannot be lifted to a two-cycle in the manifold $M_4$. If one tries for example to lift each ``meridian'' of $S^\alpha$ to $M_4$, one will see that the lift of (say) the south pole of $S^\alpha$ will become ambiguous. In other words, the $S^\alpha$ will lift to a chain $\tilde S^\alpha$ with a boundary which can be taken to be the entire U(1) fibre over the south pole of $S^\alpha$. 

On the other hand, consider a segment $s_\alpha$ joining the center $\alpha$ to the south pole of $S^\alpha$, together with the U(1) fibre over it. The resulting total space $\pi^{-1} s_\alpha$ is again a chain in $M_4$, whose boundary is the U(1) fibre over the south pole of $S^\alpha$, which exactly the same as the boundary of the chain $\tilde S^\alpha$ we described above. In other words, $\tilde S^\alpha=-\pi^{-1} s_\alpha$ in homology. 

Now observe that the sphere $S^\alpha$ can be deformed to the difference of two planes $B_\alpha-B_{\alpha-1}$. Moreover, a cycle $A_\alpha$ is a segment from center $\alpha$ to $\alpha+1$ together with the U(1) fibre over it; so it can also be considered as the difference $\pi^{-1} s_\alpha- \pi^{-1} s_{\alpha+1}$. Thus in homology we can write
\begin{align}
	A_\alpha = &\, \pi^{-1} s_{\alpha} -\pi^{-1} s_{\alpha+1}=  -\tilde S^{\alpha}+ \tilde S^{\alpha+1} 
	\cr
	= &\,	-(B_{\alpha}- B_{\alpha-1})- (B_{\alpha+1}- B_{\alpha}) = B^{\alpha-1}  - 2 B^\alpha + B^{\alpha+1}\ .
\end{align}
In other words,
\begin{equation}
	A_\alpha = C_{\alpha \beta} B^\beta\,,
\end{equation} 
where $C$ is minus the Cartan matrix for ${\rm SU}(\mu+1)$, which is of rank $\mu$. This realizes (\ref{eq:ACB}) in this particular case. (One should of course not think that $C_{\alpha \beta}$ in (\ref{eq:ACB}) is always proportional to the Cartan matrix for a Lie group.) 

It is also easy to describe the variation operator. Monodromy in this case just shuffles the $\mu+1$ centers around. So it takes $\mathfrak{m}:B^\alpha \to B^{\alpha+1}$, for $\alpha=1,\ldots, \mu-1$. For $\alpha=\mu$ we should be more careful: the plane $B_\mu$ is taken to a plane that does not go between the centers. This would seem to be a trivial cycle, but in fact it is not because of the non-triviality of the U(1) fibration over it, by consistency with the choice in footnote \ref{foot:lift}. Taking this into account, one can see that in fact $\mathfrak{m}:B_\mu \to - \sum_{\alpha=1}^\mu B_\alpha$. If we now recall the definition in (\ref{eq:var}) of ${\rm Var}\equiv \mathfrak{m}-{\rm Id}$, we see that in this case
\begin{equation}
	{\rm Var} B^\alpha = \tilde v^\alpha{}_\beta B^\beta \, ,\qquad \tilde v^\alpha{}_\beta = - \delta^\alpha_\beta + \left(\begin{array}{ccccc}
		0 & 0 & \cdots & & -1 \\
		1 & 0 & 0 &\cdots &-1 \\
		0 & 1 & 0 & \cdots &\vdots\\
		\vdots & & \ddots &  & \vdots \\
		0 &\cdots & 1& 0 &-1\\
		0 &\cdots &  0 &1 & -1 
	\end{array} \right)\,.
\end{equation}
This is not yet the matrix $v^{\alpha \beta}$ in (\ref{eq:var}): to obtain that one we need to reexpress it in terms of the $A_\alpha$. This can be done by multiplying $\tilde v^\alpha{}_\beta$ by $C^{-1}$. One thus obtains
\begin{equation}
	{\rm Var} B^\alpha = v^{\alpha\beta} C_{\beta \gamma} B^\gamma = v^{\alpha \beta} A_\beta \, ,\qquad v^{\alpha\beta} =   \left(\begin{array}{ccccc}
		1 & 1 & \cdots & 1 & 1 \\
		0 & 1 & 1 &\cdots &1 \\
		\vdots & & \ddots &  & \vdots \\
		0 &\cdots & 0& 1 &1\\
		0 &\cdots &   &0 & 1 
	\end{array} \right)\,.
\end{equation}
It then turns out that 
\begin{equation}
	v^{-1}_{\alpha \beta} = \left(\begin{array}{ccccc}
		1 & -1 & 0 &\cdots & 0  \\
		0 & 1 & -1 &\cdots & 0 \\
		\vdots & & \ddots &  & \vdots \\
		0 &\cdots & 0& 1 &-1\\
		0 &\cdots &   &0 & 1 
	\end{array} \right)\,.
\end{equation}
Hence $v^{-1}+ (v^{-1})^t = C$ (which in our case is minus the Cartan matrix of ${\rm SU}(\mu+1)$, just as stated in (\ref{eq:vvC}).


\bibliographystyle{JHEP}\bibliography{CalSas} 

\end{document}